\documentclass[aps,pra,superscriptaddress,twocolumn,floatfix,longbibliography,10pt]{revtex4-1}
\usepackage{dcolumn}
\usepackage{graphicx,color}
\usepackage{amsmath,amssymb,bm,amsfonts,dsfont,mathrsfs,amsthm}
\usepackage{braket}
\usepackage{txfonts,comment}
\usepackage[colorlinks=true, allcolors=blue!62!black]{hyperref}

\usepackage{braket}          % Bra-Ket notation
\usepackage{subcaption}
\usepackage{makecell}
\usepackage{orcidlink}       % ORCID iD links
\usepackage[most]{tcolorbox} % For boxed environments
\usepackage{lipsum}          % Dummy text
\usepackage{multirow}
\usepackage{ragged2e}
\usepackage[percent]{overpic}
\usepackage{ulem}
\usepackage{booktabs}
\usepackage{tikz}
\usepackage{cancel}
\usetikzlibrary{calc}
\newcommand{\Upomega}{\rm \Omega}

\makeatletter
\long\def\@makecaption#1#2{%
	\vskip\abovecaptionskip
	\sbox\@tempboxa{#1: #2}%
	\ifdim \wd\@tempboxa >\hsize
	{\rightskip=0pt plus 1fil\justifying\noindent #1: #2\par}
	\else
	\global\@minipagefalse
	\hb@xt@\hsize{\hfil\box\@tempboxa\hfil}%
	\fi
	\vskip\belowcaptionskip}
\makeatother



\begin{document}

	\title{ \Large
		Quantum extreme learning for forecasting experimental chaotic electronic circuits}
	%Quantum extreme learning for forecasting experimental chaotic single-transistor oscillators}
%Quantum extreme learning for experimental chaos prediction in single-transistor oscillators}
\author{Mohammad Sharifian\,\orcidlink{0000-0001-9801-239x}
}
\thanks{These authors contributed equally to this work.}
\affiliation{Institute of Fundamental and Frontier Sciences, University of Electronic Sciences and Technology of China, Chengdu 611731, China}
\affiliation{Key Laboratory of Quantum Physics and Photonic Quantum Information, Ministry of Education, University of Electronic Science and Technology of China, Chengdu 611731 , China}

\author{Qingyu Li\,\orcidlink{0000-0003-0269-8388}
}
\thanks{These authors contributed equally to this work.}
\affiliation{Institute of Fundamental and Frontier Sciences, University of Electronic Sciences and Technology of China, Chengdu 611731, China}
\affiliation{Key Laboratory of Quantum Physics and Photonic Quantum Information, Ministry of Education, University of Electronic Science and Technology of China, Chengdu 611731 , China}

\author{Chiranjib~Mukhopadhyay\,\orcidlink{0000-0002-4486-9061}
}
\affiliation{Institute of Fundamental and Frontier Sciences, University of Electronic Sciences and Technology of China, Chengdu 611731, China}
\affiliation{Key Laboratory of Quantum Physics and Photonic Quantum Information, Ministry of Education, University of Electronic Science and Technology of China, Chengdu 611731 , China}  

\author{Abolfazl Bayat\,\orcidlink{0000-0003-3852-4558}
}
\email[]{abolfazl.bayat@uestc.edu.cn}
\affiliation{Institute of Fundamental and Frontier Sciences, University of Electronic Sciences and Technology of China, Chengdu 611731, China}
\affiliation{Key Laboratory of Quantum Physics and Photonic Quantum Information, Ministry of Education, University of Electronic Science and Technology of China, Chengdu 611731 , China}
\affiliation{Shimmer Center, Tianfu Jiangxi Laboratory, Chengdu 641419, China}
%\date{\today}

\author{Ludovico Minati\,\orcidlink{0000-0002-2532-1674}
}
\email[]{lminati@uestc.edu.cn}
\affiliation{School of Life Science and Technology, University of Electronic Science and Technology of China, Chengdu 611731, China}
\affiliation{Department of Physics, University of Trento,  Trento 38123, Italy}
\affiliation{Nano Sensing Research Unit, Institute of Innovative Research, Institute of Science Tokyo, Yokohama 226-8503, Japan}

\begin{abstract}
Quantum reservoir computing and its special case, quantum extreme learning, are powerful tools for predicting classical discrete time-series with quantum devices. Forecasting continuous signals, however, requires careful discretization as oversampling artificially boosts short-term accuracy but fails over realistic intervals, while undersampling captures too few features and yields poor predictions.   Here, we experimentally generate four continuous time-series signals from a chaotic electronic circuit for forecasting with quantum extreme learning. Since simplified circuit models do not fully capture real behaviour, it is unclear whether algorithms that predict simulated time-series also work for experimental signals. We address discretization via two approaches: (i) sampling local peaks/troughs; and (ii) uniform sampling at the autocorrelation timescale. The former captures resonant behaviour, while the latter fully  captures chaotic features as well as allowing the reconstruction of chaotic attractors. Our algorithm achieves high forecasting accuracy for both simulated and experimental signals, demonstrating its power for continuous signal processing on near-term quantum simulators, and remains robust against moderate decoherence.
\end{abstract}

\maketitle

\section*{Introduction}

In the absence of fault-tolerant quantum computers, developing applications  for near-term quantum devices is a priority subject in quantum technologies~\cite{Bharti_2022}. In this context, variational quantum algorithms~\cite{peruzzo2014variational,mcclean2016theory,o2016scalable,yuan2019theory,endo2020variational,Cerezo_2021} have been developed with diverse applications in condensed matter physics~\cite{jones2019variational, Li_2023,aliverti2024can, Brozzi_2026,nakanishi2019subspace,lyu2023symmetry,han2024multilevel}, machine learning~\cite{wu2021scrambling, Ding_2024,wu2023quantum, Du_2026, wu2024randomness,Sharifian2025hybrid,ren2022experimental,chen2026ai,hu2019quantum,li2024ensemblelearning} and optimization~\cite{farhi2014quantumapproximateoptimizationalgorithm, Zhou_2020}. However, the training of such algorithms is very resource demanding  and their scalability is limited due to barren plateau~\cite{McClean_2018}. In a different approach, Quantum Reservoir computing (QRC) has emerged as a promising algorithm in which barren plateau issue is prevented by avoiding gradient-based optimizations in the training stage~\cite{fujii2017harnessing,ghosh2021quantum,kobayashi2024feedback, das2026quatum, zia2025quantum, Sannia_2026, Palacios_2024,Mujal_2023, Xia_2023,suzuki2022natural, pena2023quantum,askari2025spinnetworkquantumreservoircomputing,tariq2026quantumreservoircomputingrecent,Spagnolo_2022,paparelle2026experimental,xia2022reservoir,hou2026highaccuracy,xia2023configured}. In these algorithms, input data can be inserted into quantum state of a multi-qubit system in sequential steps, each separated by quantum evolution. Measurement is performed only at the last stage and the outcomes are used in linear regression for obtaining the quantity of interest. Quantum Extreme Learning (QEL) is a special case of  quantum reservoir computing in which quantum evolution  only happens in one stage~\cite{Innocenti_2023,xiong2025fundamental, de2025harnessing, Suprano_2024,  zia2025quantum,Solanki_2026, joly2025harnessingphotonindistinguishabilityquantum, de2026entanglement,Assil_2026,cao2026theory}. Its relative simplicity is currently generating potential applications in diverse domains \cite{wang2024application,Maier_2026}.

QRC is increasingly being viewed as an attractive algorithm for time-series forecasting~\cite{kobayashi2024feedback,Vetrano_2025,li2026quantumchaotic, li2026quantumrv,sannia2026temporalinformationprocessing4500qubit, hou2026high,xia2022reservoir,hou2026highaccuracy}. Most of the forecasting protocols have been developed for discrete time-series in which the past data can directly be encoded on individual qubits. In contrast, for continuous signals, while the underlying time-series is continuous, the encoding must be performed on a discrete set of qubits.  Therefore, a  consistent discretization protocol is necessary. While forecasting chaotic continuous signals, arising from canonical examples such as Lorenz or R\"{o}ssler systems~\cite{steinegger2025predicting}, has been performed using QRC, the discretization is typically chosen as an ad-hoc multiple of the step length used in the numerical integration of the underlying chaotic differential equation~\cite{fry2023optimizing, steinegger2025predicting}. For experimental signals, it is tempting to similarly use an ad-hoc factor of the data acquisition frequency. However, this arbitrary choice of discretization may result in poor accuracy or inefficiency.   Extremely fine-grained discretization protocols, in addition to being memory-expensive, also stand to artificially boost forecasting accuracy through oversampling. Conversely, excessively large step sizes may overlook important signal features and yield fully uncorrelated data samples, making precise forecasting infeasible. Recently, in Ref.~\cite{ahmed2025robust} a consistent discretization recipe based on the Lyapunov time of theoretical chaotic models was proposed. However, calculating Lyapunov exponent for real experimental data is extremely demanding due to complex state-space reconstruction involved. A systematic as well as simple and robust discretization procedure applicable to both theoretical and experimental signals for feeding to the quantum reservoir is therefore crucial for real-life benchmarking of forecasting accuracy of QRC or QEL.

\begin{figure*}[t]
\centering
\begin{overpic}[width=0.98\textwidth]{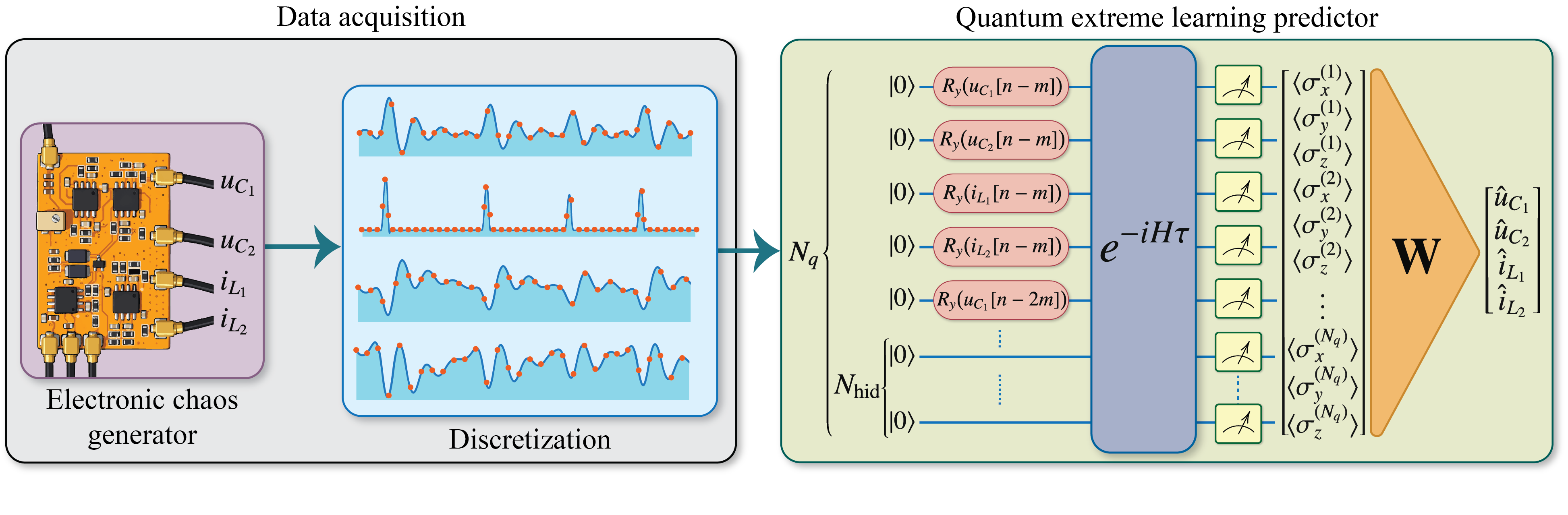}
	
\end{overpic}

\caption{
	Schematic overview of the proposed framework.
	The experimental electronic chaos generator produces four signals,
	\(u_{C_1}\), \(u_{C_2}\), \(i_{L_1}\), and \(i_{L_2}\), which are discretized using a sampling lag \(m\) determined from the first \(1/e\) crossing of the autocorrelation function of each signal.
	Selected past samples, separated by this lag, are encoded into qubits through rotation encoding, while the remaining \(N_{\mathrm{hid}}\) qubits remain in \(\lvert 0\rangle\).
	All qubits evolve under the reservoir Hamiltonian \(H\) for a time \(\tau\), followed by measurements of the local Pauli expectation values. The resulting measurement vector is used to train the linear readout and to predict the signal values. 
}
\label{fig:schematic}
\end{figure*}

Most of the QRC-based forecasting protocols focus on deterministic difference (in the case of discrete time series)~\cite{fujii2017harnessing, li2026quantumchaotic,hou2026high} or differential (in the case of continuous signals)~\cite{Sornsaeng_2024, fellner2026multivariatequantumreservoircomputing, Li2026higher} equations. Therefore, in such contexts forecasting reduces to learning a few parameters in those equations. In practice, however,  deterministic equations are only an idealized form of evolution of the systems. Electronic circuits are prime examples of such setups which have been extensively used for their nonlinear and potentially chaotic dynamics \cite{sprottbook}.  Indeed, a relatively small number of active nonlinear circuit components like transistors or memristors is enough to introduce chaotic behaviour in their dynamical evolution. The mathematical models of idealized circuits are often incomplete in an experimental setting, where various hidden generators of non-linearity and fluctuations add a separate component to the actually observed evolution. While mathematical models of such circuits are capable to generating chaotic behaviour in principle, these additional components may lead to further unexpected complex evolution \cite{sprottbook}. Predictive algorithms based on actually observed data allows us to detect them early and apply, where appropriate, suitable control schemes. Thus, we are presented with a pertinent question -- can the QRC-based predictors be used for accurately forecasting both simulated as well as actual experimental dynamics of chaotic electronic circuits?

\begin{figure*}[t]
\centering
\begin{overpic}[width=0.99\textwidth]{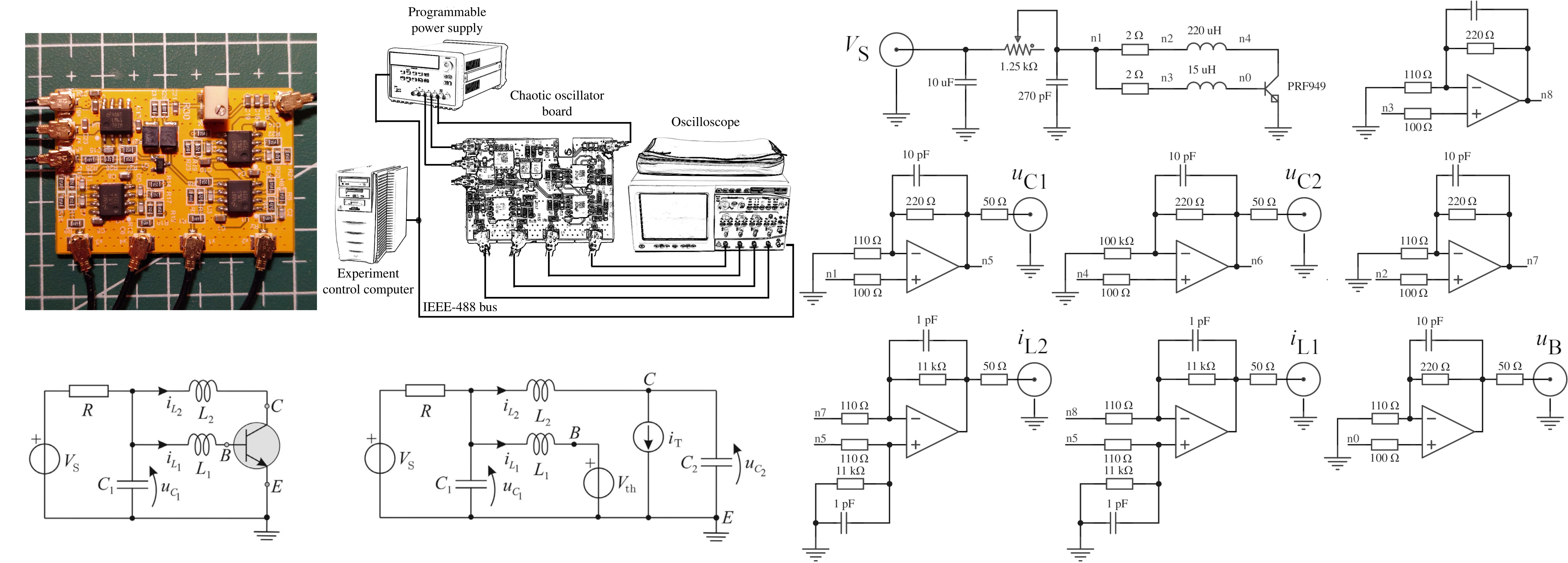}
	\put(-1,33){ {(a)}}
	\put(21,33){ {(b)}}
	\put(-1,12){ {(d)}}
	\put(21,12){ {(e)}}
	\put(50,33){ {(c)}}
\end{overpic}

\caption{
	%			Oscillator circuit topology and experimental implementation.
	%			(a) Physical AECG circuit.
	%			(b) Equivalent circuit used for the dynamical model.
	%			(c) Oscillator board.
	%			(d) Experimental measurement setup.
	%			(e) Detailed circuit implementation and signal-conditioning circuits.
	Oscillator circuit topology and experimental implementation.
	(a) Oscillator printed circuit board realization.
	(b) Experimental measurement setup.
	(c) Detailed circuit implementation and signal-conditioning circuits.
	(d) Physical AECG circuit.
	(e) Equivalent circuit used for the dynamical model.
}
\label{fig:Circuit}
\end{figure*}

In this paper, we experimentally implement an electronic circuit which shows chaotic behavior in four selected signals, namely two voltages and two currents, mathematically modeled by four coupled nonlinear differential equations. In addition to numerically integrating the differential equations to simulate the evolution of the four signals, we also experimentally acquire the signals with a temporal resolution of 3.2~\(n\)s, which is effectively continuous as this is far below the evolution timescale (${\sim} \mu$s).  To forecast both simulated and experimental signals, we design a QEL protocol using up to 11 qubits. We address the problem of discretization through two different approaches: (i) by only considering the magnitude of the signals at their local peaks/troughs (maplike discretization); and (ii) by uniformly sampling the underlying signal (uniform discretization). In the former, owing to the aperiodicity of chaotic oscillations, the signal is not uniformly sampled and local features other than chaotic resonances are missed. In the latter, these local features are captured as  we identify the autocorrelation timescale of each signal as the key discretization timestep to ensure a meaningful test of predictive accuracy. This discretization timescale is comparable for both simulated and experimental data and is at least seven times larger than the data acquisition timescale of 3.2 \(n\)s. This indicates that when QEL encodings are performed at the autocorrelation timescale, they avoid both trivial oversampling and excessive coarse-graining while yielding high forecasting accuracy. We provide a side-by-side comparison of QEL performance in forecasting both simulated and experimental signals. We show that, even with a very small quantum reservoir, QEL achieves highly accurate predictions for both signal types and can reconstruct  the chaotic attractor diagrams. The proposed quantum reservoir  requires only nearest-neighbour interacting spins, which can be readily implemented on current quantum platforms. Finally, our simulations also demonstrate robust performance under moderate decoherence and dissipation. Our results provide a systematic approach for continuous signal processing and forecasting with near-term quantum simulators.

%%%%%%%%%%%%%%%%%%%%%%%%%%%%%%%%%%%%%%%%%%%%%%%%%%%%%%%%

\section*{RESULTS} 
The main results of this work are presented in the following manner. We begin by describing the QEL technique and the Autonomous Electronic Chaos Generator (AECG) system used in this study. Subsequently, we present forecasting results for both theoretical and experimental implementations of the AECG circuit using QEL. This is reported for two different frameworks employing complementary ways of discretizing the underlying chaotic flow.

\subsection*{Quantum Extreme Learning}

QEL, as a special case of QRC, is a machine-learning algorithm suitable to run with as little individual control over qubits as possible during its runtime. The algorithm can be broken down into four stages that are schematically shown in Fig.~\ref{fig:schematic} -- \textbf{(i) Encoding :} $N_q$ blank qubits are each prepared in the state $|0\rangle$ of the Pauli-Z, i.e., computational basis. A classical $c$-dimensional time-series signal, represented at discrete time $n$ by $\mathbf{x}[n]{\in}\mathbb{R}^{c}$, is then encoded into one set of blank qubits.  Each element of $\mathbf{x}[n]$ is then encoded through rotation encoding with
\(R_Y(\theta){=}\exp(-i\theta\sigma_y/2)\), giving \(		R_Y(\theta)|0\rangle
{=}\cos(\theta/2)|0\rangle{+}\sin(\theta/2)|1\rangle\),  where \(\theta\) is obtained by remapping the \(z\)-score normalized input value to the interval \([0,\pi]\) (the details of data preparation are explained in the Methods section).	For the $n$-th input window, the input vectors being encoded are
\begin{eqnarray}\label{eq:input}
\bm{\mathcal{X}}[n]
=
\left[
\mathbf{x}^{\mathsf{T}}[n-m],
\mathbf{x}^{\mathsf{T}}[n-2m],
\cdots,
\mathbf{x}^{\mathsf{T}}[n-N_{\mathrm{steps}}m]
\right]^{\mathsf{T}},
\end{eqnarray}
where \(\bm{\mathcal{X}}[n]{\in}\mathbb{R}^{cN_{\mathrm{steps}}}\), corresponding to \(N_{\mathrm{steps}}\) instances of the time series \(\mathbf{x}[n]\), separated by a lag of \(m\) samples. Thus, \(cN_{\mathrm{steps}}\) qubits are encoded in this way.  The other set of blank qubits, called hidden qubits and numbering $N_{\textrm{hid}}{=}N_q{-}cN_{\textrm{steps}}$, are not encoded.  \textbf{(ii) Evolution :} All the  $cN_{\textrm{steps}}$ encoded qubits as well as $N_{\textrm{hid}}$ hidden qubits evolve under an interacting many-body Hamiltonian $H$ for time $\tau$. The form of this Hamiltonian is specified in advance, individual interaction strengths are randomly initialized but fixed during evolution. This step scrambles information about the original classical signal from the originally encoded qubits into all qubits. \textbf{(iii) Readout:} Post evolution, all the qubits are measured in their respective local Pauli bases to yield a vector $\mathbf{m}$ recording expectation values of measurement outcomes for each qubit. \textbf{(iv) Training :} We generate a forecast $\tilde{\mathbf{x}}[n]$ for an input window $n$ by left-multiplying the readout vector $\mathbf{m}[n]$ with a weight matrix $\mathbf{W}$, i.e., $\tilde{\mathbf{x}}[n] {=} \mathbf{W} \mathbf{m}[n]$. Let the true value for this window $n$ be $\mathbf{x}[n]$ against which we benchmark the forecast $\tilde{\mathbf{x}}[n]$. Now, we consider many such input windows $\lbrace n \rbrace$ such that the forecasts $\mathbf{\tilde{X}} {=} \left[\tilde{\mathbf{x}}[1], \ \tilde{\mathbf{x}}[2], \cdots, \tilde{\mathbf{x}}[n], \cdots\right]$ are generated by left-multiplying the weight matrix $\mathbf{W}$ with corresponding readout vectors $\mathbf{M} {=} \left[\mathbf{m}[1], \ \mathbf{m}[2], \cdots, \mathbf{m}[n], \cdots\right]$, i.e., $\mathbf{\tilde{X}} {=} \mathbf{WM}$. Our goal in training is to find the best possible $\mathbf{W}$ so that $\mathbf{\tilde{X}}$  be collectively as close to the corresponding true values of these sequences $\mathbf{X} {=} [\mathbf{x}[1],\ \mathbf{x}[2], \cdots, \mathbf{x}[n], \cdots] $ as possible. To achieve this we define the loss function as
\begin{eqnarray}
\mathcal{L}(\mathbf W)
=
\|\mathbf X-\mathbf W\mathbf M\|_F^2, %+		\epsilon\|\mathbf W\|_F^2,
\end{eqnarray}
where \(\|\cdot\|_F\) denotes the Frobenius norm. Then, we end up with the following ridge-regression formula for the optimal weight matrix $\mathbf{W}^{*}$ minimizing the above loss function~\cite{fujii2017harnessing, ghosh2021quantum,kobayashi2024feedback}
\begin{equation}\label{eq:weigh_matrix}
\mathbf{W}^{*}
=
\mathbf{X}\mathbf{M}^{\mathsf{T}}
\left(
\mathbf{M}\mathbf{M}^{\mathsf{T}}
+
\epsilon\mathbf{I}
\right)^{-1}.
\end{equation} \noindent  where $\epsilon{>}0$ is a small number which is used to avoid possible singularity and in this paper we take it as $\epsilon{=}10^{-7}$. Importantly, the training happens solely on the final readout layer treating the evolution stage as a black box, in stark contrast to variational quantum algorithms. This makes  QEL a far simpler  technique that does not rely on gradient-based optimization and thus preventing barren plateau issue.

\begin{figure*}[t]
\centering

% ==================== Shared legend ====================
\includegraphics[
width=0.62\textwidth,
trim=0 0 0 0,
clip
]{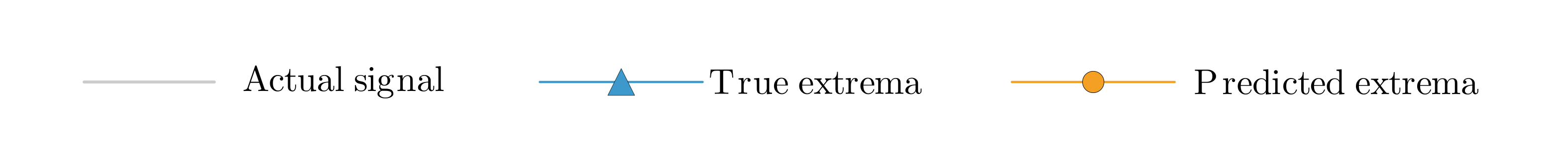}

\vspace{-23pt}

% ==================== Left column: Theory ====================
\begin{minipage}[t]{0.485\textwidth}
	\vspace{0pt}
	\begin{tcolorbox}[
		enhanced,
		colback=white,
		colframe=black,
		boxrule=0.6pt,
		arc=0pt,
		left=3pt,
		right=3pt,
		top=7pt,
		bottom=-2pt,
		title=Theory,
		fonttitle=\small\bfseries,
		coltitle=black,
		colbacktitle=white,
		attach boxed title to top left={
			xshift=8pt,
			yshift=-2mm
		},
		boxed title style={
			colback=white,
			colframe=white,
			boxrule=0pt,
			left=2pt,
			right=2pt,
			top=0pt,
			bottom=0pt
		}
		]
		\centering
		
		\begin{overpic}[
			width=\textwidth
			]{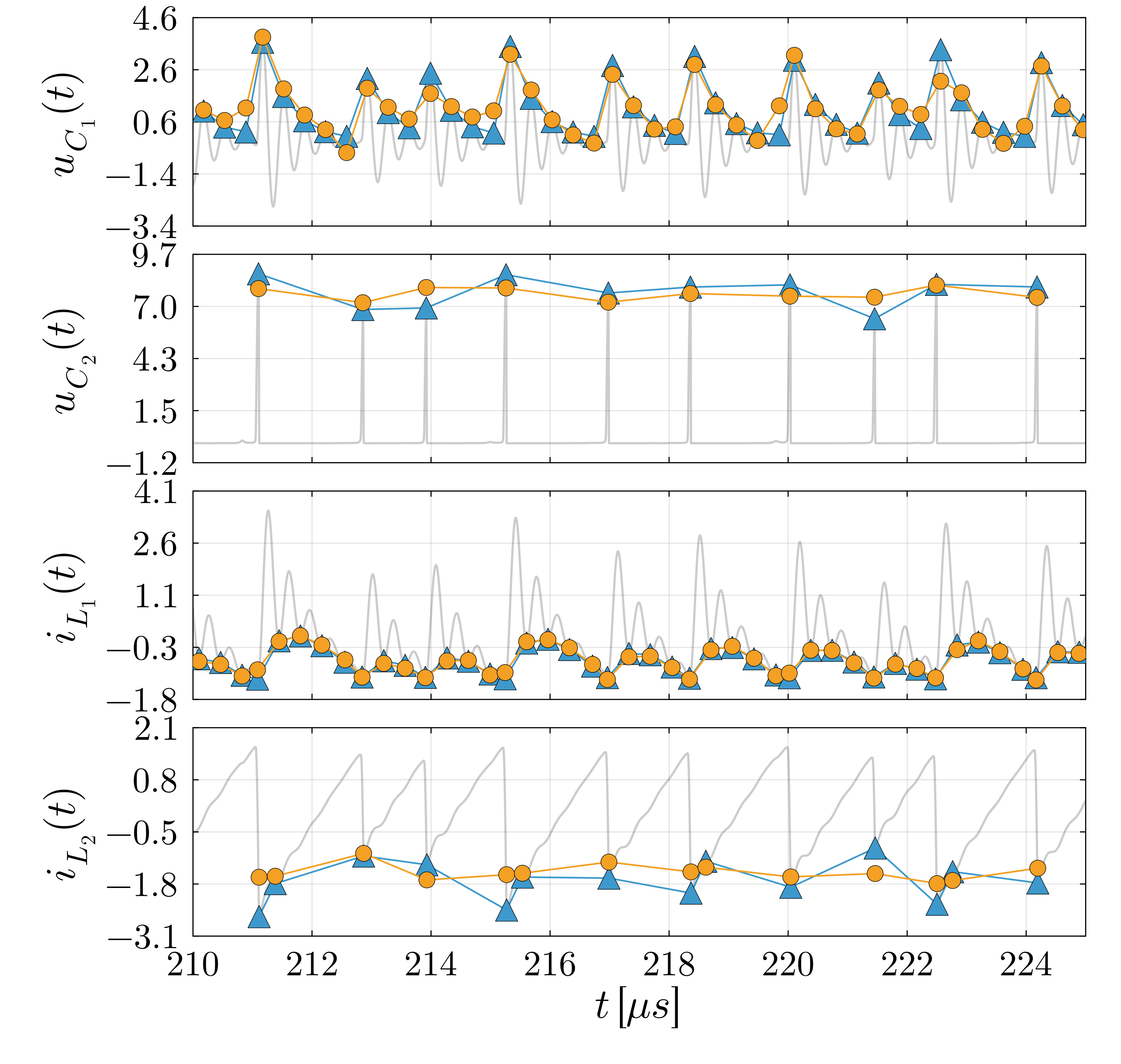}
			
			% First subplot: u_C1
			\put(-2,92){{(a)}}
			
			% Second subplot: u_C2
			\put(-2,70){{(c)}}
			
			% Third subplot: i_L1
			\put(-2,49){{(e)}}
			
			% Fourth subplot: i_L2
			\put(-2,28){{(g)}}
		\end{overpic}
	\end{tcolorbox}
\end{minipage}
\hfill
% ==================== Right column: Experiment ====================
\begin{minipage}[t]{0.485\textwidth}
	\vspace{0pt}
	\begin{tcolorbox}[
		enhanced,
		colback=white,
		colframe=black,
		boxrule=0.6pt,
		arc=0pt,
		left=3pt,
		right=3pt,
		top=7pt,
		bottom=-2pt,
		title=Experiment,
		fonttitle=\small\bfseries,
		coltitle=black,
		colbacktitle=white,
		attach boxed title to top left={
			xshift=8pt,
			yshift=-2mm
		},
		boxed title style={
			colback=white,
			colframe=white,
			boxrule=0pt,
			left=2pt,
			right=2pt,
			top=0pt,
			bottom=0pt
		}
		]
		\centering
		
		\begin{overpic}[
			width=\textwidth
			]{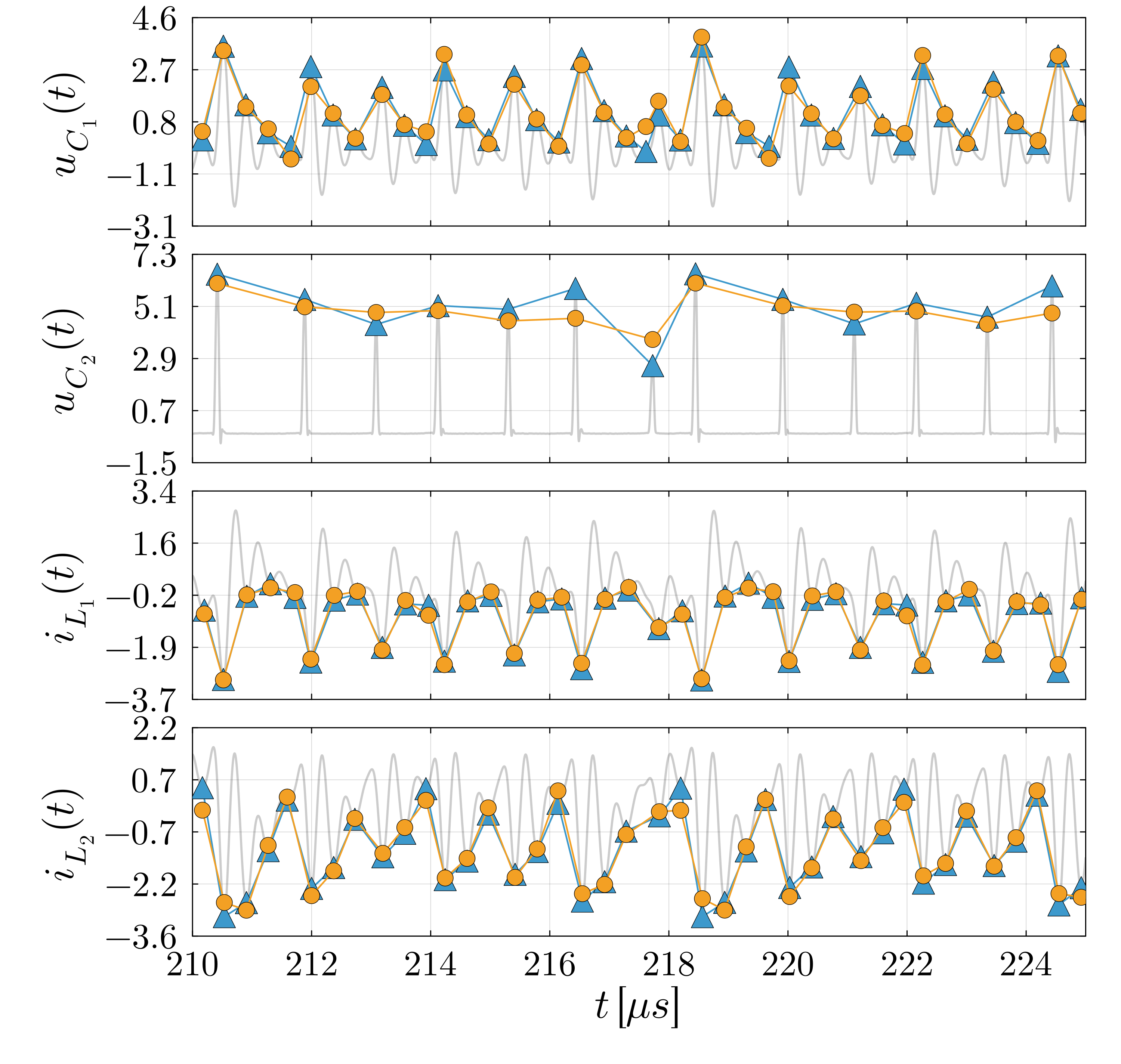}
			
			% First subplot: u_C1
			\put(-2,92){{(b)}}
			
			% Second subplot: u_C2
			\put(-2,70){{(d)}}
			
			% Third subplot: i_L1
			\put(-2,49){{(f)}}
			
			% Fourth subplot: i_L2
			\put(-2,28){{(h)}}
		\end{overpic}
	\end{tcolorbox}
\end{minipage}

\caption{
	Prediction results of maplike discretization. The left and right columns show
	the theory and experiment results, respectively:
	(a,b) $u_{C_1}$,
	(c,d) $u_{C_2}$,
	(e,f) $i_{L_1}$,
	and (g,h) $i_{L_2}$.
	The gray curves represent the actual continuous signals,
	while the triangles and circles represent the true and
	predicted extrema, respectively.
}
\label{fig:maplike1to1}
\end{figure*}

\subsection*{Autonomous Electronic Chaos Generator}

We use the Minati--Frasca oscillator, an autonomous electronic chaos generator which comprises a single NPN bipolar-junction transistor, two inductors, one capacitor, and a bias resistor connected to a constant voltage source~\cite{minati2022mechanisms}. Unlike analog circuits that are designed to integrate a prescribed differential equation system using operational amplifiers and multipliers, this circuit directly exploits the transistor nonlinear device characteristics and is therefore considerably smaller. Figure~\ref{fig:Circuit}(a) shows the printed circuit
board used to realize the chaotic oscillator, while
Fig.~\ref{fig:Circuit}(b) presents the experimental
measurement setup used to acquire the chaotic signals
simultaneously. The detailed circuit implementation,
including the signal-conditioning stages, is shown
in Fig.~\ref{fig:Circuit}(c). The physical topology and its reduced equivalent circuit are shown in Figs.~\ref{fig:Circuit}(d) and \ref{fig:Circuit}(e), respectively. The transistor emitter is grounded, while its base and collector are connected through inductors $L_1$ and $L_2$, respectively, to a common node. This node is connected to ground through $C_1$ and to the DC supply $V_\textrm{S}$ through $R$. Its chaotic dynamics arise from the mutual interaction between a relaxation process and a damped resonance. These unfold in two subcircuits that are partially overlapping: relaxation events excite the resonant branch, whose response in turn modifies subsequent reset events~\cite{minati2022mechanisms}. The detailed component specifications, shunt resistors, operational amplifiers, and acquisition settings are described in the Methods section.

\begin{figure*}[t]
\centering

% =========================================================
% Theory group
% =========================================================
\begin{tcolorbox}[
	enhanced,
	colback=white,
	colframe=black,
	boxrule=0.6pt,
	arc=0pt,
	left=3pt,
	right=3pt,
	top=1pt,
	bottom=-8pt,
	title=Theory,
	fonttitle=\small\bfseries,
	coltitle=black,
	colbacktitle=white,
	attach boxed title to top left={
		xshift=8pt,
		yshift=-2mm
	},
	boxed title style={
		colback=white,
		colframe=white,
		boxrule=0pt,
		left=2pt,
		right=2pt,
		top=0pt,
		bottom=0pt
	}
	]
	\centering
	
	\begin{subfigure}{.235\linewidth}
		\centering
		\begin{overpic}[width=\textwidth]{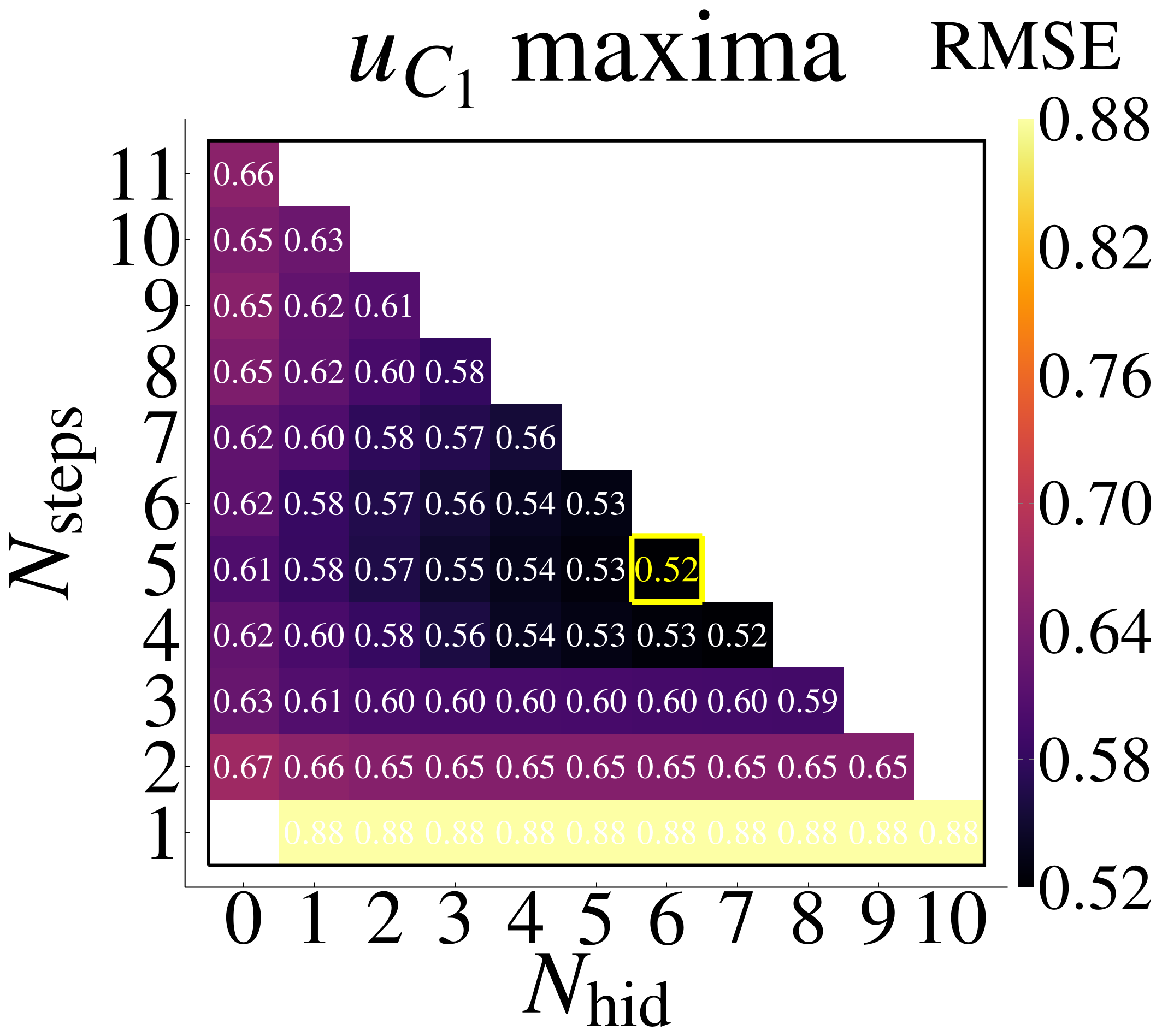}
			\put(-2,84){{(a)}}
		\end{overpic}
		\label{fig:u1heatmap1to1theory_maplike}
	\end{subfigure}
	\hfill
	\begin{subfigure}{.235\linewidth}
		\centering
		\begin{overpic}[width=\textwidth]{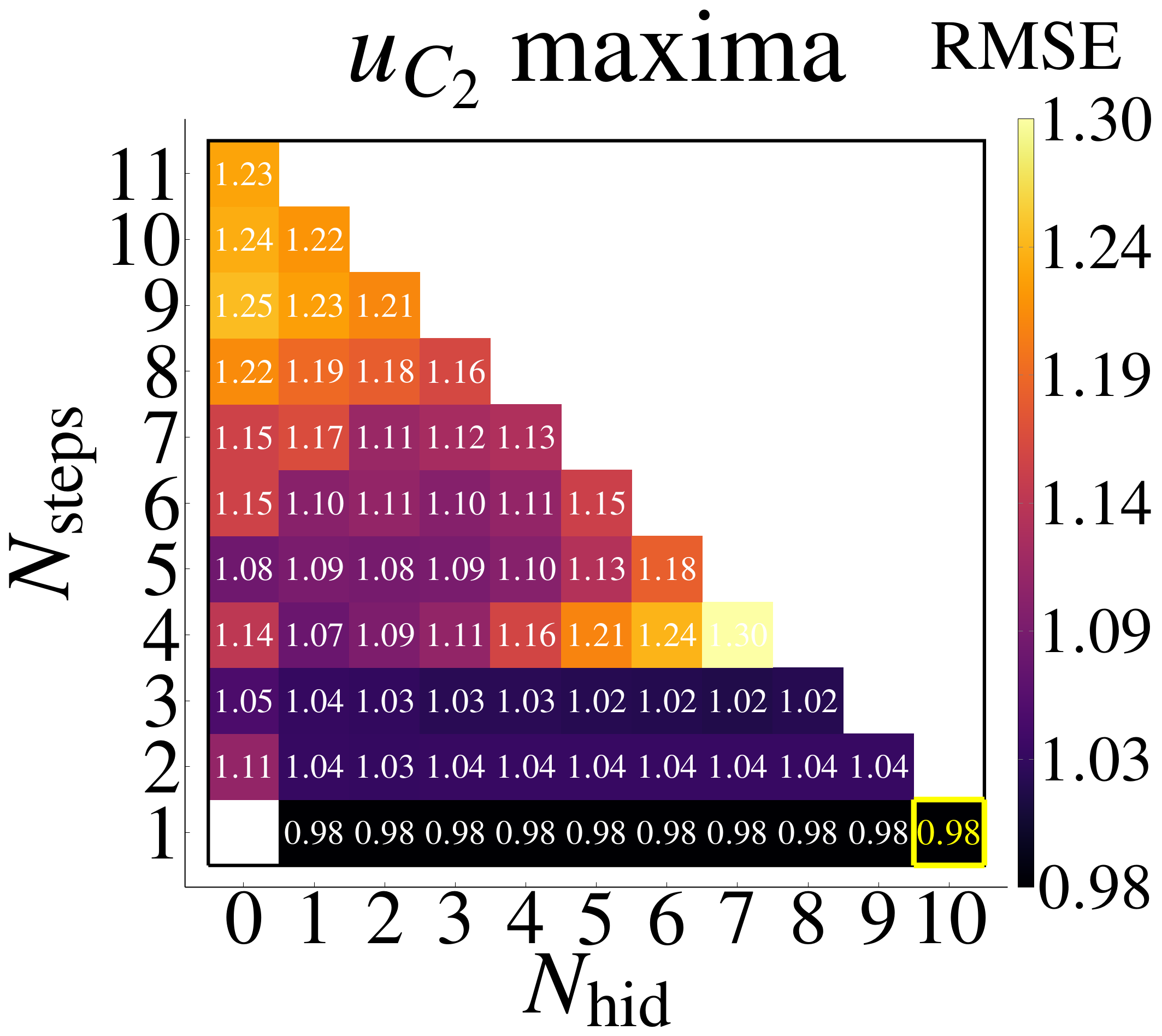}
			\put(-2,84){{(c)}}
		\end{overpic}
		\label{fig:u2heatmap1to1theory_maplike}
	\end{subfigure}
	\hfill
	\begin{subfigure}{.235\linewidth}
		\centering
		\begin{overpic}[width=\textwidth]{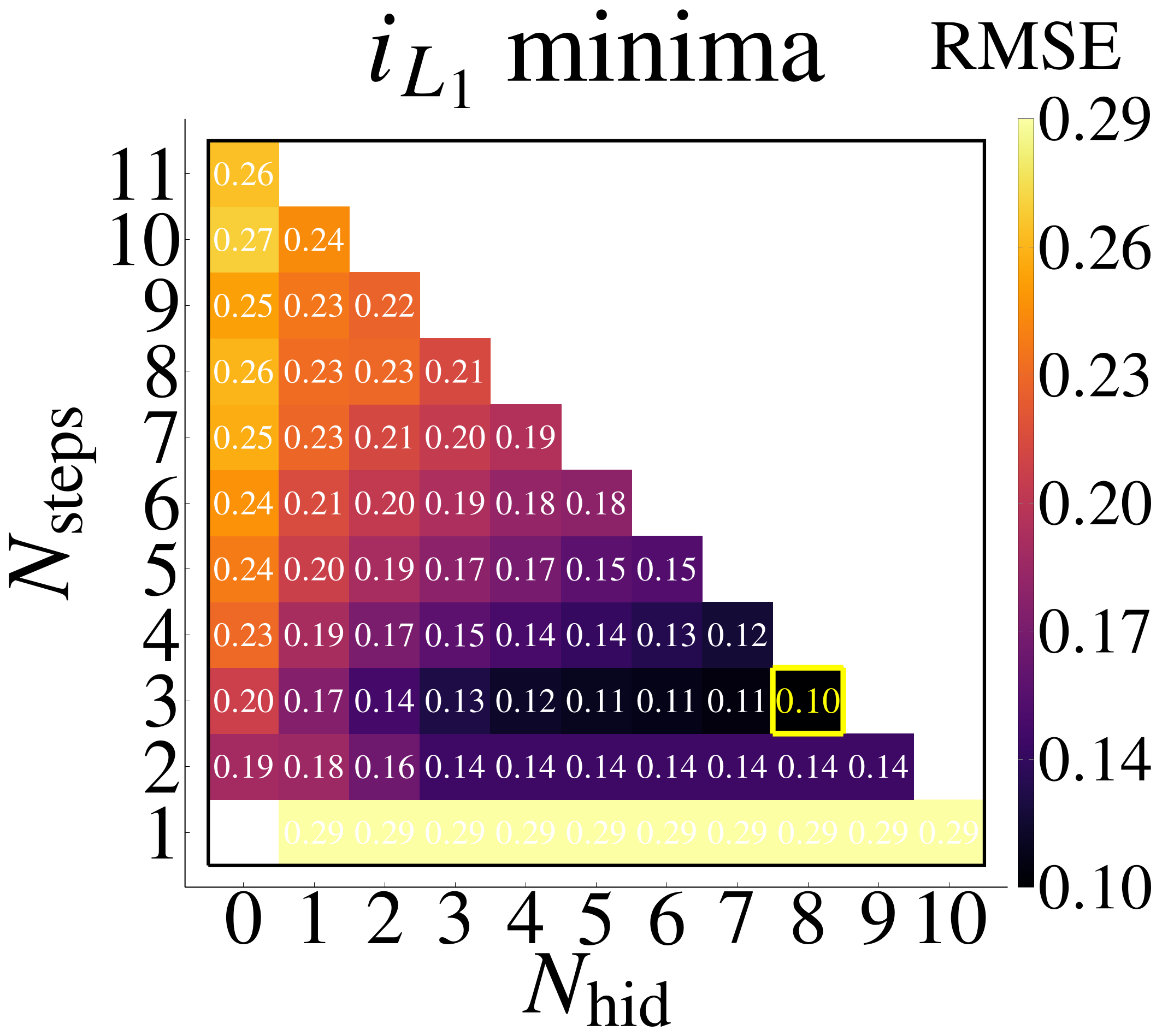}
			\put(-2,84){{(e)}}
		\end{overpic}
		\label{fig:i1heatmap1to1theory_maplike}
	\end{subfigure}
	\hfill
	\begin{subfigure}{.235\linewidth}
		\centering
		\begin{overpic}[width=\textwidth]{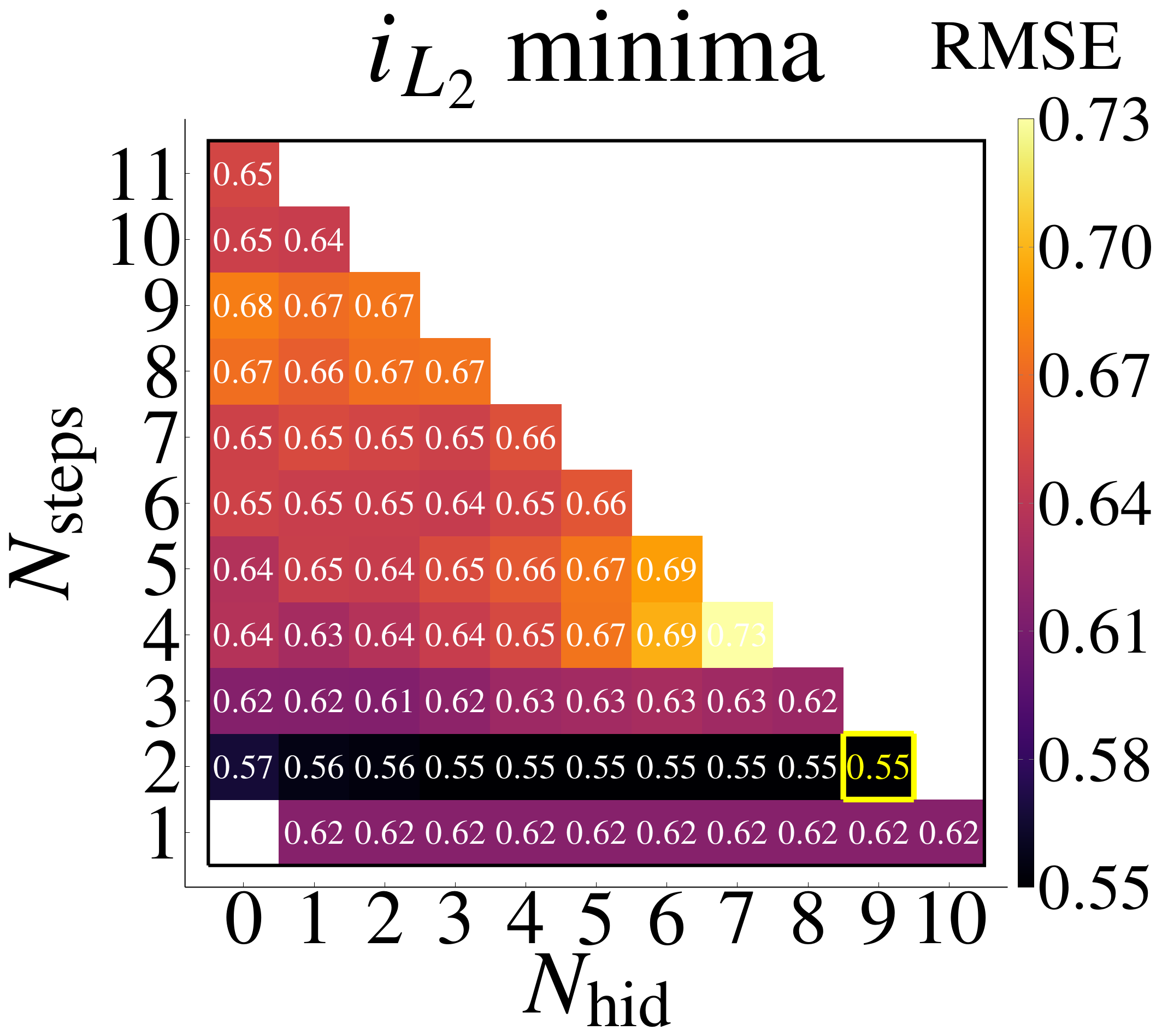}
			\put(-2,84){{(g)}}
		\end{overpic}
		\label{fig:i2heatmap1to1theory_maplike}
	\end{subfigure}
\end{tcolorbox}

\vspace{-0.25cm}

% =========================================================
% Experiment group
% =========================================================
\begin{tcolorbox}[
	enhanced,
	colback=white,
	colframe=black,
	boxrule=0.6pt,
	arc=0pt,
	left=3pt,
	right=3pt,
	top=1pt,
	bottom=-8pt,
	title=Experiment,
	fonttitle=\small\bfseries,
	coltitle=black,
	colbacktitle=white,
	attach boxed title to top left={
		xshift=8pt,
		yshift=-2mm
	},
	boxed title style={
		colback=white,
		colframe=white,
		boxrule=0pt,
		left=2pt,
		right=2pt,
		top=0pt,
		bottom=0pt
	}
	]
	\centering
	
	\begin{subfigure}{.235\linewidth}
		\centering
		\begin{overpic}[width=\textwidth]{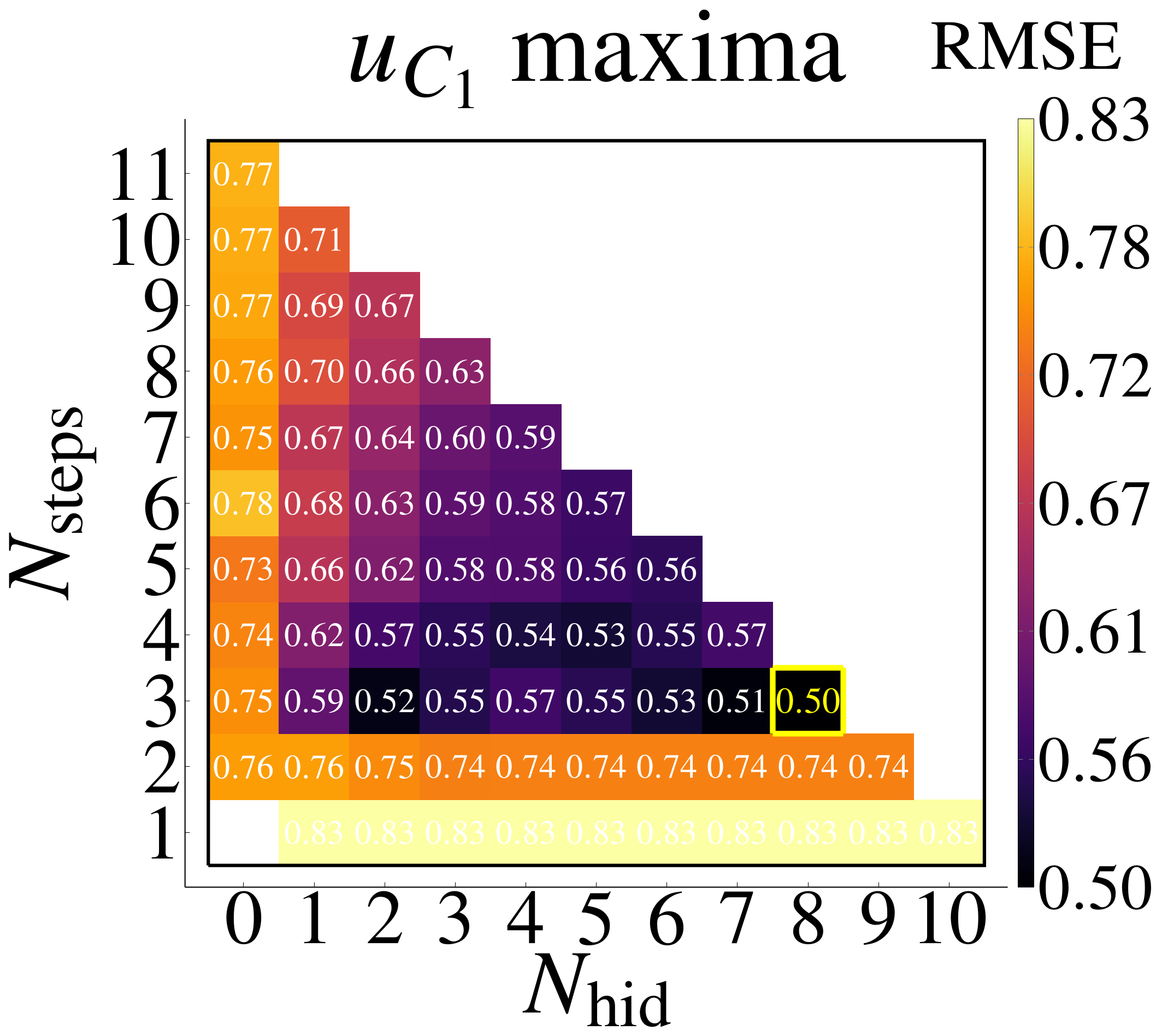}
			\put(-2,84){{(b)}}
		\end{overpic}
		\label{fig:u1heatmap1to1experiment_maplike}
	\end{subfigure}
	\hfill
	\begin{subfigure}{.235\linewidth}
		\centering
		\begin{overpic}[width=\textwidth]{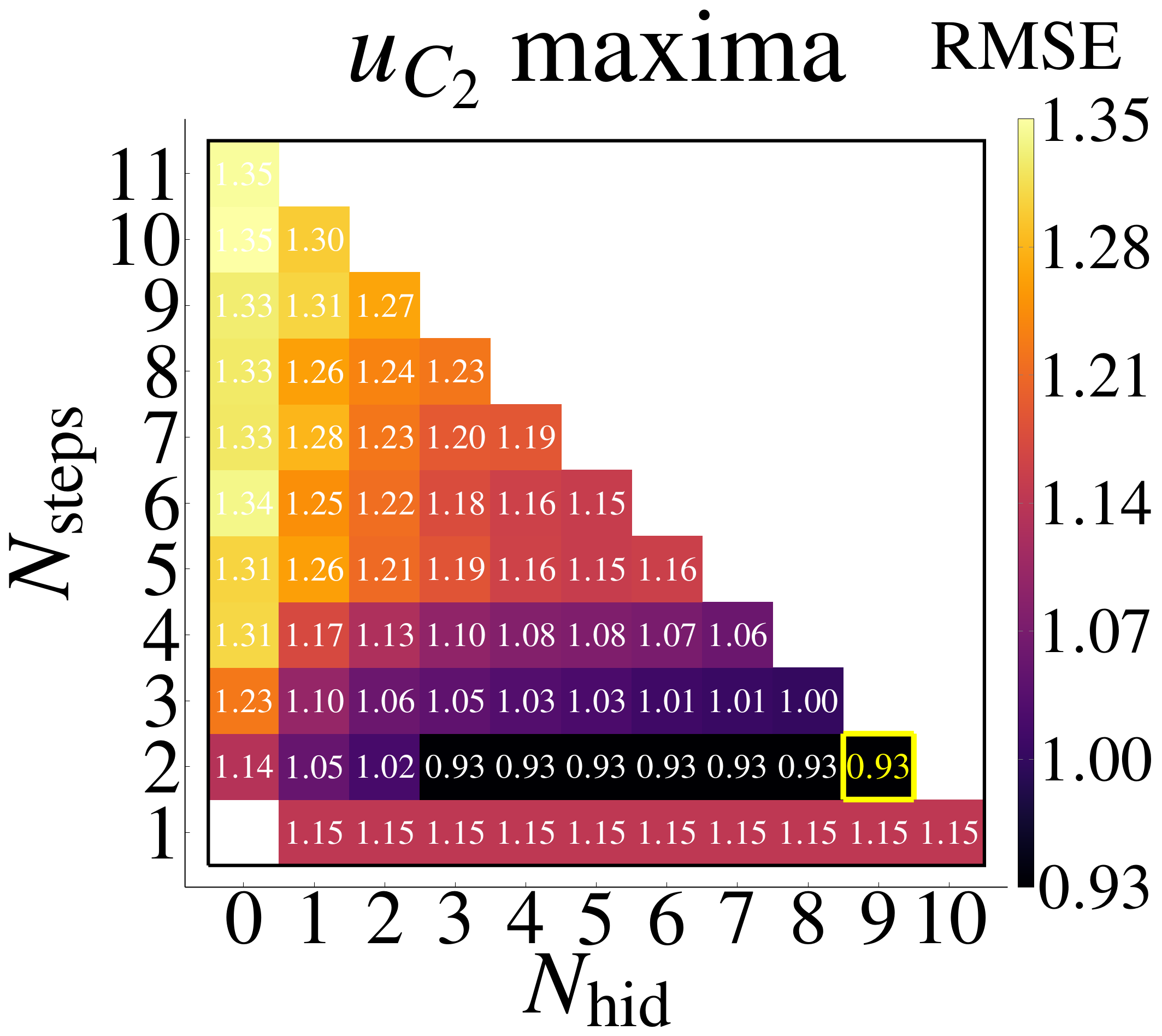}
			\put(-2,84){{(d)}}
		\end{overpic}
		\label{fig:u2heatmap1to1experiment_maplike}
	\end{subfigure}
	\hfill
	\begin{subfigure}{.235\linewidth}
		\centering
		\begin{overpic}[width=\textwidth]{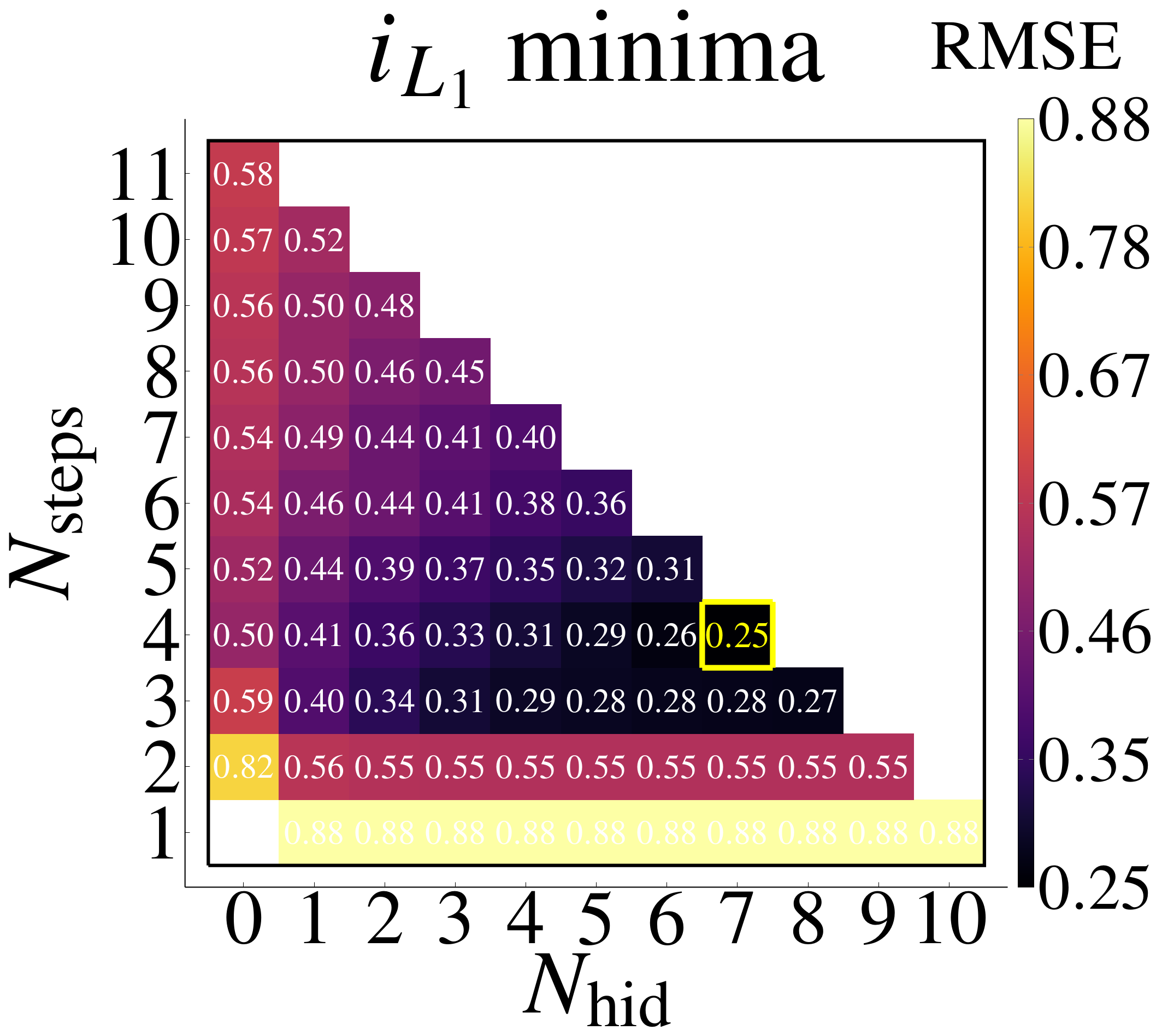}
			\put(-2,84){{(f)}}
		\end{overpic}
		\label{fig:i1heatmap1to1experiment_maplike}
	\end{subfigure}
	\hfill
	\begin{subfigure}{.235\linewidth}
		\centering
		\begin{overpic}[width=\textwidth]{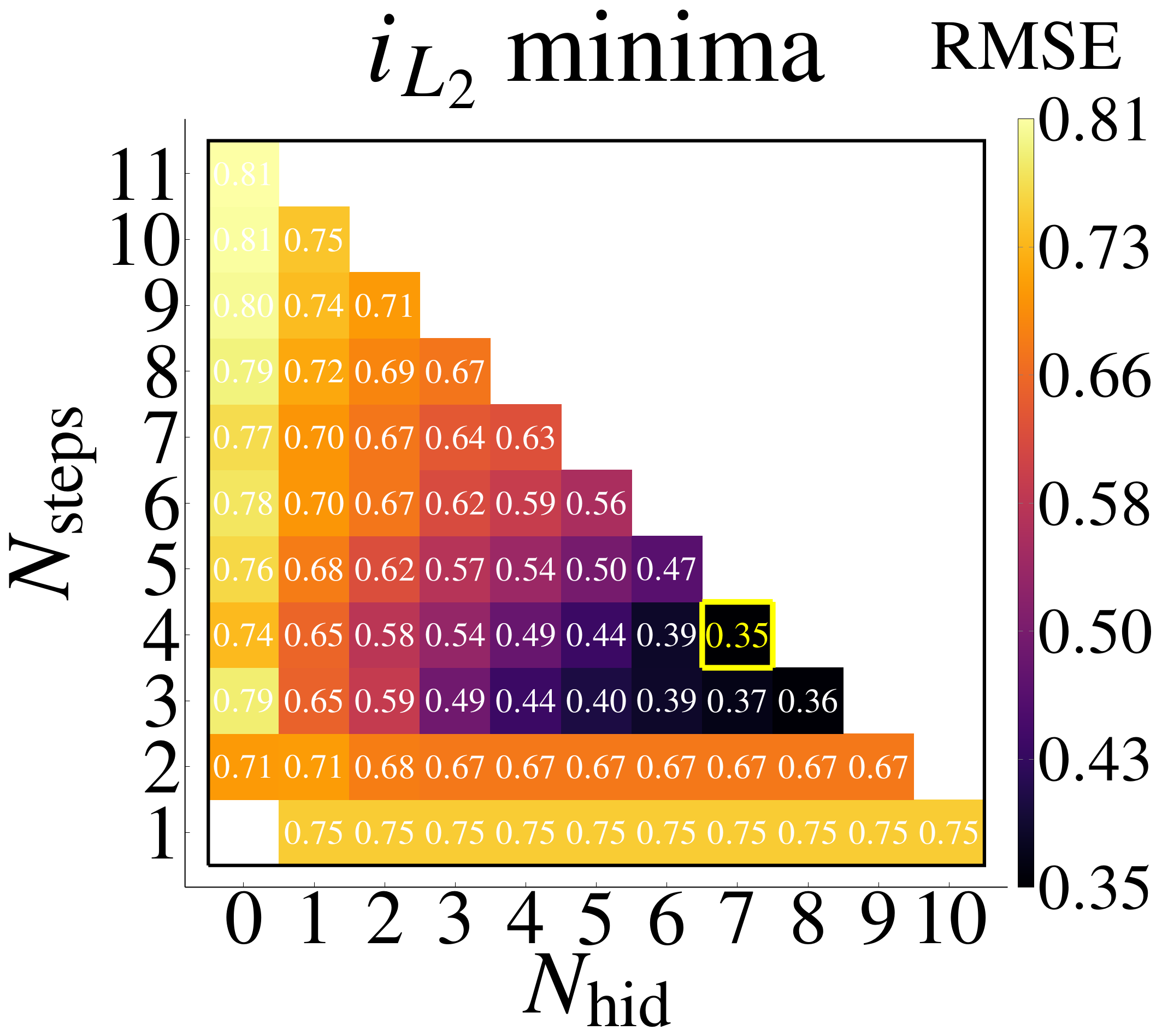}
			\put(-2,84){{(h)}}
		\end{overpic}
		\label{fig:i2heatmap1to1experiment_maplike}
	\end{subfigure}
\end{tcolorbox}

\caption{
	Performance heatmap for maplike discretization.
	The top row shows the theoretical results and the bottom row shows the experimental results.
	Panels (a,b), (c,d), (e,f), and (g,h) correspond to $u_{C_1}$, $u_{C_2}$, $i_{L_1}$, and $i_{L_2}$, respectively.
}
\label{fig:heatmap1to1_maplike}
\end{figure*}

\begin{figure}[t]
\centering

\begin{overpic}[width=0.49\textwidth]
	{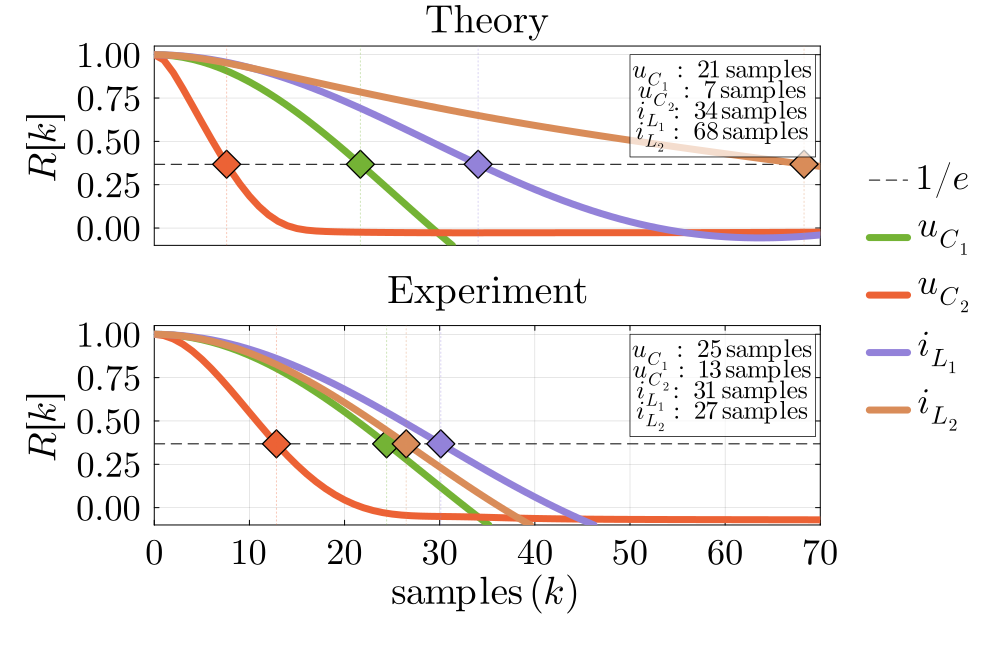}
	\put(0,59){{(a)}}
	\put(0,31){{(b)}}
\end{overpic}

\caption{Autocorrelation functions of the signals. (a) Theoretical signals. (b) Experimental signals. The dashed horizontal line denotes the (1/e) threshold used to determine the characteristic lag of each signal, and the markers identify the corresponding first crossings of this threshold.}

\label{fig:ACF}
\end{figure}

In the numerical simulations, we retained the reduced model introduced in Ref.~\cite{minati2022mechanisms} and shown in {Fig.~\ref{fig:Circuit}(e). Briefly, we make three simplifying approximations to obtain a viable and compact phenomenological description. The base--emitter voltage is assumed to be constant at $V_{\mathrm{th}}$, and the internal collector--emitter capacitance also is taken to be a constant $C_2$.
The transistor conduction current is then written as $i_T{=}\beta\Gamma(i_{L_1})\tanh[u_{C_2}/(2V_{\mathrm{th}})]$. Here, $\beta$ represents the current gain, while the ramp function $\Gamma(x){=}\max(x,0)$ prohibits the amplification of negative base currents. The hyperbolic tangent approximates the output characteristic as a function of the collector--emitter voltage, with $V_{\mathrm{th}}$ serving as a voltage scaling factor. Based on these assumptions, Kirchhoff's laws give
\begin{equation}
	\left\{
	\begin{aligned}
		\frac{\textrm{d}u_{C_1}}{\textrm{d}t}
		&=\frac{V_\textrm{S}-u_{C_1}}{RC_1}-\frac{i_{L_1}+i_{L_2}}{C_1},
		\\[4pt]
		\frac{\textrm{d}u_{C_2}}{\textrm{d}t}
		&=\frac{i_{L_2}-\beta\Gamma(i_{L_1})
			\tanh\!\left(\frac{u_{C_2}}{2V_{\mathrm{th}}}\right)}{C_2},
		\\[4pt]
		\frac{\textrm{d}i_{L_1}}{\textrm{d}t}
		&=\frac{u_{C_1}-V_{\mathrm{th}}}{L_1},
		\\[4pt]
		\frac{\textrm{d}i_{L_2}}{\textrm{d}t}
		&=\frac{u_{C_1}-u_{C_2}}{L_2}.
	\end{aligned}
	\right.
	\label{eq:AECG}
\end{equation}

The currents $i_{L_1}$ and $i_{L_2}$ are defined as positive from the common node towards the transistor. The voltages $u_{C_1}$ and $u_{C_2}$ denote the common-node and collector potentials, respectively, relative to ground. Throughout the below, the parameter were set to $V_\textrm{S} {=} 5~\mathrm{V}$, $R {=} 1~\mathrm{k}\Upomega$, $C_1 {=} 220~\mathrm{pF}$, $C_2 {=}5~\mathrm{ pF}$, $L_1 {=} 15~\mu\mathrm{H}$, $L_2 {=} 220~\mu\mathrm{H}$, $\beta {=} 200$, $V_{\mathrm{th}} {=} 0.6~\mathrm{V}$. Eq.~\eqref{eq:AECG} was integrated with a Runge--Kutta method, until $t_\textrm{max}{=}300~\mu\mathrm{s}$. The solutions were sampled every $\Delta t{=}3.2~\mathrm{ns}$, matching the experimental sampling interval. Representative time series over $210$--$225~\mu\mathrm{s}$ are shown as gray curves in Fig.~\ref{fig:maplike1to1}, with simulations on the left and recordings on the right. The reduced model accounts for the main chaos-generating mechanism, but omits multiple sources of additional dynamical complexity that are inherent in the physical realization of the circuit, most notably, higher-order transistor non-linearities, inductor self-resonances and dissipation, loading by the readout amplifiers, as well as uncertainties in all the parameters.

\begin{figure*}[t]
	\centering
	
	% ==================== Shared legend ====================
	\includegraphics[
	width=0.22\textwidth,
	trim=0 0 0 0,
	clip
	]{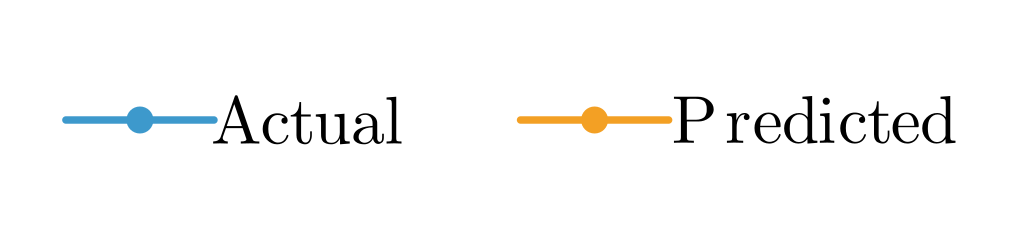}
	
	\vspace{-13pt}
	
	% ==================== Left column: Theory ====================
	\begin{minipage}[t]{0.485\textwidth}
		\vspace{0pt}
		\begin{tcolorbox}[
			enhanced,
			colback=white,
			colframe=black,
			boxrule=0.6pt,
			arc=0pt,
			left=3pt,
			right=3pt,
			top=7pt,
			bottom=-2pt,
			title=Theory,
			fonttitle=\small\bfseries,
			coltitle=black,
			colbacktitle=white,
			attach boxed title to top left={
				xshift=8pt,
				yshift=-2mm
			},
			boxed title style={
				colback=white,
				colframe=white,
				boxrule=0pt,
				left=2pt,
				right=2pt,
				top=0pt,
				bottom=0pt
			}
			]
			\centering
			
			\begin{overpic}[
				width=\textwidth
				]{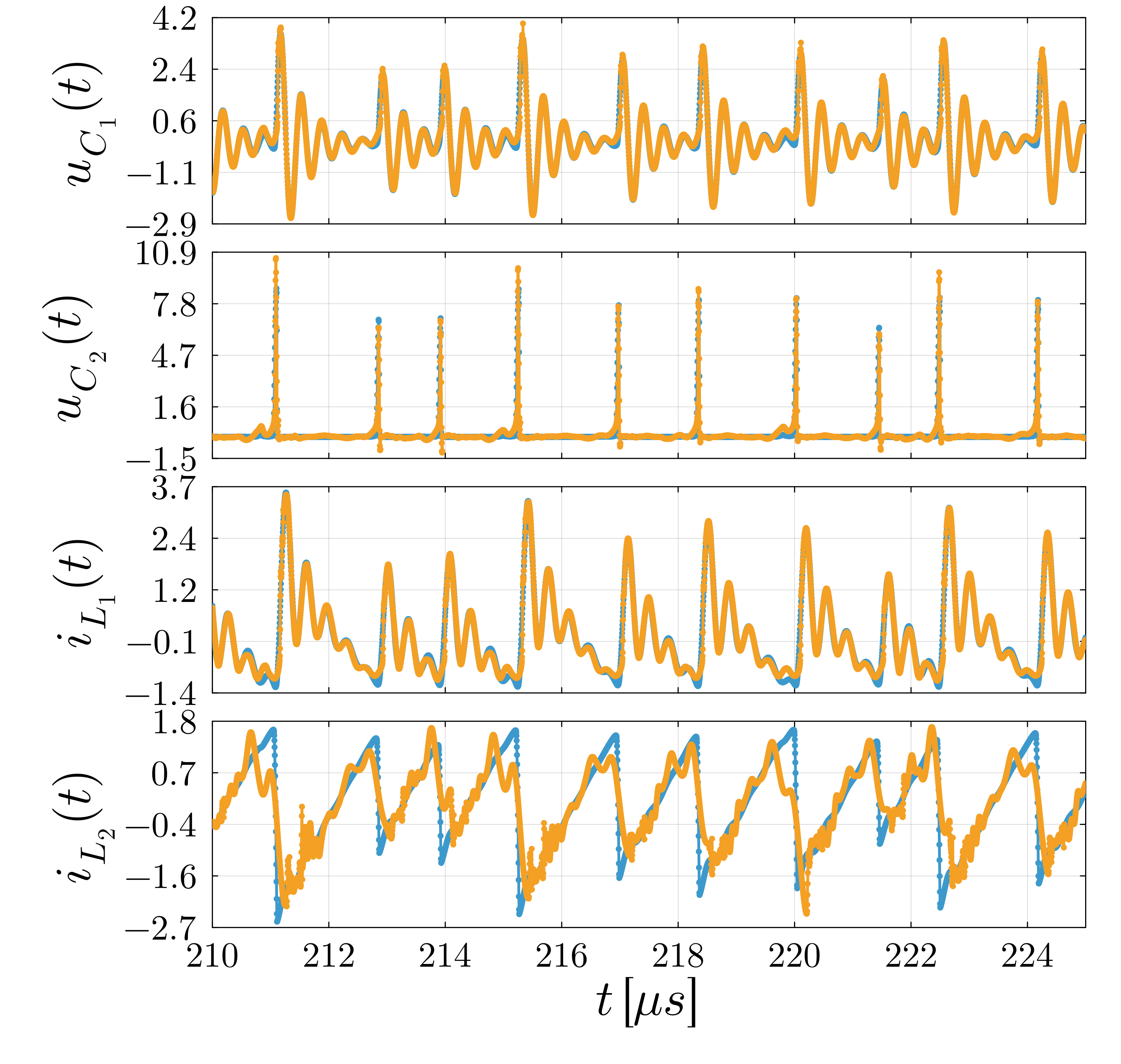}
				
				% First subplot: u_C1
				\put(-2,92){{(a)}}
				
				% Second subplot: u_C2
				\put(-2,72){{(c)}}
				
				% Third subplot: i_L1
				\put(-2,50){{(e)}}
				
				% Fourth subplot: i_L2
				\put(-2,29){{(g)}}
				
			\end{overpic}
			
			\label{fig:timeseries4to1theory}
		\end{tcolorbox}
	\end{minipage}
	\hfill
	% ==================== Right column: Experiment ====================
	\begin{minipage}[t]{0.485\textwidth}
		\vspace{0pt}
		\begin{tcolorbox}[
			enhanced,
			colback=white,
			colframe=black,
			boxrule=0.6pt,
			arc=0pt,
			left=3pt,
			right=3pt,
			top=7pt,
			bottom=-2pt,
			title=Experiment,
			fonttitle=\small\bfseries,
			coltitle=black,
			colbacktitle=white,
			attach boxed title to top left={
				xshift=8pt,
				yshift=-2mm
			},
			boxed title style={
				colback=white,
				colframe=white,
				boxrule=0pt,
				left=2pt,
				right=2pt,
				top=0pt,
				bottom=0pt
			}
			]
			\centering
			
			\begin{overpic}[
				width=\textwidth
				]{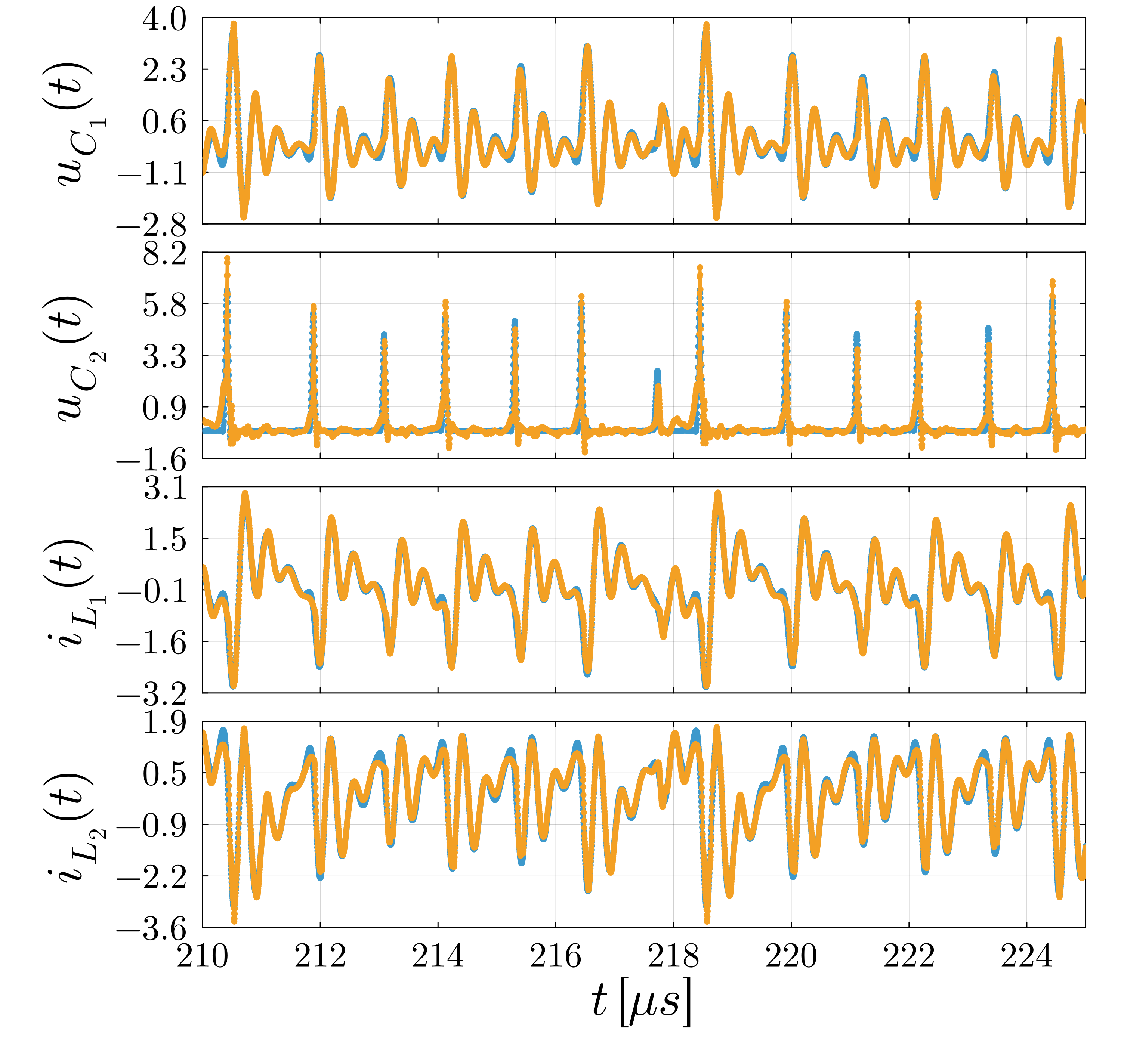}
				
				% First subplot: u_C1
				\put(-2,92){{(b)}}
				
				% Second subplot: u_C2
				\put(-2,72){{(d)}}
				
				% Third subplot: i_L1
				\put(-2,50){{(f)}}
				
				% Fourth subplot: i_L2
				\put(-2,29){{(h)}}
				
			\end{overpic}
			
			\label{fig:timeseries4to1experiment}
		\end{tcolorbox}
	\end{minipage}
	
	\caption{
		Prediction results of uniform discretization with four-to-one prediction setting.
		The left and right columns show the theoretical and
		experimental results, respectively:
		(a,b) $u_{C_1}$,
		(c,d) $u_{C_2}$,
		(e,f) $i_{L_1}$, and
		(g,h) $i_{L_2}$.
	}
	\label{fig:timeseries4to1}
\end{figure*}

\subsection*{Forecasting Results}

The electronic-circuit above physically generates continuous signals. However, QEL framework outlined above pre-supposes a discrete time-series signal. Thus, unlike inherently discrete time-series datasets like maps or financial-data studied in QRC literature, one has to specifically design this discretization. We consider two such designs in this paper. (i)~Maplike~discretization. We notice that the signals emanating from the AECG  have pronounced time-local peaks at various intervals. In this design, we create a discrete time-series dataset by extracting only the past peak signal values, then use the QEL framework to predict the future peak signal value. (ii)~Uniform~discretization. We sample the signal at defined intervals. However, in this situation, oversampling makes the problem resource-demanding and boost the prediction accuracy artificially. On the other hand, undersampling makes the prediction impossible. To have the right sampling rate, we propose to keep the discretization at the uniform interval determined by the autocorrelation timescale of the original signal. The autocorrelation of a signal \(u[n]\) is defined as
\begin{eqnarray}
	R[k]
	=
	\frac{
		\displaystyle
		\sum_{i=1}^{N_\mathrm{train}-k}
		\left(u[i]-\mu_{u}\right)
		\left(u[i+k]-\mu_{u}\right)
	}{
		\displaystyle
		\sum_{i=1}^{N_\mathrm{train}}
		\left(u[i]-\mu_{u}\right)^2
	}, \label{eq:ACF}
\end{eqnarray}
where \(k\) is the lag in samples, \(N_\mathrm{train}\) is the number of samples in the training interval, and
\(\mu_{u}
{=}
\frac{1}{N}
\sum_{i=1}^{N}u[i]\) is the average value of the training signal.
The autocorrelation $R[k]$ ranges from 0, indicating no correlation between signal values separated by a time-lag $k$, to 1, indicating full correlation between them. 
% We define the sampling interval $k^*$  as the time-lag at which the autocorrelation function of the given signal decays to  $1/e$, namely $R(k^*)=1/e$.

In all prediction frameworks, we use $5$--$210~\mu\mathrm{s}$ for training and $210$--$300~\mu\mathrm{s}$ for testing. The evolution Hamiltonian $H$ for the QEL is an open nearest-neighbor XYZ spin chain,
\begin{equation}
	H
	=
	\sum_{j=1}^{N_q-1}
	\left(
	J_j^{x}\sigma_x^{(j)}\sigma_x^{(j+1)}
	+
	J_j^{y}\sigma_y^{(j)}\sigma_y^{(j+1)}
	+
	J_j^{z}\sigma_z^{(j)}\sigma_z^{(j+1)}
	\right).
\end{equation}
The coupling coefficients are independently sampled from uniform distributions,
$\widetilde{J}_j^{\alpha}{\sim}\mathcal{U}(0,1)$, with
$\alpha{\in}\{x,y,z\}$. We vary $N_{\mathrm{steps}}$ and $N_{\mathrm{hid}}$ while restricting the total reservoir size to a maximum of $N_q{=}11$ qubits. The Hamiltonian evolution lasts for a time $\tau{=}N_q$, chosen to allow the encoded information to propagate throughout the entire qubit reservoir. The final measurement vector for one input window is
\begin{eqnarray}
	\mathbf{m}
	=
	\left[
	\langle\sigma_x^{(1)}\rangle,
	\langle\sigma_y^{(1)}\rangle,
	\langle\sigma_z^{(1)}\rangle,
	\ldots,
	\langle\sigma_x^{(N_q)}\rangle,
	\langle\sigma_y^{(N_q)}\rangle,
	\langle\sigma_z^{(N_q)}\rangle
	\right]^{\mathsf T}. 
\end{eqnarray}
To quantify the prediction performance of QEL, we use the root mean square error (RMSE) between the predicted and true values,
\begin{equation}
	\mathrm{RMSE}
	=
	\sqrt{
		\frac{1}{N_{\mathrm{test}}}
		\sum_{n=1}^{N_{\mathrm{test}}}
		\left(
		\tilde{u}[n]-u[n]
		\right)^2
	},
\end{equation}
where \(u[n]\) and \(\tilde{u}[n]\) denote the true and predicted values of a signal, respectively, and \(N_{\mathrm{test}}\) is the number of prediction samples in the testing interval. The RMSE is evaluated on the standardized (\(z\)-score) signal scale, with lower values indicating better prediction accuracy. For each reservoir configuration, 100 independent realizations of the coupling coefficients $\{\widetilde{J}_j^{\alpha}\}$ are generated, and the RMSE is averaged over these realizations to compare the predictive performance of different reservoir sizes.

Below, we discuss the forecasting performance of the QEL method with both maplike and uniform discretization methods.

\begin{figure*}[t]
	\centering
	
	% Prevent small unwanted spaces around subfigures
	\setlength{\fboxsep}{0pt}
	
	% =========================================================
	% Theory group
	% =========================================================
	\begin{tcolorbox}[
		enhanced,
		colback=white,
		colframe=black,
		boxrule=0.6pt,
		arc=0pt,
		left=3pt,
		right=3pt,
		top=-1pt,
		bottom=-8pt,
		title=Theory,
		fonttitle=\small\bfseries,
		coltitle=black,
		colbacktitle=white,
		attach boxed title to top left={
			xshift=8pt,
			yshift=-2mm
		},
		boxed title style={
			colback=white,
			colframe=white,
			boxrule=0pt,
			left=2pt,
			right=2pt,
			top=0pt,
			bottom=0pt
		}
		]
		\centering
		
		\begin{subfigure}{.235\linewidth}
			\centering
			\begin{overpic}[width=\textwidth]{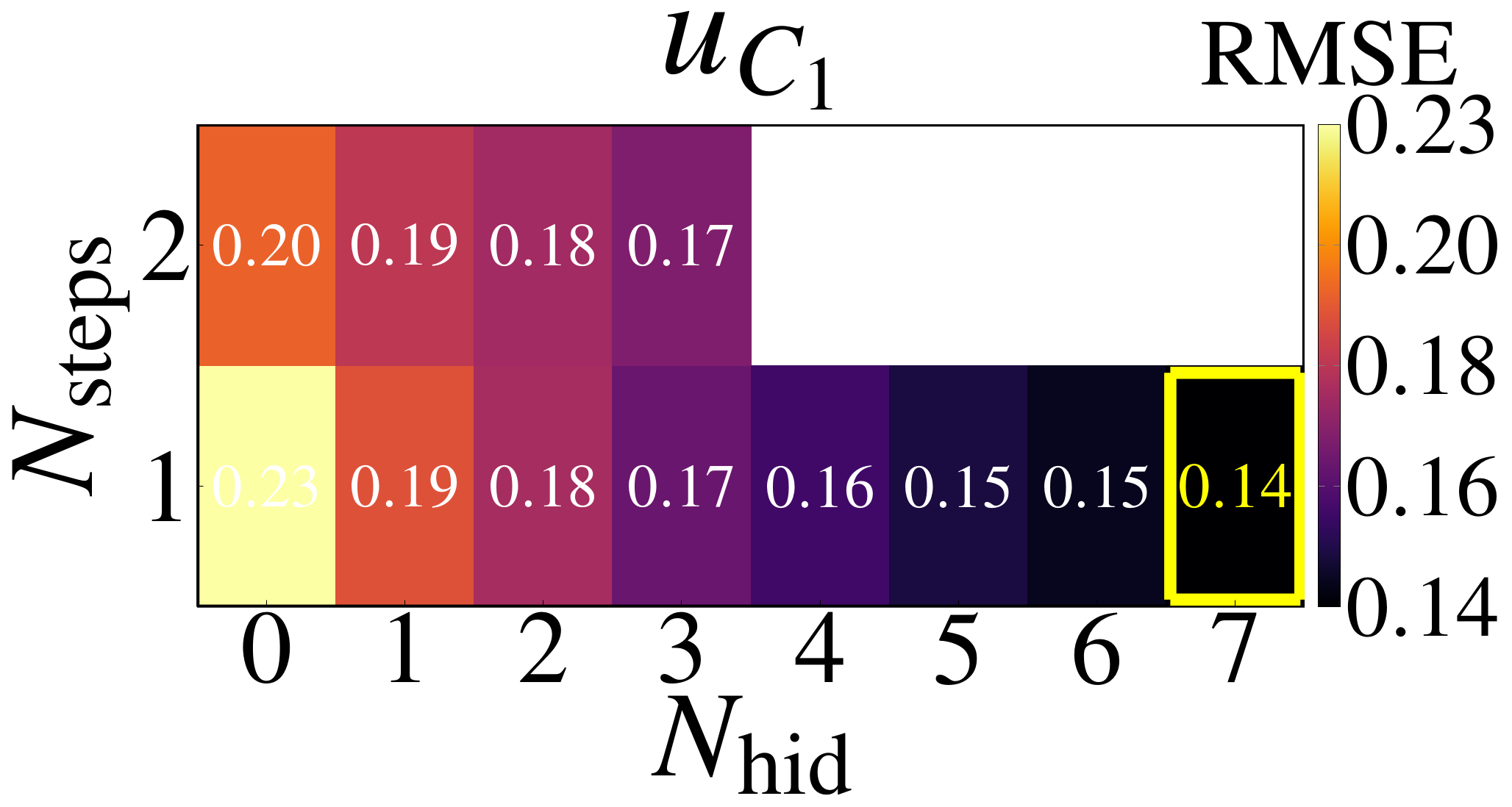}
				\put(-4,47){{(a)}}
			\end{overpic}
			\label{fig:u1heatmap4to1theory}
		\end{subfigure}
		\hfill
		\begin{subfigure}{.235\linewidth}
			\centering
			\begin{overpic}[width=\textwidth]{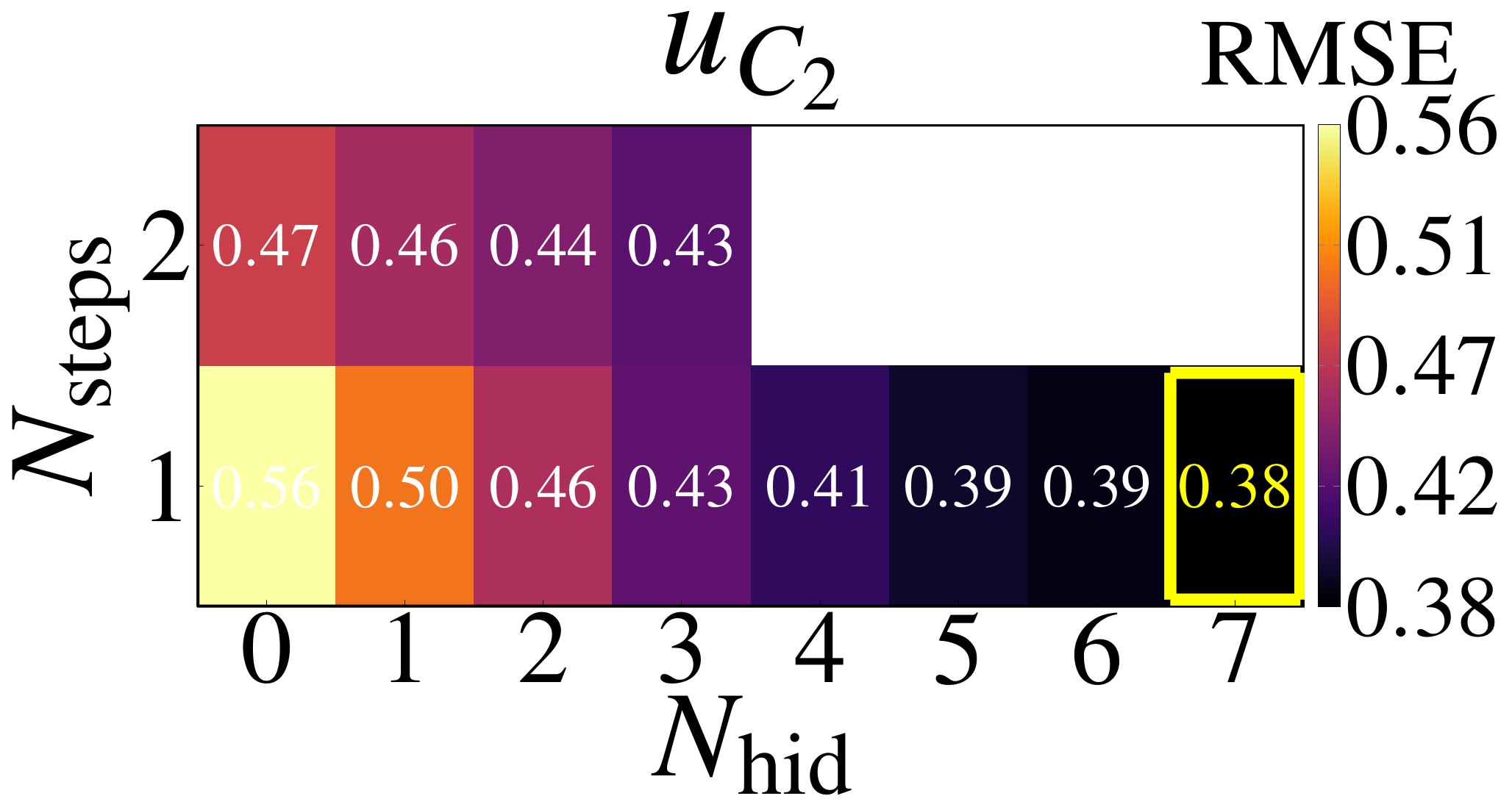}
				\put(-4,47){{(c)}}
			\end{overpic}
			\label{fig:u2heatmap4to1theory}
		\end{subfigure}
		\hfill
		\begin{subfigure}{.235\linewidth}
			\centering
			\begin{overpic}[width=\textwidth]{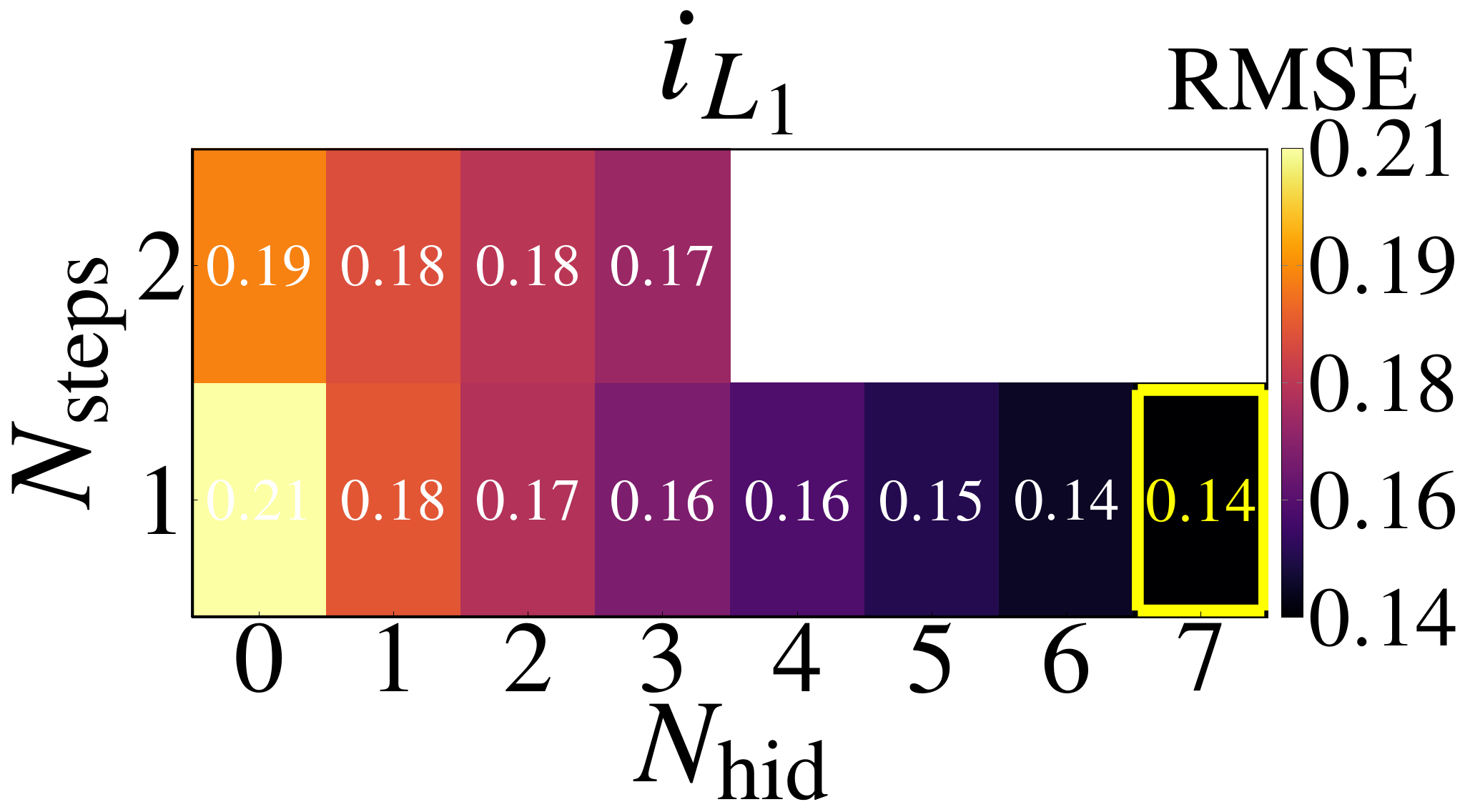}
				\put(-4,49){{(e)}}
			\end{overpic}
			\label{fig:i1heatmap4to1theory}
		\end{subfigure}
		\hfill
		\begin{subfigure}{.235\linewidth}
			\centering
			\begin{overpic}[width=\textwidth]{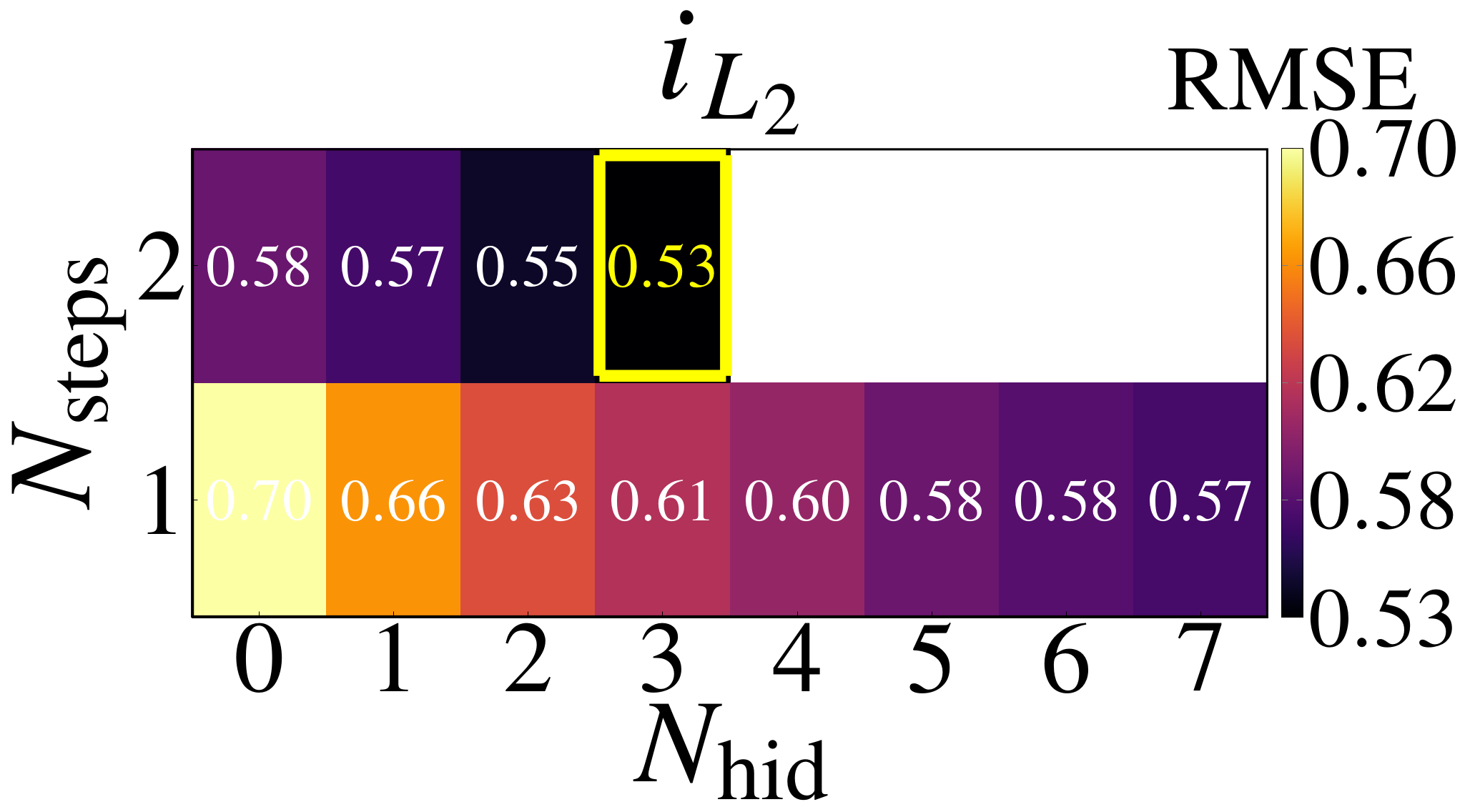}
				\put(-4,49){{(g)}}
			\end{overpic}
			\label{fig:i2heatmap4to1theory}
		\end{subfigure}
	\end{tcolorbox}
	
	\vspace{-0.25cm}
	
	% =========================================================
	% Experiment group
	% =========================================================
	\begin{tcolorbox}[
		enhanced,
		colback=white,
		colframe=black,
		boxrule=0.6pt,
		arc=0pt,
		left=3pt,
		right=3pt,
		top=-1pt,
		bottom=-8pt,
		title=Experiment,
		fonttitle=\small\bfseries,
		coltitle=black,
		colbacktitle=white,
		attach boxed title to top left={
			xshift=8pt,
			yshift=-2mm
		},
		boxed title style={
			colback=white,
			colframe=white,
			boxrule=0pt,
			left=2pt,
			right=2pt,
			top=0pt,
			bottom=0pt
		}
		]
		\centering
		
		\begin{subfigure}{.235\linewidth}
			\centering
			\begin{overpic}[width=\textwidth]{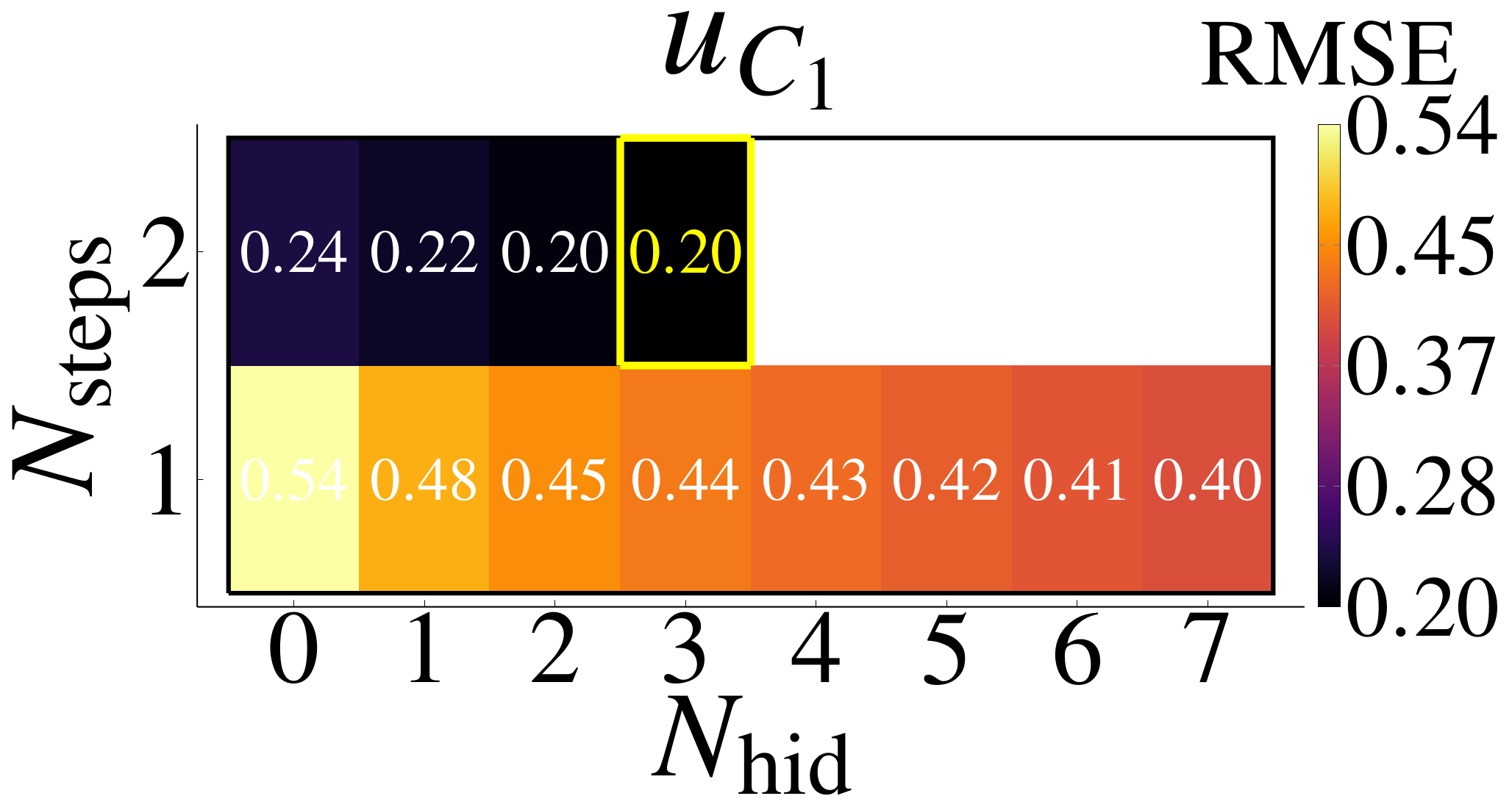}
				\put(-4,47){{(b)}}
			\end{overpic}
			\label{fig:u1heatmap4to1experiment}
		\end{subfigure}
		\hfill
		\begin{subfigure}{.235\linewidth}
			\centering
			\begin{overpic}[width=\textwidth]{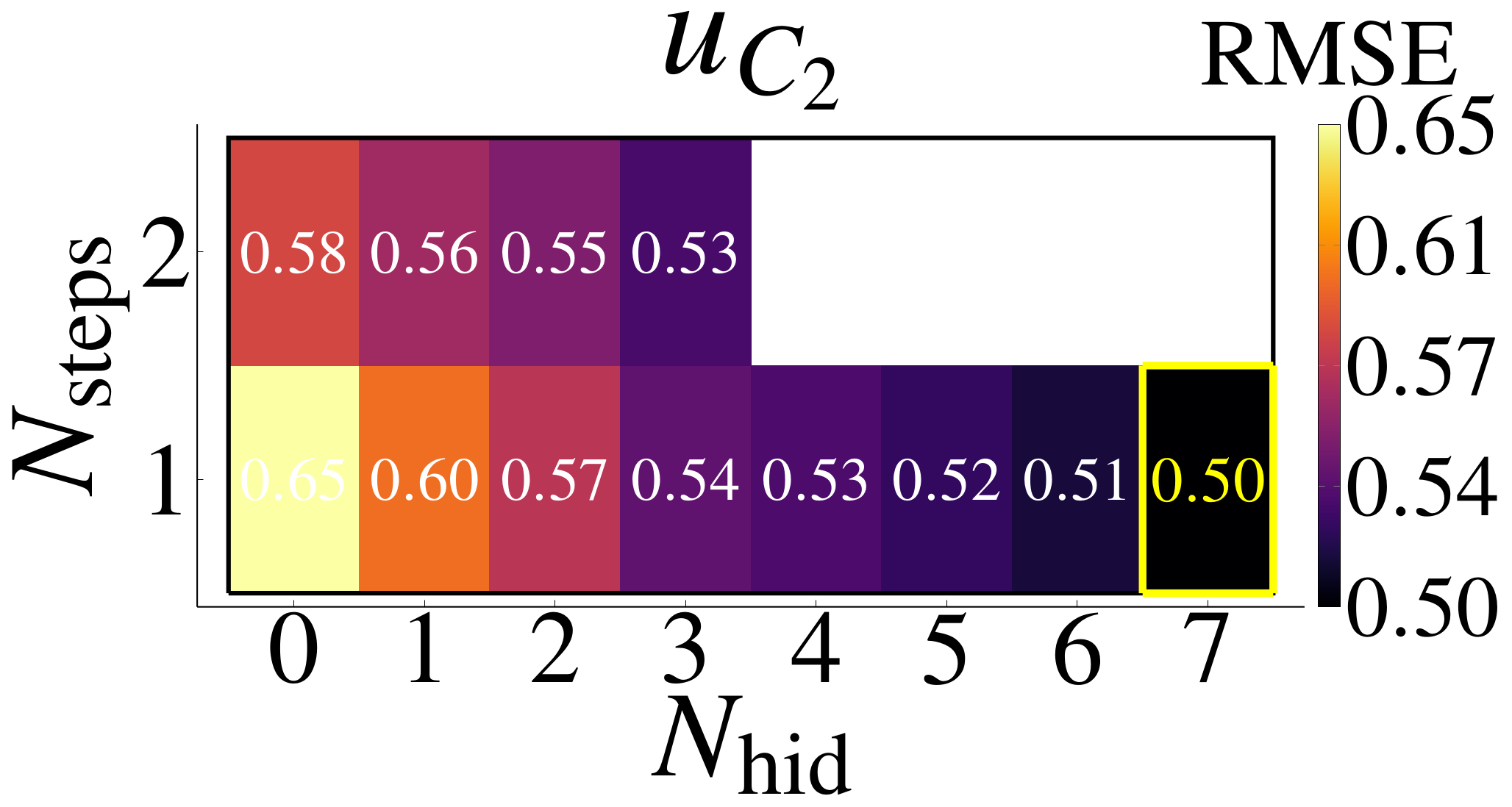}
				\put(-4,47){{(d)}}
			\end{overpic}
			\label{fig:u2heatmap4to1experiment}
		\end{subfigure}
		\hfill
		\begin{subfigure}{.235\linewidth}
			\centering
			\begin{overpic}[width=\textwidth]{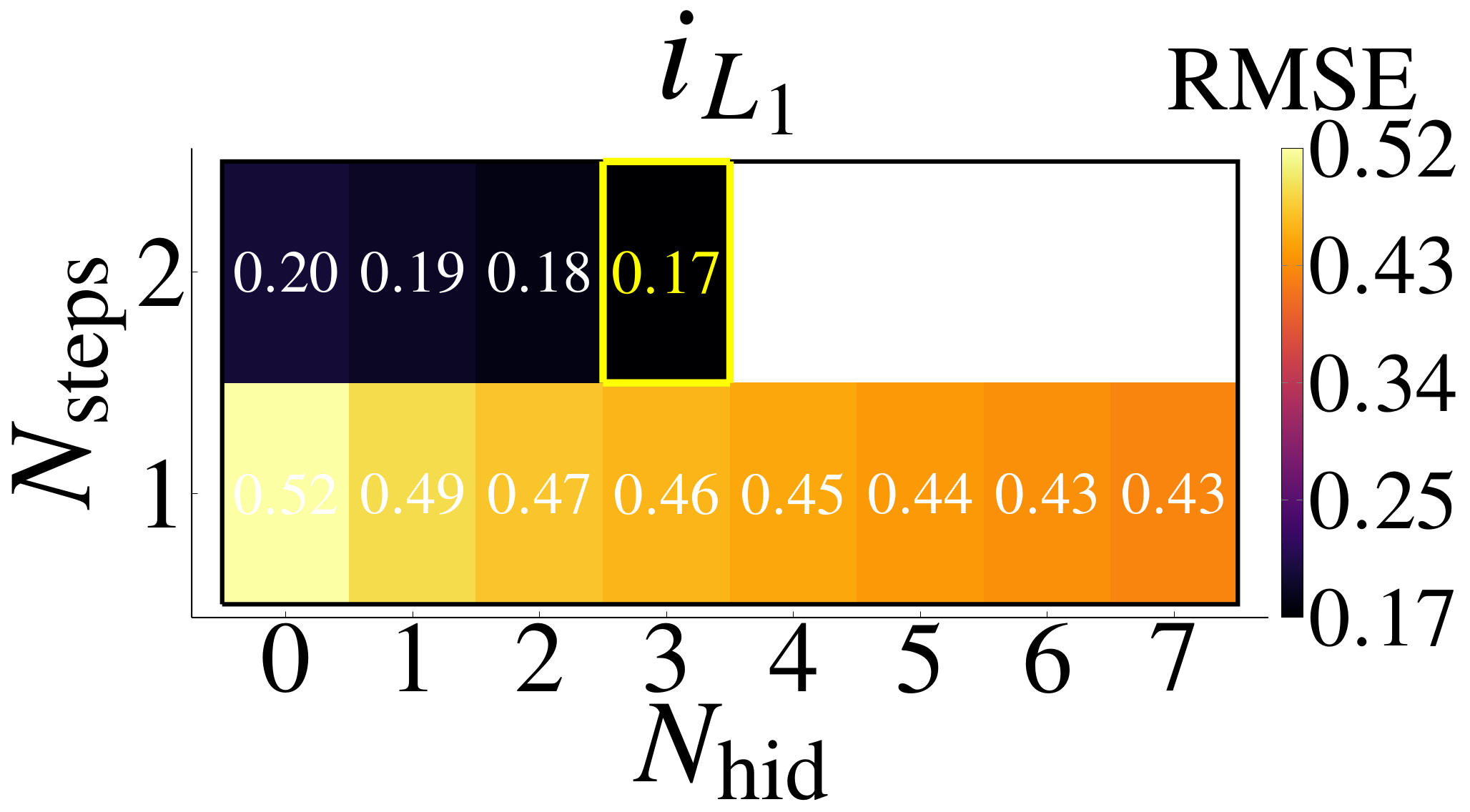}
				\put(-4,49){{(f)}}
			\end{overpic}
			\label{fig:i1heatmap4to1experiment}
		\end{subfigure}
		\hfill
		\begin{subfigure}{.235\linewidth}
			\centering
			\begin{overpic}[width=\textwidth]{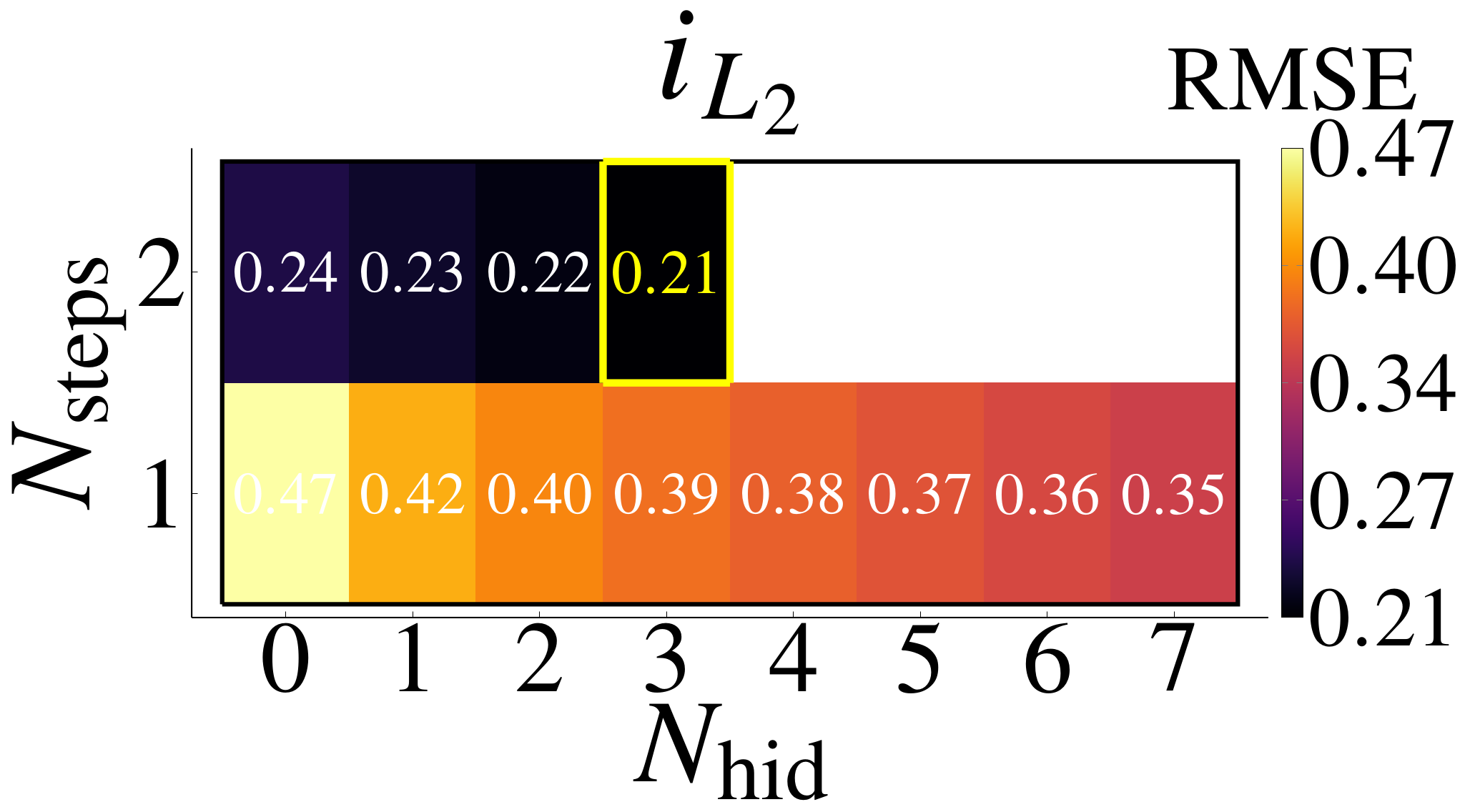}
				\put(-4,49){{(h)}}
			\end{overpic}
			\label{fig:i2heatmap4to1experiment}
		\end{subfigure}
	\end{tcolorbox}
	
	\caption{Performance heatmap for uniform discretization with  four-to-one prediction setting.
		The top row shows the theoretical results and the bottom row shows the experimental results.
		Panels (a,b), (c,d), (e,f), and (g,h) correspond to $u_{C_1}$, $u_{C_2}$, $i_{L_1}$, and $i_{L_2}$, respectively.}
	\label{fig:heatmap4to1}
\end{figure*}

\subsubsection*{I. Maplike Discretization}

The dynamical variables of AECG considered, viz., the currents $i_{L_1}, i_{L_2}$, and the voltages $u_{C_1}, u_{C_2}$ -- all show aperiodic chaotic oscillatory behaviour, as depicted in Fig.~\ref{fig:maplike1to1} by gray curves. This naturally lends itself to the maplike discretization scheme, where we sample the signals at successive peaks (or troughs) and predict the signal value at the next peak (or trough). Since the oscillations are \textit{not} periodic, this implies the sampling is over unequal intervals. The choice between peaks and troughs for discretization is governed by which exhibits greater variance. For example, since the peak values of $u_{C_2}$ show greater variation than the trough values, the peaks are selected for the maplike discretization of $u_{C_2}$. The prediction setting corresponds to one-to-one prediction, that is, we encode the past peak (or trough) values of each signal to predict the future peak (or trough) value of that particular signal and the other signals play no role in the prediction. This naturally implies that $N_q{=}N_\textrm{steps}{+}N_\textrm{hid}$.

As discussed above, we sweep $N_{\mathrm{steps}}$ and $N_{\mathrm{hid}}$ while restricting the total number of qubits to a maximum of $N_q{=}11$. For each reservoir configuration, 100 independent Hamiltonian realizations are considered. For each signal, the realization yielding the lowest RMSE is selected, and the corresponding prediction is shown in Fig.~\ref{fig:maplike1to1}. Panels (a,b), (c,d), (e,f), and (g,h) show the results for \(u_{C_1}\), \(u_{C_2}\), \(i_{L_1}\), and \(i_{L_2}\), respectively, with the left panels corresponding to the theoretical signals obtained from the AECG differential equations and the right panels to the experimental signals. The results visually indicate that QEL provides reasonably accurate predictions for all signals. To illustrate how the prediction performance varies with the number of qubits, Fig.~\ref{fig:heatmap1to1_maplike} shows the RMSE for different combinations of $N_{\mathrm{steps}}$ and $N_{\mathrm{hid}}$. For the theoretical signals, the lowest RMSE is obtained at $(N_{\mathrm{hid}},N_{\mathrm{steps}}){=}(6,5)$ for $u_{C_1}$, $(N_{\mathrm{hid}},N_{\mathrm{steps}}){=}(10,1)$ for $u_{C_2}$, $(N_{\mathrm{hid}},N_{\mathrm{steps}}){=}(8,3)$ for $i_{L_1}$, and $(N_{\mathrm{hid}},N_{\mathrm{steps}}){=}(9,2)$ for $i_{L_2}$. For the experimental signals, the corresponding optimal configurations are $(8,3)$, $(9,2)$, $(7,4)$, and $(7,4)$, respectively. One common feature is that, in all the four signals, increasing the total number of qubits $N_q$ improves the accuracy, though the optimal partitioning between $N_{\rm hid}$ and $N_{\rm steps}$ qubits may be signal dependent.  This clearly shows that scaling up the algorithm enhances the accuracy and yields  better performance. 

\begin{table*}[t]
	\centering
	\renewcommand{\arraystretch}{1.18}
	\setlength{\tabcolsep}{6pt}
	
	\begin{tabular}{
			!{\vrule width 0.9pt}c
			!{\vrule width 0.9pt}c
			!{\vrule width 0.9pt}ccc
			!{\vrule width 0.9pt}cccc
			!{\vrule width 0.9pt}
		}
		
		\noalign{\hrule height 0.9pt}
		
		\multirow{2}{*}{\textbf{Data}}
		&
		\multirow{2}{*}{\textbf{Target}}
		&
		\multicolumn{3}{c!{\vrule width 0.9pt}}{\textbf{QEL}}
		&
		\multicolumn{4}{c!{\vrule width 0.9pt}}{\textbf{LSTM}}
		\\
		
		\cline{3-5}
		\cline{6-9}
		
		&
		&
		$N_{\mathrm{steps}}$
		& $N_{\mathrm{hid}}$
		& RMSE
		& $N^{\mathrm{LSTM}}_{\mathrm{steps}}$
		& $d_{\mathrm{hid}}$
		& $N^{\mathrm{LSTM}}_{\mathrm{par}}$
		& RMSE
		\\
		
		\noalign{\hrule height 0.9pt}
		
		% ==================== Theory ====================
		
		\multirow{4}{*}{Theory}
		& $u_{C_1}$
		& 1 & 7 & $0.14 \pm 0.003$
		& 1 & 2 & 67
		& $0.14 \pm 0.01$
		\\
		
		& $u_{C_2}$
		& 1 & 7 & $0.38 \pm 0.004$
		& 1 & 1 & 30
		& $0.18 \pm 0.01$
		\\
		
		& $i_{L_1}$
		& 1 & 7 & $0.14 \pm 0.003$
		& 1 & 2 & 67
		& $0.13 \pm 0.02$
		\\
		
		& $i_{L_2}$
		& 2 & 3 & $0.53 \pm 0.01$
		& 1 & 1 & 30
		& $0.45 \pm 0.02$
		\\
		
		\noalign{\hrule height 0.8pt}
		
		% ==================== Experiment ====================
		
		\multirow{4}{*}{Experiment}
		& $u_{C_1}$
		& 2 & 3 & $0.20 \pm 0.01$
		& 2 & 2 & 67
		& $0.19 \pm 0.01$
		\\
		
		& $u_{C_2}$
		& 1 & 7 & $0.50 \pm 0.02$
		& 1 & 1 & 30
		& $0.39 \pm 0.01$
		\\
		
		& $i_{L_1}$
		& 2 & 3 & $0.17 \pm 0.01$
		& 2 & 3 & 112
		& $0.12 \pm 0.01$
		\\
		
		& $i_{L_2}$
		& 2 & 3 & $0.21 \pm 0.02$
		& 2 & 3 & 112
		& $0.16 \pm 0.02$
		\\
		
		\noalign{\hrule height 0.9pt}
		
	\end{tabular}
	
	\caption{
		Results for the uniform discretization task with the four-to-one prediction setting using QEL reservoirs of up to \(N_q{=}11\) qubits, together with the LSTM results. The QEL RMSE is reported as the mean $\pm$ standard deviation over 100 Hamiltonian realizations, whereas the LSTM RMSE is reported as the mean $\pm$ standard deviation over five random seeds. Among the LSTM architectures whose mean RMSE reaches or outperforms the corresponding QEL result, the architecture with the smallest number of trainable parameters is reported. The LSTM temporal depth is restricted to $N^{\mathrm{LSTM}}_{\mathrm{steps}}{\leq}2$ to match the maximum temporal depth considered in in the four-to-one QEL analysis.
	}
	\label{tab:LSTM}
\end{table*}

\subsubsection*{II. Uniform Discretization}

We have shown in the previous part that QEL with maplike discretization based encoding is broadly successful in term of prediction. Now, we report the results for uniform discretization. As we noted above, the sampling interval is determined on the basis of the decay of the corresponding autocorrelation function. The autocorrelation functions are  depicted in Fig.~\ref{fig:ACF}(a) for the theoretical simulations and Fig.~\ref{fig:ACF}(b) for the experimental data. Specifically, for each target signal, we choose the lag \(m\) as the first sample lag at which its autocorrelation function decreases below \(1/e\). Thus, the input samples in Eq.~\eqref{eq:input} are separated by \(m\) acquisition intervals, corresponding to a temporal separation of \(m\Delta t\), with \(\Delta t{=}3.2~\mathrm{ns}\). This operation alleviates the issues of the strong correlation that exists between temporally neighboring samples, while retaining a sufficient level of temporal predictability for forecasting. The corresponding values of \(m\) are determined separately for each theoretical and experimental target signal from its training data. While the samples within a given input window are separated by \(m\), successive input windows are advanced by only one acquisition sample at a time.

Now we note an important fact -- with maplike discretization, each signal has peak values at different times. Therefore with maplike discretization we were forced to resort to one-to-one prediction, i.e., for forecasting any signal we could not feed the past data of other signals into the model, potentially depriving the QEL from another information source. 
However, uniform discretization enables four-to-one prediction,
in which the past values of all four signals, represented by
\(\mathbf{x}[n]
{=}
\left[
u_{C_1}[n],
u_{C_2}[n],
i_{L_1}[n],
i_{L_2}[n]
\right]^{\mathsf{T}}\),
are jointly encoded to predict the future value of any one
of these signals. For completeness, we also evaluate the uniform discretization scheme in the one-to-one setting, where each signal is predicted solely from its own past values, considering reservoir sizes up to $N_q{=}11$. These results are provided in the Supplementary Material.  

\begin{figure}[t]
	\centering
	
	\begin{overpic}[width=0.49\textwidth]
		{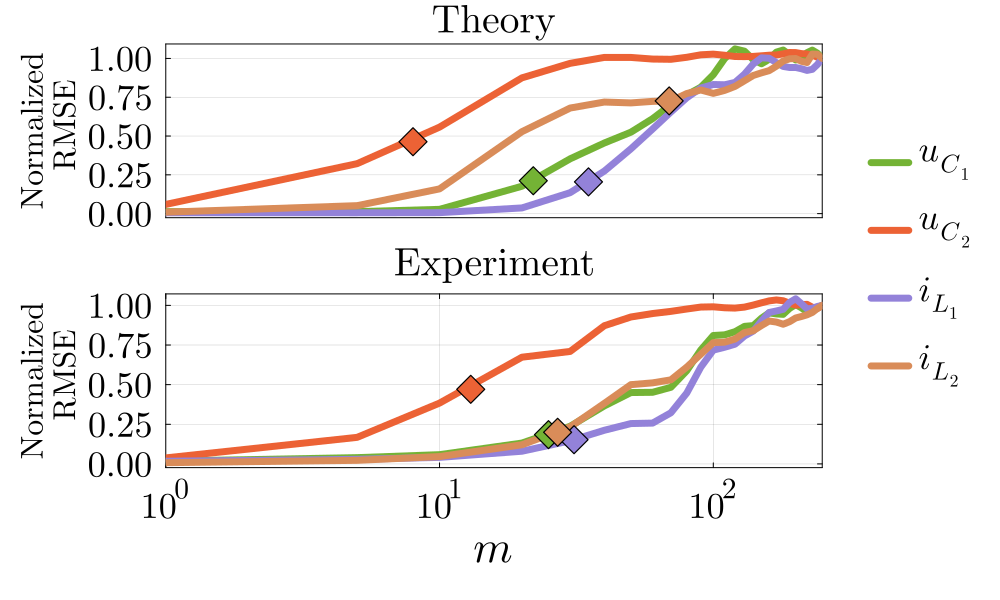}
		\put(0,57){{(a)}}
		\put(0,32){{(b)}}
	\end{overpic}
	\vspace{-25pt}
	\caption{Prediction error versus input lag. (a) Theoretical signals and (b) experimental signals. For each signal, the RMSE is normalized by its value at \(m{=}250\), so that the normalized RMSE equals unity at this large lag. The markers indicate the lag selected from the \(1/e\) crossing of the corresponding autocorrelation function (see Fig.~\ref{fig:ACF}). }
	\label{fig:RMSE_lag}
\end{figure}

As before, we first illustrate the QEL-based prediction using uniform discretization in Fig.~\ref{fig:timeseries4to1}, where the best configuration is selected by sweeping the other model parameters while restricting the reservoir size to $N_q{=}11$ qubits. The left column depicts predictions for the theoretical data obtained by numerically solving the AECG differential equations, while the right column depicts predictions for the experimental data. Panels (a,b), (c,d), (e,f), and (g,h) correspond to $u_{C_1}$, $u_{C_2}$, $i_{L_1}$, and $i_{L_2}$, respectively. The results visually illustrate several things. (i) Generally, the fit between prediction and ground truth in uniform discretization appears much tighter than the maplike discretization.
(ii) QEL succeeds in almost perfectly predicting the  chaotic oscillations with a pronounced oscillatory behaviour of $u_{C_1}$ and $i_{L_1}$ and has remarkable success in predicting the large amplitude resonances of $u_{C_2}$ for both theoretical simulated as well as experimental datasets. (iii) The ``worst" performance of QEL occurs for prediction of theoretically simulated time-series of $i_{L_2}$ exhibiting sawtooth-like chaotic oscillations. However, even in this case, the prediction is potentially valuable as it serves as a leading, rather than a lagging indicator of the broad temporal trend of sharp periodic drops of current $i_{L_2}$. Moreover, as the experimental data smoothens out these chaotic oscillations in $i_{L_2}$, the prediction becomes remarkably more accurate.

\begin{figure*}[t]
	\centering
	
	% ==================== Left column: Theory ====================
	\begin{minipage}[t]{0.485\textwidth}
		\vspace{0pt}
		\begin{tcolorbox}[
			enhanced,
			colback=white,
			colframe=black,
			boxrule=0.6pt,
			arc=0pt,
			left=3pt,
			right=3pt,
			top=5pt,
			bottom=-5pt,
			title=Theory,
			fonttitle=\small\bfseries,
			coltitle=black,
			colbacktitle=white,
			attach boxed title to top left={xshift=8pt,yshift=-2mm},
			boxed title style={
				colback=white,
				colframe=white,
				boxrule=0pt,
				left=2pt,
				right=2pt,
				top=0pt,
				bottom=0pt
			}
			]
			\centering
			
			% ==================== Row 1 ====================
			\begin{subfigure}[t]{\linewidth}
				\centering
				\begin{overpic}[width=\textwidth]{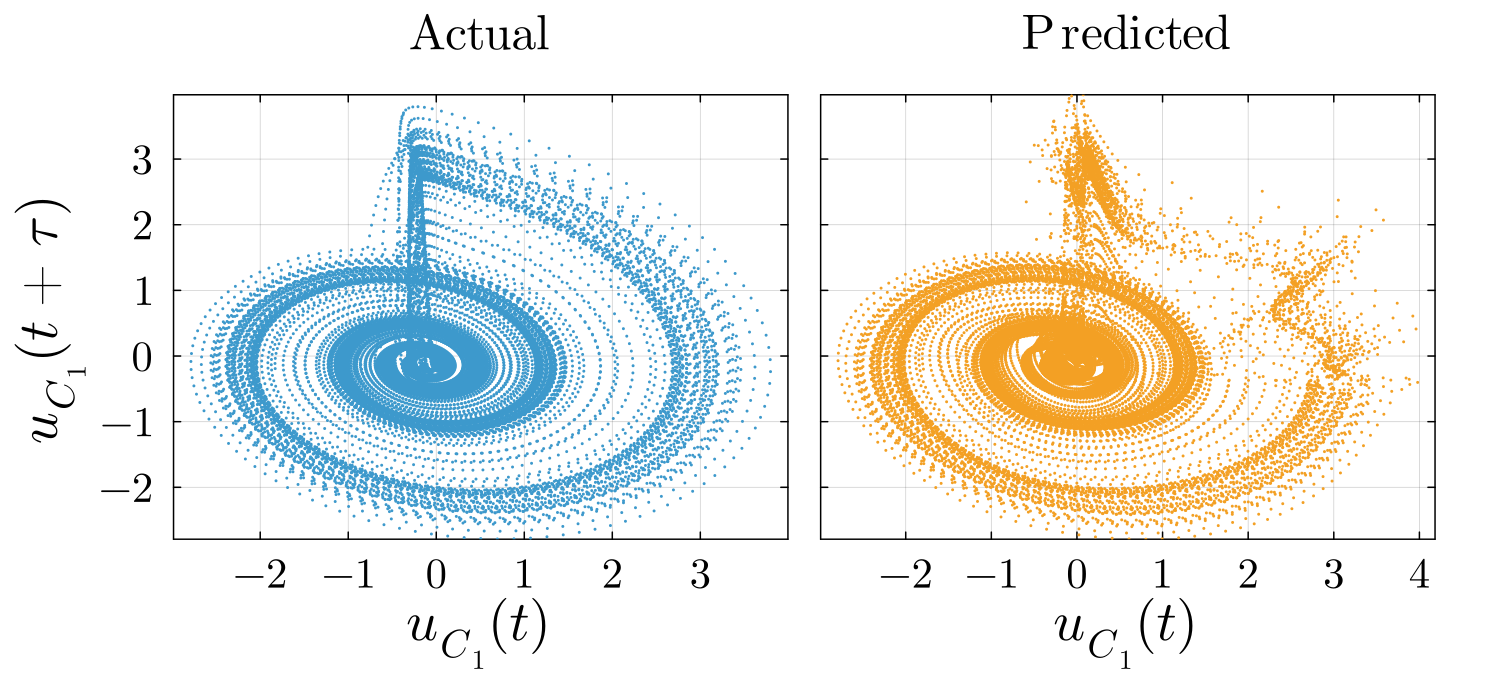}
					\put(-1,37){{(a)}}
				\end{overpic}
				\label{fig:u1attractor_theory}
			\end{subfigure}
			
			\vspace{-12pt}
			
			% ==================== Row 2 ====================
			\begin{subfigure}[t]{\linewidth}
				\centering
				\begin{overpic}[width=\textwidth]{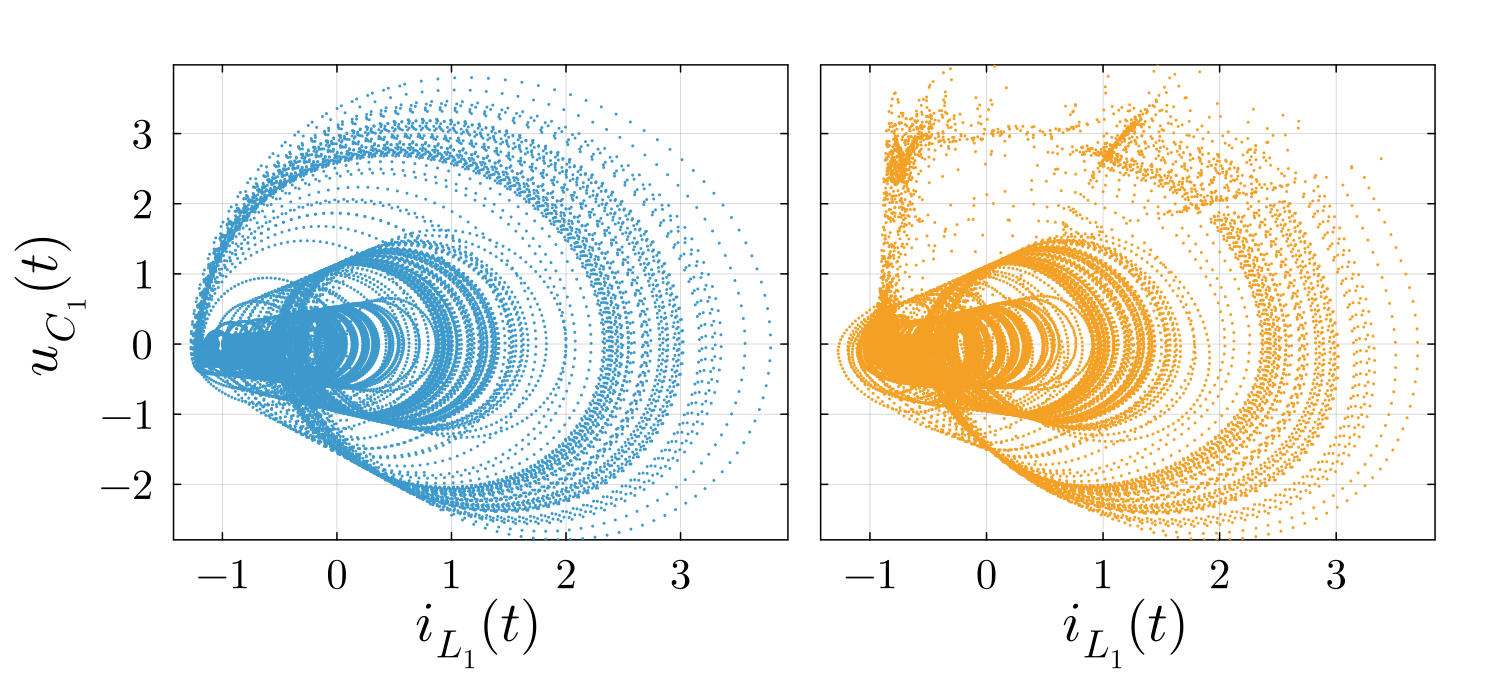}
					\put(-1,39){{(c)}}
				\end{overpic}
				\label{fig:i1u1attractor_theory}
			\end{subfigure}
			
			\vspace{-12pt}
			
			% ==================== Row 3 ====================
			\begin{subfigure}[t]{\linewidth}
				\centering
				\begin{overpic}[width=\textwidth]{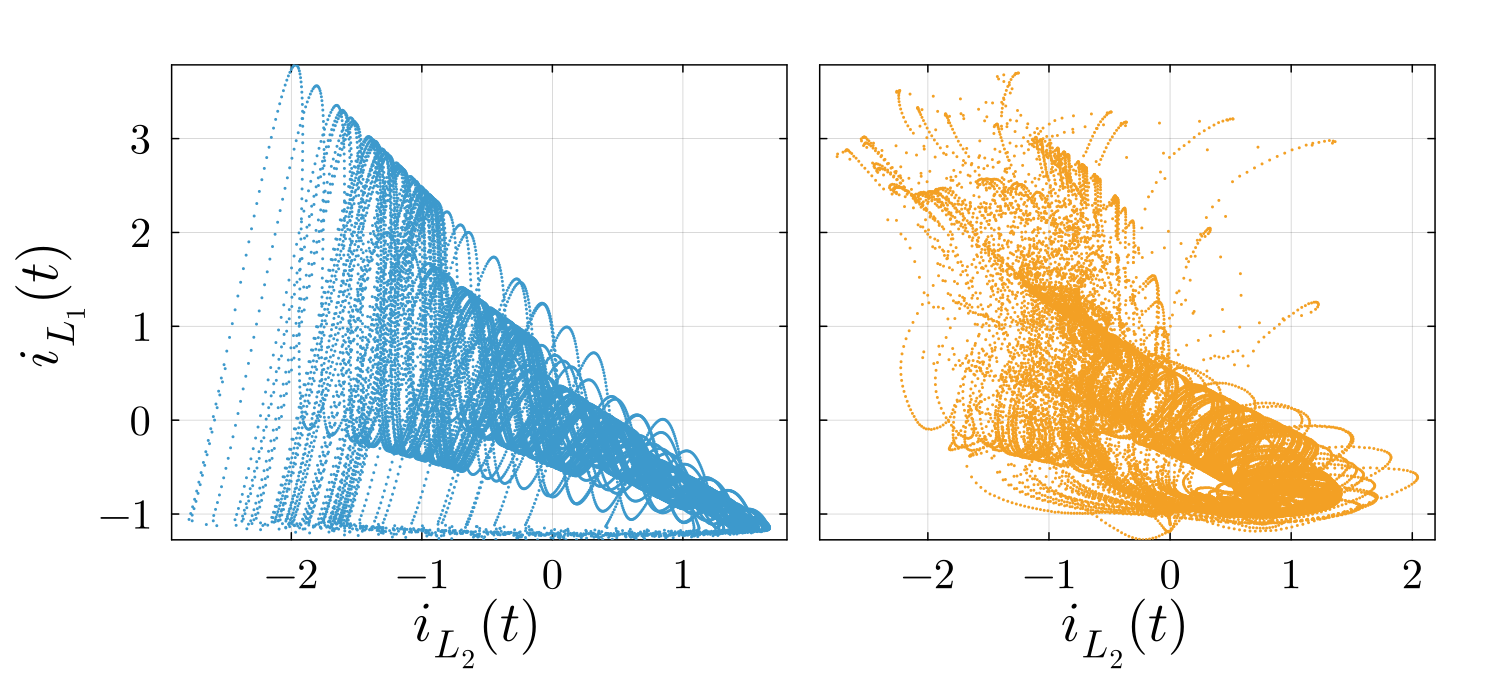}
					\put(-1,39){{(e)}}
				\end{overpic}
				\label{fig:i2i1attractor_theory}
			\end{subfigure}
		\end{tcolorbox}
	\end{minipage}
	\hfill
	% ==================== Right column: Experiment ====================
	\begin{minipage}[t]{0.485\textwidth}
		\vspace{0pt}
		\begin{tcolorbox}[
			enhanced,
			colback=white,
			colframe=black,
			boxrule=0.6pt,
			arc=0pt,
			left=3pt,
			right=3pt,
			top=5pt,
			bottom=-5pt,
			title=Experiment,
			fonttitle=\small\bfseries,
			coltitle=black,
			colbacktitle=white,
			attach boxed title to top left={xshift=8pt,yshift=-2mm},
			boxed title style={
				colback=white,
				colframe=white,
				boxrule=0pt,
				left=2pt,
				right=2pt,
				top=0pt,
				bottom=0pt
			}
			]
			\centering
			
			% ==================== Row 1 ====================
			\begin{subfigure}[t]{\linewidth}
				\centering
				\begin{overpic}[width=\textwidth]{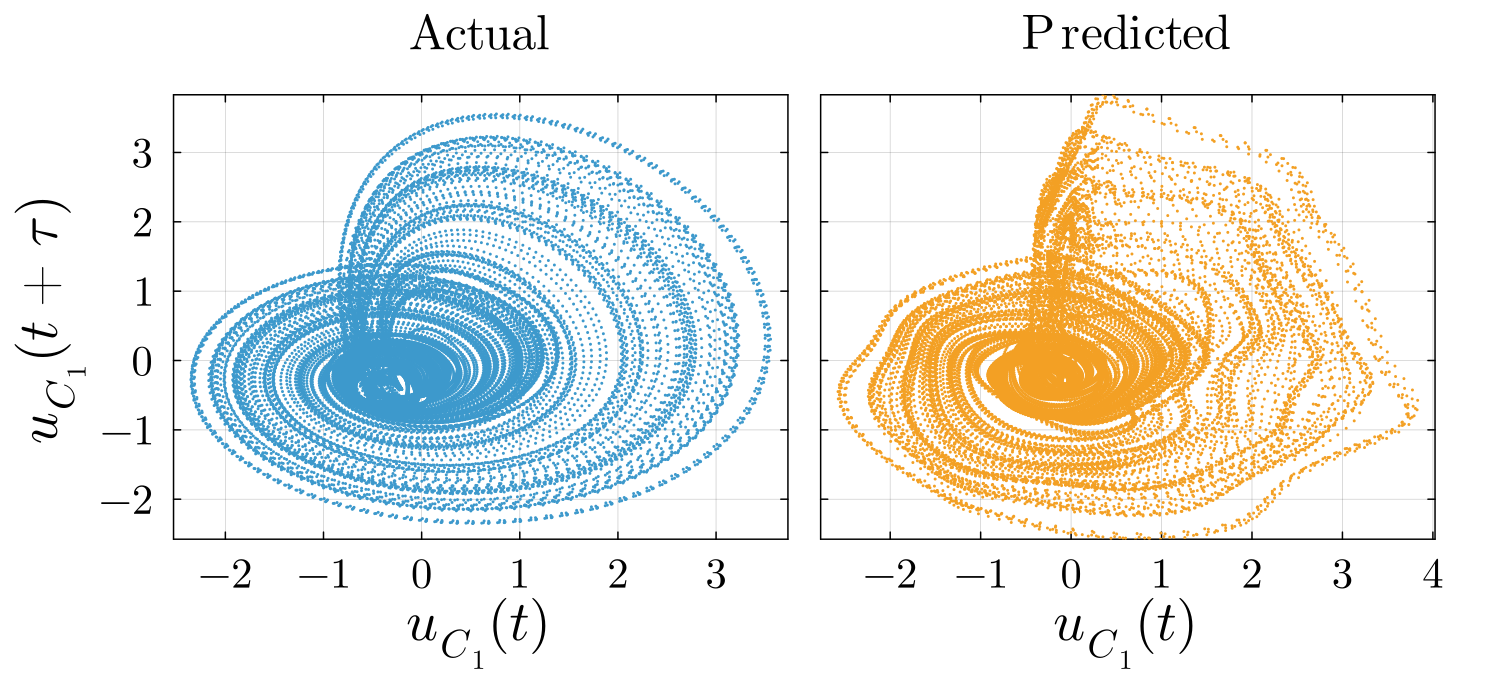}
					\put(-1,37){{(b)}}
				\end{overpic}
				\label{fig:u1attractor_experiment}
			\end{subfigure}
			
			\vspace{-12pt}
			
			% ==================== Row 2 ====================
			\begin{subfigure}[t]{\linewidth}
				\centering
				\begin{overpic}[width=\textwidth]{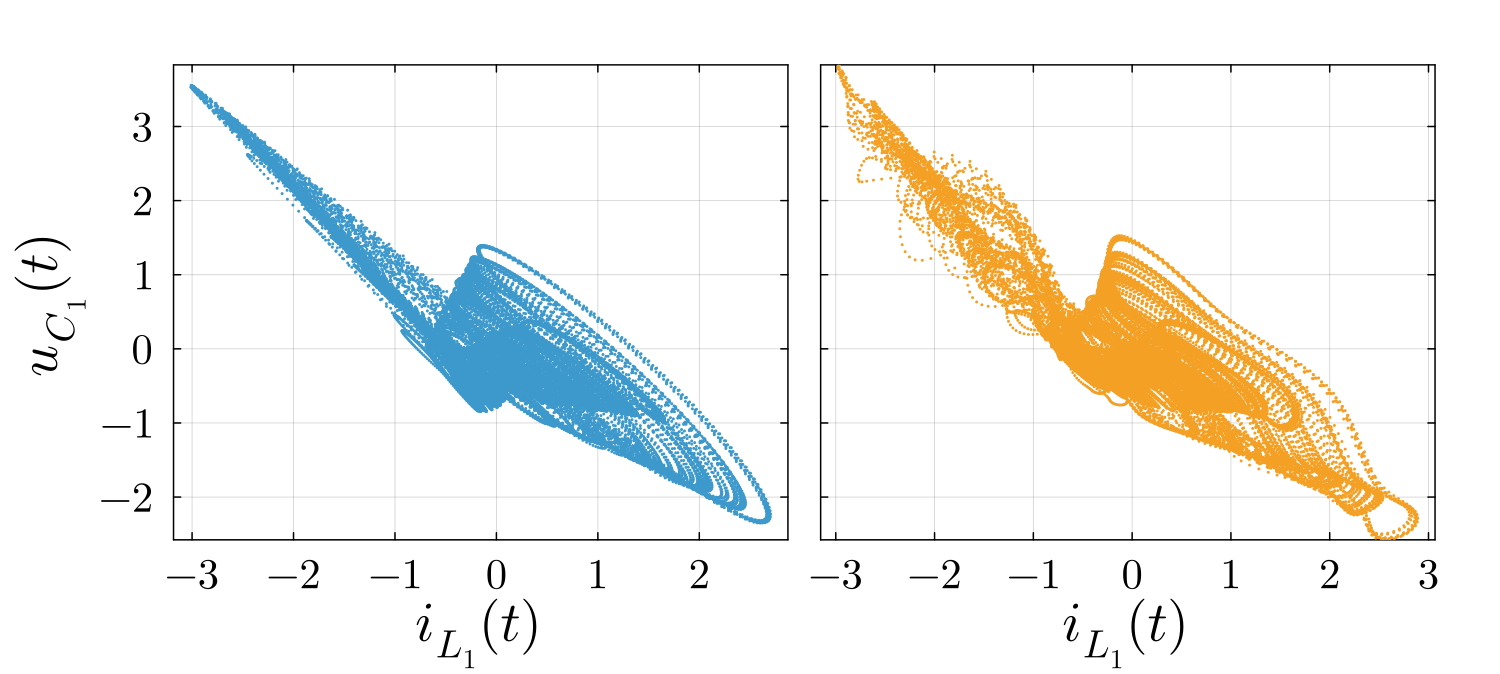}
					\put(-1,39){{(d)}}
				\end{overpic}
				\label{fig:i1u1attractor_experiment}
			\end{subfigure}
			
			\vspace{-12pt}
			
			% ==================== Row 3 ====================
			\begin{subfigure}[t]{\linewidth}
				\centering
				\begin{overpic}[width=\textwidth]{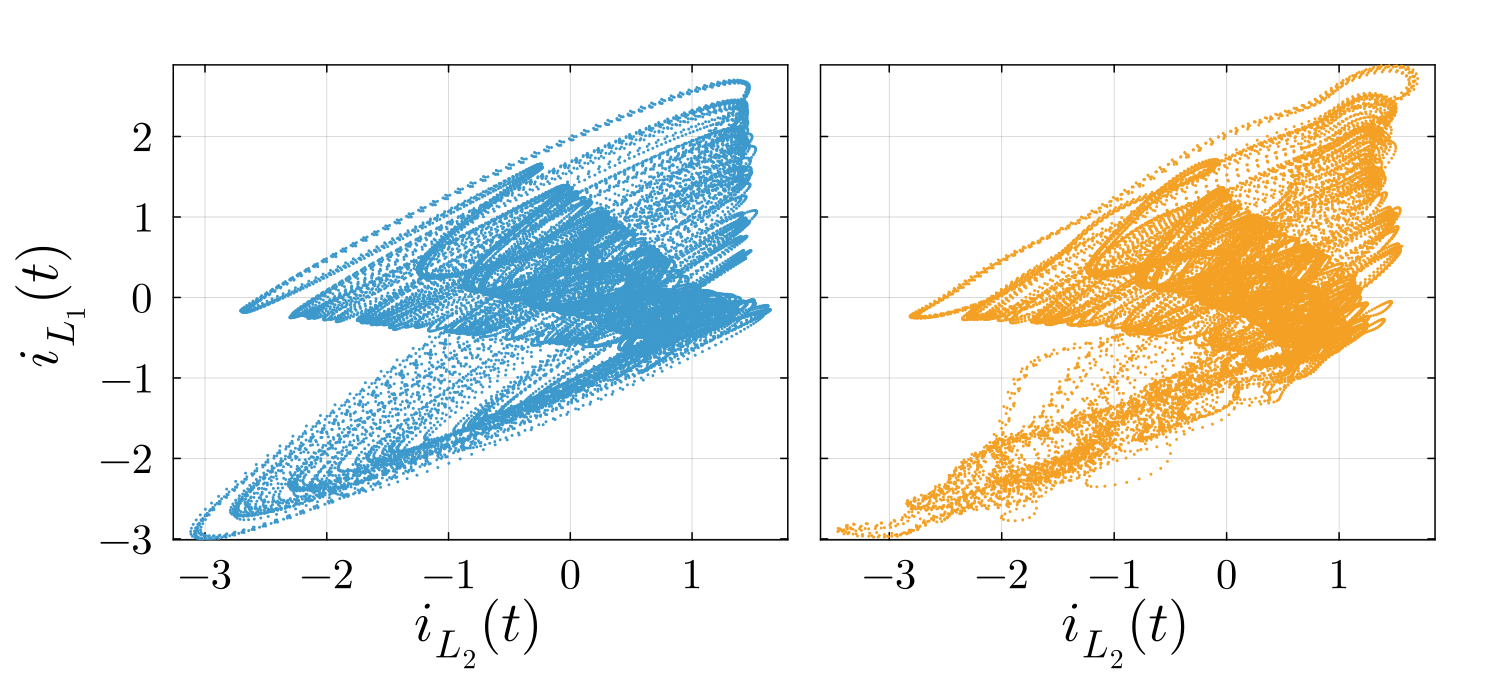}
					\put(-1,39){{(f)}}
				\end{overpic}
				\label{fig:i2i1attractor_experiment}
			\end{subfigure}
		\end{tcolorbox}
	\end{minipage}
	
	\caption{Attractor reconstruction for the uniform discretization prediction task. The left and right columns show the theoretical and experimental results, respectively, with the actual and predicted attractors displayed side by side in each panel. Panels (a,b) show the self-attractor of $u_{C_1}$, constructed using a delay-coordinate embedding with $\tau{=}30$ samples ($96\,\mathrm{ns}$). Panels (c,d) show the $u_{C_1}$--$i_{L_1}$ projections for the simulated and experimental data, respectively. Panels (e,f) show the corresponding $i_{L_1}$--$i_{L_2}$ projections. The cross attractors are constructed from simultaneous samples and therefore use no time delay.}
	\label{fig:attractor}
\end{figure*}

Following this visual illustration, we now report the quantitative aspects of prediction accuracy, as quantified by RMSE-error, with uniform discretization.The results are summarized in Fig.~\ref{fig:heatmap4to1}. The top row shows the theoretical results and the bottom row shows the experimental results. Panels (a,b), (c,d), (e,f), and (g,h) correspond to $u_{C_1}$, $u_{C_2}$, $i_{L_1}$, and $i_{L_2}$, respectively. Note that the total number of qubits used in QEL is $N_q{=}4N_{\textrm{steps}}{+}N_{\textrm{hid}}$, as we consider the four-to-one prediction setting in each case. For the theoretical signals, the lowest RMSE is obtained at $(N_{\mathrm{hid}},N_{\mathrm{steps}}){=}(7,1)$ for $u_{C_1}$, $u_{C_2}$, and $i_{L_1}$, and at $(N_{\mathrm{hid}},N_{\mathrm{steps}}){=}(3,2)$ for $i_{L_2}$. For the experimental signals, the corresponding optimal configurations are
$(3,2)$, $(7,1)$, $(3,2)$, and $(3,2)$ for
$u_{C_1}$, $u_{C_2}$, $i_{L_1}$, and $i_{L_2}$, respectively.
Notably, all of these optimal configurations use the maximum reservoir size
considered, 	$N_q{=}N_{\mathrm{hid}}{+}4N_{\mathrm{steps}}{=}11$. This again shows that the algorithm improves by scaling up the total number of qubits.    
The best QEL results obtained from these parameter sweeps, together with
their corresponding RMSE values, are summarized in
Table~\ref{tab:LSTM}.

Three trends emerge from the above results. (i) The prediction accuracy is indeed generally higher (RMSE-error lower) in all cases compared to the maplike discretization with the same QEL setting, as visually suggested by Fig.~\ref{fig:timeseries4to1} previously. (ii) As a general rule, increasing the number of hidden qubits always improves predictive accuracy, thus validating the core logic of QEL of scrambling the encoded information into a bigger reservoir through evolution. (iii) Increasing the input history, indexed by $N_{\mathrm{steps}}$, was more beneficial for several experimental signals than for the simulated ones. One reason may be attributed to the inherent difference between the model state and the measured signals. In Eq.~\eqref{eq:AECG}, the four simultaneous values fully specify the state. Under fixed parameters, they determine the future trajectory, and hence the past points become irrelevant in the absence of noise. A finite predictor may nevertheless benefit from them. In experiments, the set of observed variables does not necessarily resolve all the degrees of freedom of the oscillator and associated readout electronics. They are, furthermore, inevitably subject to noise. Higher-order aspects of the transistor dynamics omitted from the model, further parasitic elements and finite-bandwidth amplification are possible sources of unresolved dynamics. Previous observations can aid in identifying states that look identical in one moment but evolve diﬀerently afterwards, hence alleviating measurement uncertainties. The present results do not isolate a particular nonideality, such as inductor self-resonance and dissipation, as the cause.

As mentioned above, to determine an appropriate temporal spacing between the input samples, we use the first \(1/e\) crossing of the autocorrelation function of each signal as the corresponding lag \(m\) in Eq.~\eqref{eq:input}. For both the theoretical and experimental signals, one unit of lag corresponds to one sampling interval of \(3.2~\mathrm{ns}\). Fig.~\ref{fig:RMSE_lag} shows the prediction error as a function of \(m\) for (a) the theoretical signals and (b) the experimental signals, with the selected autocorrelation lags indicated by stars on the corresponding curves. For each signal, the optimal reservoir configuration and the corresponding best Hamiltonian realization identified from the parameter sweep are kept fixed, and only the lag $m$ is varied. The selected autocorrelation lags are indicated by the markers on the corresponding curves. For each signal, the RMSE is normalized by its value at \(m{=}250\), such that the normalized RMSE equals unity at \(m{=}250\). This large lag is taken as a reference corresponding to a regime in which short-range temporal correlations have substantially decayed. In general, smaller values of \(m\) can lead to a lower prediction error, simply reflecting the fact that neighboring samples are strongly correlated. In this regime, a substantial portion of the apparent predictive performance is actually supported by the short-time correlations, rather than actually reflecting the model learning the underlying longer-range dynamical evolution. As \(m\) is increased, this confounding correlation decreases in its intensity, and the prediction task consequently becomes more demanding, since the reservoir must infer the dynamics from more weakly correlated observations. We therefore choose the lag at which the autocorrelation function drops under \(1/e\) as a physically motivated compromise, which provides sufficient temporal separation reducing the influence of trivial short-range correlations while retaining enough dynamical dependence between successive input samples to support forecasting. Accordingly, as Fig.~\ref{fig:RMSE_lag} clearly depicts, the normalized RMSE initially shows a relative stable plateau-like behavior until $m$ reaches the autocorrelation time after which it rises quite rapidly. This demonstrates that our choice of discretization time, determined by autocorrelation, provides an optimal threshold between accuracy and sampling efficiency.

So far, we have demonstrated the impressive predictive accuracy of the QEL with uniform discretization.  However, this still leaves open the possibility that this is an artifact of the specific time-periods chosen for study. The full reconstruction of the attractor diagrams of the AECG is thus essential to  validate the success of QEL for prediction. To do so, we employ the standard delay-coordinate embedding based on Takens' theorem \cite{takens2006detecting}. For $u_{C_1}(t)$, we use a delay of 30 samples, with the resulting attractors shown in Figs.~\ref{fig:attractor}(a) and \ref{fig:attractor}(b) for the theoretical and experimental data, respectively. We further examine the joint dynamics through cross-attractor projections constructed from simultaneously sampled signals. The cross attractor between $u_{C_1}(t)$ and $i_{L_1}(t)$ is shown in Figs.~\ref{fig:attractor}(c) and \ref{fig:attractor}(d) for the theoretical and experimental data, respectively. Similarly, Figs.~\ref{fig:attractor}(e) and \ref{fig:attractor}(f) show the cross attractor between $i_{L_1}(t)$ and $i_{L_2}(t)$ for the theoretical and experimental data, respectively. The results illustrate the success of QEL in reproducing the attractors. 
%	\sout{Especially striking is the illustration in Fig.~\ref{fig:attractor} that the attractor reconstruction is  more accurate for actual experimental data. This is especially apparent for the current $i_{L_2}$ where the chaotic sawtooth oscillations in the theoretical simulations are smoothened out in the experimental data, where the resulting attractor diagram is reproduced by QEL with remarkable fidelity. }
For the theoretical signals, the predictions preserve the central
spiral in the voltage-delay projection (Fig.~\ref{fig:attractor}(a))
and the nested loops in the $u_{C_1}$--$i_{L_1}$ projection
(Fig.~\ref{fig:attractor}(c)). The agreement is less satisfactory
along the outer excursions, where the predicted branches are
displaced, locally distorted and more dispersed. For the experimental signals, the elongated,
negatively sloping voltage--current pattern
(Fig.~\ref{fig:attractor}(d)) and the oblique, folded arrangement of
the two currents (Fig.~\ref{fig:attractor}(f)) are both recognizable
in the predictions. In particular, the current--current geometry is
retained more clearly than in the theoretical case.  The experimental
voltage-delay plot (Fig.~\ref{fig:attractor}(b)) the central rotation is retained, but the surrounding
loops become flatter and more angular, with appreciable dispersion
along the upper excursions. The
comparison supports partial recovery of the phase-space organization,
that can be enhanced with larger QEL systems.

%	\subsection*{Robustness of Prediction}

\subsection*{Benchmarking with a classical predictor}
Quantum prediction tools are still at their infancy and hardware limitations compelled us to simulate the effects of a small quantum computer on our classical computing devices. In contrast, classical prediction is a long-established discipline with numerous techniques. We benchmark the predictive accuracy of QEL against a popular classical machine-learning based prediction tool - namely, Long Short-Term Memory (LSTM) neural networks~\cite{hochreiter1997long}. We focus on the four-to-one uniform discretization prediction task and use the same training and testing intervals, signal normalization, and target-specific sampling lags as in the QEL analysis. To ensure a comparable architecture for the two predictors, the LSTM input history is restricted to at most two previous time steps, \(N^{\mathrm{LSTM}}_{\mathrm{steps}}{\leq}2\), matching the maximum \(N_{\mathrm{steps}}\) considered in the four-to-one QEL analysis, and a single recurrent layer is used to reflect the single reservoir-evolution stage of QEL.

Table~\ref{tab:LSTM} compares the optimal QEL results with the smallest tested LSTM architectures whose five-seed average RMSE reaches or outperforms the corresponding QEL error criterion. For the LSTM, model complexity is characterized by $N^{\mathrm{LSTM}}_{\mathrm{steps}}$, the hidden-state dimension, and the resulting number of trainable parameters. The details of the LSTM architecture are provided in the Methods section.

\begin{figure}[t]
	\centering
	
	\begin{tikzpicture}
		
		% ==================== Decoherence figure ====================
		\node[inner sep=0pt] (img) {
			\includegraphics[
			width=0.48\textwidth
			]{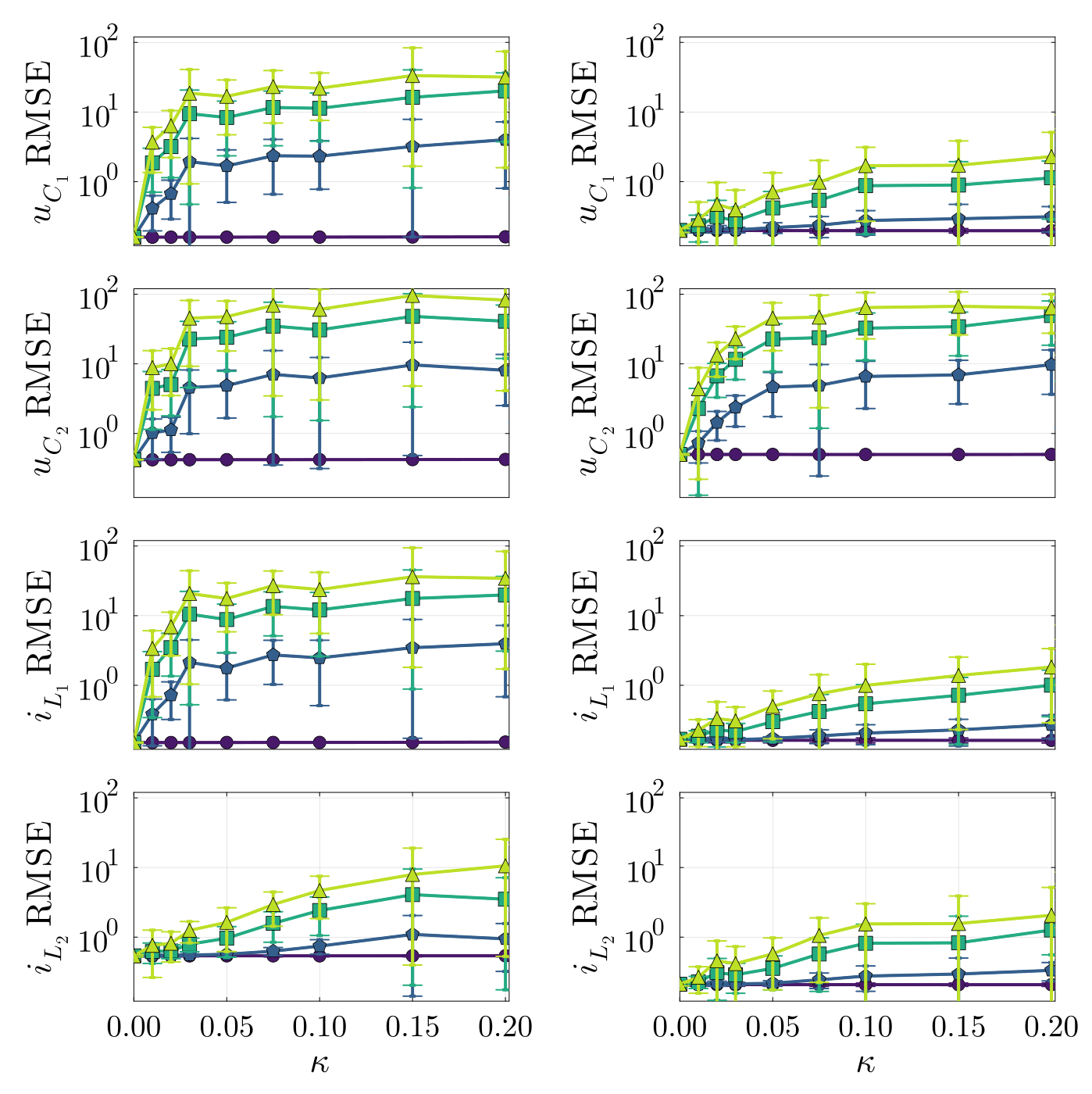}
		};
		
		% ==================== Legend above ====================
		\node[
		anchor=south,
		inner sep=0pt
		]
		at ($(img.north)+(0,0.25cm)$) {
			\includegraphics[
			width=0.49\textwidth
			]{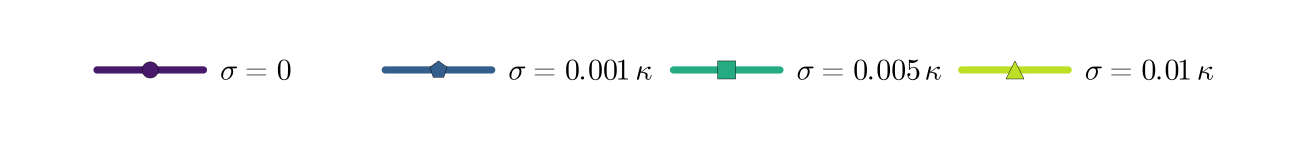}
		};
		
		% ==================== Left border: Theory ====================
		\draw[line width=0.6pt]
		($(img.north west)+(0.0cm,0.2cm)$) --
		($(img.north)+(0cm,0.2cm)$);
		
		\draw[line width=0.6pt]
		($(img.north west)+(0.0cm,0.2cm)$) --
		($(img.south west)+(0cm,0cm)$) --
		($(img.south)+(0cm,0cm)$);
		
		% ==================== Middle border ====================
		\draw[line width=0.6pt]
		($(img.north)+(0cm,0.2cm)$) --
		($(img.south)+(0cm,0cm)$);
		%			
		% ==================== Right border: Experiment ====================
		\draw[line width=0.6pt]
		($(img.north)+(0cm,0.2cm)$) --
		($(img.north east)+(0.0cm,0.2cm)$);
		
		\draw[line width=0.6pt]
		($(img.north east)+(0.0cm,0.2cm)$) --
		($(img.south east)+(0.0cm,0cm)$) --
		($(img.south)+(0cm,0cm)$);
		
		% ==================== Theory title above border ====================
		\node[
		fill=white,
		inner xsep=3pt,
		inner ysep=1pt,
		font=\small\bfseries
		]
		at ($(img.north west)!0.5!(img.north)+(0,0.25cm)$)
		{Theory};
		
		% ==================== Experiment title above border ====================
		\node[
		fill=white,
		inner xsep=3pt,
		inner ysep=1pt,
		font=\small\bfseries
		]
		at ($(img.north)!0.5!(img.north east)+(0,0.25cm)$)
		{Experiment};
		
		% ==================== Two Column Panel labels ====================
		
		% Left column
		\node[font=\normalfont]
		at ($(img.north west)+(0.25cm,-0.3cm)$) {(a)};
		
		\node[font=\normalfont]
		at ($(img.north west)+(0.25cm,-2.3cm)$) {(c)};
		
		\node[font=\normalfont]
		at ($(img.north west)+(0.25cm,-4.3cm)$) {(e)};
		
		\node[font=\normalfont]
		at ($(img.north west)+(0.25cm,-6.3cm)$) {(g)};
		
		% Right column
		\node[font=\normalfont]
		at ($(img.north)+(0.25cm,-0.3cm)$) {(b)};
		
		\node[font=\normalfont]
		at ($(img.north)+(0.25cm,-2.3cm)$) {(d)};
		
		\node[font=\normalfont]
		at ($(img.north)+(0.25cm,-4.3cm)$) {(f)};
		
		\node[font=\normalfont]
		at ($(img.north)+(0.25cm,-6.3cm)$) {(h)};
		
	\end{tikzpicture}
	
	\caption{
		Effect of decoherence on the prediction performance of the quantum reservoir.
		The left and right columns show the theoretical and experimental results,
		respectively:
		(a,b) $u_{C_1}$,
		(c,d) $u_{C_2}$,
		(e,f) $i_{L_1}$, and
		(g,h) $i_{L_2}$. Error bars indicate the standard deviation of the RMSE across reservoir batches
		at the same mean decoherence rate $\kappa$, with independently sampled
		qubit-specific rates $\kappa_k$ in each batch.
	}
	\label{fig:decoherence}
\end{figure}

\subsection*{Decoherence effect}
An important source of decoherence is amplitude damping, which describes the loss of excitation from a quantum system to its environment. For a qubit, this process drives the excited state $\ket{1}$ toward the ground state $\ket{0}$. We model amplitude damping using the Lindblad master equation
\begin{equation}
	\frac{\mathrm{d}\rho}{\mathrm{d}\tau}
	=
	-i[H,\rho]
	+
	\sum_{k=1}^{N_q}
	\kappa_k
	\left(
	\sigma_-^{(k)}\rho\sigma_+^{(k)}
	-
	\frac{1}{2}
	\left\{
	\sigma_+^{(k)}\sigma_-^{(k)},\rho
	\right\}
	\right),
	\label{eq:amplitude_damping}
\end{equation}
where 
$\sigma_-{=}\ket{0}\bra{1}$ and $\sigma_+{=}\ket{1}\bra{0}$ are the lowering and raising operators, respectively, and $\kappa_k$ is the amplitude-damping rate of the $k$-th qubit expressed in units of the inverse reservoir-evolution time. For the decoherence analysis, we use, for each signal and for both the theoretical and experimental data, the corresponding optimal reservoir configuration identified in Fig.~\ref{fig:heatmap4to1}. In all cases, these optimal configurations correspond to the maximum reservoir size considered, $N_q{=}11$.

To account for variations in the decoherence rates of individual physical qubits, the amplitude-damping rate of each qubit is independently sampled from a Gaussian distribution with mean $\kappa$ and standard deviation $\sigma$. We consider several levels of qubit-to-qubit variation, namely $\sigma/\kappa{=}0$, $0.001$, $0.005$, and $0.01$. Independent realizations of the damping rate distribution are generated for training and testing, such that the reservoir encountered during testing is not identical to that used during training. Figure~\ref{fig:decoherence} shows the resulting prediction performance as a function of the mean amplitude-damping rate \(\kappa\) for different values of \(\sigma\). For spatially uniform damping (\(\sigma{=}0\)), the damping rates are identical during training and testing, and the prediction error varies only slightly over the range considered. For small nonzero values of \(\sigma\), the qubit-to-qubit variations remain weak, allowing the readout to effectively learn the influence of decoherence during training.

\section*{Discussion}

While QRC and QEL are strong tools for predicting temporal signals, their performance have been mainly investigated on discrete time-series. To use  these algorithms for continuous dynamical processes, one has to first devise a meaningful discretization  approach to avoid both oversampling as well as undersampling. In this paper, we generate four continuous time-series signals by implementing an experimental chaotic electronic circuit. The simplified model reproduces the main chaos-generating mechanism,
but does not describe all aspects of the physical circuit. Quantitative discrepancies may reflect the approximate transistor model, component variability, further parasitic dynamics such as inductor self-resonances and dissipation, and the finite bandwidth of the readout circuitry. Therefore, we develop a QEL algorithm for predicting time-series generated by both the mathematical model as well as the experimental circuit. To address the discretization problem, we proposed two strategies: (i)  maplike sampling of local peaks/troughs 
which captures resonant behaviour but misses the detailed dynamical features; and (ii) uniform sampling at the signal autocorrelation timescale which fully retains these features and can be used to reconstruct the entire family of chaotic attractor diagrams. In particular, the introduction of autocorrelation timescale as a unit of discretization is shown to represent a suitable compromise avoiding both over and undersampling.  This establishes QEL as an efficient predictive algorithm which achieves high forecasting accuracy with an operationally motivated choice of sampling frequency. In the maplike strategy for discretization, since each signal peaks at different times, the sampling is inherently non-uniform in time and occurs with different intervals across signals. This prevents  us from the use of cross information in other signals for forecasting, thus our prediction for each signal exclusively used only the past data of that particular signal. In contrast, the uniform sampling strategy enables us to exploit the information of all four signals together for predicting future value of each individual signal. Despite using a very small quantum reservoir of up to 11 qubits with only nearest-neighbour interactions, QEL achieved high forecasting accuracy for both simulated and experimental signals.
The protocol shows consistent trend of improving the forecasting accuracy with increasing the size of the reservoir which shows its potential for significant improvement as the reservoir size is scaled up. Moreover, the protocol demonstrates robustness against moderate decoherence and dissipation, making it suitable for realizing on near-term quantum devices.  Our results provide a systematic approach for dealing with continuous signal processing and paves the way towards real-time forecasting of  signals on quantum hardware.

\section*{Methods}

\subsection*{Experimental circuit and signal acquisition} \label{sec:experimental_circuit}

The oscillator was fabricated in a printed-circuit-board implementation using surface-mount elements (Fig.~\ref{fig:Circuit}(a)). For the acquisition of all four dynamical variables, it also contained voltage and current reading amplifiers; these were arranged to enable voltage recording with minimal loading, and current recording by means of observing the potential differences across two negligibly small load resistors. Additional filters supported the achievement of high signal-to-noise ratio in the recordings    
(Fig.~\ref{fig:Circuit}(c)). The transistor $Q_1$ was type PRF949 or equivalent. In line with the numerical simulations, the base branch comprised a $15~\mu\mathrm{H}$ inductor (type NLFV32T-150K-EF or equivalent), while the collector branch comprised a $220~\mu\mathrm{H}$ inductor (type NLFV32T-221K-EF or equivalent). The capacitor had value $C_1{=}270~\mathrm{pF}$. The supply voltage was set to $V_\textrm{S}{=}7.450~\mathrm{V}$ and the series resistance was $R{=}1.25~\mathrm{k}\Upomega$. A $10~\mu\mathrm{F}$ capacitor provided supply decoupling upstream of this resistor. 

Signal conditioning used type LM6172 or equivalent dual high-speed voltage-feedback operational amplifiers. The voltage channels measured nodes $n1$ and $n4$, corresponding to $u_{C_1}$ and $u_{C_2}$. To measure the inductor currents, we inserted a $2~\Upomega$ shunt in each branch. The voltage drops between $n1$ and $n3$, and between $n1$ and $n2$, were therefore proportional to $i_{L_1}$ and $i_{L_2}$, respectively. Buffered copies of the potentials at either side of each shunt were fed to a differential amplifier. The four conditioned signals were simultaneously digitized using an oscilloscope at $312.5~\mathrm{MSa/s}$ on every channel, giving a sampling interval of $3.2~\mathrm{ns}$ (Fig.~\ref{fig:Circuit}(b)). Acquisitions sweeping $V_\textrm{S}$ from 2 V to 10 V in steps of 50 mV were performed under computer control, and all experimental time series data have been made freely available \cite{zenodo}.\par

\subsection*{Data preparation}
Here, we describe the data-preparation procedure applied to each signal. Since the data acquisition of the experimental circuit was with 3.2\(~n\mathrm{s}\) steps, after solving the theoretical equations, we use the same timesteps for the theoretical data as well. The experimental signals were low-pass filtered to suppress high frequency measurement noise. A low-pass filter with a cutoff frequency of \(5~\mathrm{MHz}\) was applied to \(u_{C_1}\), \(i_{L_1}\), and \(i_{L_2}\), while a higher cutoff frequency of \(20~\mathrm{MHz}\) was used for \(u_{C_2}\) to better preserve its sharp peaks. Before encoding, the signals are normalized using statistics calculated exclusively from the training interval. Specifically, each signal is transformed using z-score normalization \(u'{=}(u{-}\mu_{u})/\sigma_{u}\),
where \(\mu_{u}\) and \(\sigma_{u}\) are the mean and standard deviation of \(u\) over the training interval, respectively. The same parameters are then applied unchanged to the testing interval. The normalized values are subsequently remapped to the interval \([0,\pi]\) according to \(\theta{=}\pi(u'{-}u'_{\min})/(u'_{\max}{-}u'_{\min})\),
where \(u'_{\min}\) and \(u'_{\max}\) denote the minimum and maximum normalized values in the training time span, respectively. Thus, \(u'_{\min}\) and \(u'_{\max}\) are mapped to \(0\) and \(\pi\), respectively, and \(\theta\) is used as the rotation angle for quantum-circuit encoding.
To predict the signal value \(\tilde{u}[n]\), the model receives \(N_{\mathrm{steps}}\) previous samples. As described in the results section, the maplike discretization and uniform discretization have different approaches to select the previous samples.

For the maplike discretization, the relevant extrema are extracted from the normalized signal. The extrema are identified using MATLAB-based peak detection, implemented through the \verb*|findpeaks| function. This method characterizes a peak by its prominence, which measures how strongly the peak stands out from its surrounding waveform, while a minimum peak-distance condition prevents closely spaced fluctuations from being counted as separate extrema. Feature-specific values of these parameters are used for $u_{C_1}$, $u_{C_2}$, $i_{L_1}$, and $i_{L_2}$ to account for their different waveform characteristics. Depending on the selected mode, either local maxima or local minima are retained. 

For the uniform discretization, the lag value \(m\) in Eq.~\eqref{eq:input} shouldn't be low otherwise the interval data are highly correlated. Therefore, we calculate the autocorrelation function of the signals by Eq.~\eqref{eq:ACF}. Thus, the autocorrelation lags are determined exclusively from the
training data, without using the final testing interval. We identify the first lag $k$ at which $R[k]$ decreases below $1/e$. As shown in Fig.~\ref{fig:ACF}, the corresponding $1/e$ crossing lags for $u_{C_1}$, $u_{C_2}$, $i_{L_1}$, and $i_{L_2}$ are 21, 7, 34, and 68 samples, respectively, for the theoretical signals, and 25, 13, 31, and 27 samples, respectively, for the experimental signals. These values are used as the corresponding lags \(m\) in the QEL model. Importantly, \(m\) determines the temporal separation between samples within each input window, whereas successive input windows are shifted forward by only one original sampling step, corresponding to \(3.2~\mathrm{ns}\). Thus, the input window slides through the signal sample by sample, while the elements within each window remain separated by \(m{\times}3.2~\mathrm{ns}\).
This approach prevents the
model from relying mainly on strongly correlated neighboring samples and enforces it to learn the longer-range and nonlinear temporal relationships in the signals. Finally, the output RMSE is calculated on the standardized (\(z\)-score) signal scale, using normalization parameters determined from the training data.

\subsection*{LSTM benchmarking}

As a classical benchmark, we employ an LSTM
network for the uniformly discretized chaotic
time series. To maintain consistency with the QEL analysis, the LSTM uses
the same training and testing time intervals that are \(5\text{--}210~\mu\mathrm{s}\) and
\(210\text{--}300~\mu\mathrm{s}\), respectively.
Each signal is normalized by a \(z\)-score transformation, with the mean and standard deviation calculated exclusively from the training interval and subsequently applied unchanged to the testing interval. Before constructing the LSTM input sequences, all four signals are sampled
using the characteristic lag of the target signal, as determined from
the first \(1/e\) crossing of its autocorrelation function in
Fig.~\ref{fig:ACF}. 
Defining
\(
\mathbf{x}[n]
{=}
\left[
u_{C_1}[n],
u_{C_2}[n],
i_{L_1}[n],
i_{L_2}[n]
\right]^{\mathsf T}\),
for a selected scalar target, take \(\tilde{u}_{C_1}\), the LSTM input is
\begin{equation}
	\bm{\mathcal{X}}_{\mathrm{LSTM}}[n]
	=
	\begin{bmatrix}
		\mathbf{x}^{\mathsf{T}}[n-N^{\mathrm{LSTM}}_{\mathrm{steps}}m] \\
		\vdots \\
		\mathbf{x}^{\mathsf{T}}[n-2m] \\
		\mathbf{x}^{\mathsf{T}}[n-m]
	\end{bmatrix}
	\longmapsto
	\tilde{u}_{C_1}[n].
\end{equation}
Here, \(n\) and \(m\) are sample indices and lags on the original
acquisition grid, and
\(N_{\mathrm{steps}}^{\mathrm{LSTM}}\) denotes the number of previous
time steps supplied to the LSTM. To provide a temporal history comparable
to that used by QEL, we restrict the LSTM to
\(N_{\mathrm{steps}}^{\mathrm{LSTM}}{\in}\{1,2\}\).

The prediction model consists of a single LSTM layer followed by a fully connected linear readout. At each time step, the LSTM receives the four-dimensional input vector \(\mathbf{x}[n]\) and has hidden-state dimension \(d_{\mathrm{hid}}\). The hidden state produced at the final element of the input sequence is passed through a dropout layer with probability \(0.1\), followed by a linear layer producing the scalar prediction \(\tilde{u}_{C_1}[n]\).

For completeness, the trainable size of each LSTM architecture is evaluated
explicitly. Let \(c\) denote the input dimension (4 in our four-to-one scenario) and \(d_{\mathrm{hid}}\) the hidden-state dimension. A standard LSTM layer contains four gates and therefore contains
\(	N_{\mathrm{par}}^{\mathrm{LSTM}}
{=}
4d_{\mathrm{hid}}(c{+}d_{\mathrm{hid}}{+}2)
{+}
d_{\mathrm{hid}}{+}1,\)
trainable parameters, where the two bias contributions within the LSTM correspond to the input-hidden and hidden-hidden biases used by the \verb*|PyTorch| implementation, and \(d_{\mathrm{hid}}{+}1\) accounts for the final scalar linear readout. 

The LSTM parameters are optimized by minimizing the mean-squared prediction
error using the Adam optimizer with a learning rate of \(10^{-3}\) and a
mini-batch size of 256. Every architecture is trained for a fixed
\(200\) epochs. After the \(200\) training
epochs, the network is placed in evaluation mode and its predictive
performance is calculated on the testing interval. The architecture sweep uses \(N_{\mathrm{steps}}^{\mathrm{LSTM}}{\in}\{1,2\}\), and \(d_{\mathrm{hid}}{\in}\{1,2,3\}\), with a single recurrent LSTM layer used throughout. Each architecture is independently trained using five random seeds.

For comparison with QEL, the test RMSE is first averaged over the five
random seeds for every LSTM architecture. An architecture is considered
to reach the QEL prediction accuracy when
\(\mathrm{RMSE}_{\mathrm{LSTM}}
{\leq}
\mathrm{RMSE}_{\mathrm{QEL}}\).
Among all architectures satisfying this condition, the one with the
smallest value of \(N_{\mathrm{par}}^{\mathrm{LSTM}}\) is reported in Table~\ref{tab:LSTM}.

\section*{Data Availability}

The experimental data recorded from the analog circuits are downloadable from Ref. \cite{zenodo}. 

\section*{Code Availability}
The code scripts analyzing the study are available on the project page at https://github.com/mohammadshariifian/QEL-for-experimental-chaos-prediction.

\section*{Acknowledgments}
AB acknowledges support
from the National Natural Science Foundation of China
(grants No. W2541020, No. 12274059, No. 12574528, and
No. 1251101297). MS acknowledges support
from the National Natural Science Foundation of China
(grant No. W2633018). CM acknowledges support from National Natural Science Foundation of China (grant no. 12374482) and a  talent introduction scholarship. LM gratefully acknowledges the support of the ‘‘Hundred Talents’’ program of the University of Electronic Science and Technology of China, of the ‘‘Outstanding Young Talents Program (Overseas)’’ program of the National Natural Science Foundation of China, and of the talent programs of the Sichuan province and Chengdu municipality. The funders played no role in study design, data collection, analysis and interpretation of data, or the
writing of this manuscript. The authors are grateful to Manyu Zhao for initial discussions regarding the estimation and the influence of temporal autocorrelation.

%==============================

\nocite{apsrev41Control} 
\bibliographystyle{apsrev4-1}
\bibliography{refs.bib,refcontrol.bib}

%merlin.mbs apsrev4-1.bst 2010-07-25 4.21a (PWD, AO, DPC) hacked
%Control: key (0)
%Control: author (8) initials jnrlst
%Control: editor formatted (1) identically to author
%Control: production of article title (0) allowed
%Control: page (1) range
%Control: year (1) truncated
%Control: production of eprint (0) enabled
\begin{thebibliography}{72}%
\makeatletter
\providecommand \@ifxundefined [1]{%
 \@ifx{#1\undefined}
}%
\providecommand \@ifnum [1]{%
 \ifnum #1\expandafter \@firstoftwo
 \else \expandafter \@secondoftwo
 \fi
}%
\providecommand \@ifx [1]{%
 \ifx #1\expandafter \@firstoftwo
 \else \expandafter \@secondoftwo
 \fi
}%
\providecommand \natexlab [1]{#1}%
\providecommand \enquote  [1]{``#1''}%
\providecommand \bibnamefont  [1]{#1}%
\providecommand \bibfnamefont [1]{#1}%
\providecommand \citenamefont [1]{#1}%
\providecommand \href@noop [0]{\@secondoftwo}%
\providecommand \href [0]{\begingroup \@sanitize@url \@href}%
\providecommand \@href[1]{\@@startlink{#1}\@@href}%
\providecommand \@@href[1]{\endgroup#1\@@endlink}%
\providecommand \@sanitize@url [0]{\catcode `\\12\catcode `\$12\catcode
  `\&12\catcode `\#12\catcode `\^12\catcode `\_12\catcode `\%12\relax}%
\providecommand \@@startlink[1]{}%
\providecommand \@@endlink[0]{}%
\providecommand \url  [0]{\begingroup\@sanitize@url \@url }%
\providecommand \@url [1]{\endgroup\@href {#1}{\urlprefix }}%
\providecommand \urlprefix  [0]{URL }%
\providecommand \Eprint [0]{\href }%
\providecommand \doibase [0]{http://dx.doi.org/}%
\providecommand \selectlanguage [0]{\@gobble}%
\providecommand \bibinfo  [0]{\@secondoftwo}%
\providecommand \bibfield  [0]{\@secondoftwo}%
\providecommand \translation [1]{[#1]}%
\providecommand \BibitemOpen [0]{}%
\providecommand \bibitemStop [0]{}%
\providecommand \bibitemNoStop [0]{.\EOS\space}%
\providecommand \EOS [0]{\spacefactor3000\relax}%
\providecommand \BibitemShut  [1]{\csname bibitem#1\endcsname}%
\let\auto@bib@innerbib\@empty
%</preamble>
\bibitem [{\citenamefont {Bharti}\ \emph {et~al.}(2022)\citenamefont {Bharti},
  \citenamefont {Cervera-Lierta}, \citenamefont {Kyaw}, \citenamefont {Haug},
  \citenamefont {Alperin-Lea}, \citenamefont {Anand}, \citenamefont {Degroote},
  \citenamefont {Heimonen}, \citenamefont {Kottmann}, \citenamefont {Menke},
  \citenamefont {Mok}, \citenamefont {Sim}, \citenamefont {Kwek},\ and\
  \citenamefont {Aspuru-Guzik}}]{Bharti_2022}%
  \BibitemOpen
  \bibfield  {author} {\bibinfo {author} {\bibfnamefont {K.}~\bibnamefont
  {Bharti}}, \bibinfo {author} {\bibfnamefont {A.}~\bibnamefont
  {Cervera-Lierta}}, \bibinfo {author} {\bibfnamefont {T.~H.}\ \bibnamefont
  {Kyaw}}, \bibinfo {author} {\bibfnamefont {T.}~\bibnamefont {Haug}}, \bibinfo
  {author} {\bibfnamefont {S.}~\bibnamefont {Alperin-Lea}}, \bibinfo {author}
  {\bibfnamefont {A.}~\bibnamefont {Anand}}, \bibinfo {author} {\bibfnamefont
  {M.}~\bibnamefont {Degroote}}, \bibinfo {author} {\bibfnamefont
  {H.}~\bibnamefont {Heimonen}}, \bibinfo {author} {\bibfnamefont {J.~S.}\
  \bibnamefont {Kottmann}}, \bibinfo {author} {\bibfnamefont {T.}~\bibnamefont
  {Menke}}, \bibinfo {author} {\bibfnamefont {W.-K.}\ \bibnamefont {Mok}},
  \bibinfo {author} {\bibfnamefont {S.}~\bibnamefont {Sim}}, \bibinfo {author}
  {\bibfnamefont {L.-C.}\ \bibnamefont {Kwek}}, \ and\ \bibinfo {author}
  {\bibfnamefont {A.}~\bibnamefont {Aspuru-Guzik}},\ }\bibfield  {title}
  {\enquote {\bibinfo {title} {Noisy intermediate-scale quantum algorithms},}\
  }\href {\doibase 10.1103/revmodphys.94.015004} {\bibfield  {journal}
  {\bibinfo  {journal} {Reviews of Modern Physics}\ }\textbf {\bibinfo {volume}
  {94}} (\bibinfo {year} {2022}),\ 10.1103/revmodphys.94.015004}\BibitemShut
  {NoStop}%
\bibitem [{\citenamefont {Peruzzo}\ \emph {et~al.}(2014)\citenamefont
  {Peruzzo}, \citenamefont {McClean}, \citenamefont {Shadbolt}, \citenamefont
  {Yung}, \citenamefont {Zhou}, \citenamefont {Love}, \citenamefont
  {Aspuru-Guzik},\ and\ \citenamefont {O’Brien}}]{peruzzo2014variational}%
  \BibitemOpen
  \bibfield  {author} {\bibinfo {author} {\bibfnamefont {A.}~\bibnamefont
  {Peruzzo}}, \bibinfo {author} {\bibfnamefont {J.}~\bibnamefont {McClean}},
  \bibinfo {author} {\bibfnamefont {P.}~\bibnamefont {Shadbolt}}, \bibinfo
  {author} {\bibfnamefont {M.-H.}\ \bibnamefont {Yung}}, \bibinfo {author}
  {\bibfnamefont {X.-Q.}\ \bibnamefont {Zhou}}, \bibinfo {author}
  {\bibfnamefont {P.~J.}\ \bibnamefont {Love}}, \bibinfo {author}
  {\bibfnamefont {A.}~\bibnamefont {Aspuru-Guzik}}, \ and\ \bibinfo {author}
  {\bibfnamefont {J.~L.}\ \bibnamefont {O’Brien}},\ }\bibfield  {title}
  {\enquote {\bibinfo {title} {A variational eigenvalue solver on a photonic
  quantum processor},}\ }\href {\doibase 10.1038/ncomms5213} {\bibfield
  {journal} {\bibinfo  {journal} {Nature Communications}\ }\textbf {\bibinfo
  {volume} {5}} (\bibinfo {year} {2014}),\ 10.1038/ncomms5213}\BibitemShut
  {NoStop}%
\bibitem [{\citenamefont {McClean}\ \emph {et~al.}(2016)\citenamefont
  {McClean}, \citenamefont {Romero}, \citenamefont {Babbush},\ and\
  \citenamefont {Aspuru-Guzik}}]{mcclean2016theory}%
  \BibitemOpen
  \bibfield  {author} {\bibinfo {author} {\bibfnamefont {J.~R.}\ \bibnamefont
  {McClean}}, \bibinfo {author} {\bibfnamefont {J.}~\bibnamefont {Romero}},
  \bibinfo {author} {\bibfnamefont {R.}~\bibnamefont {Babbush}}, \ and\
  \bibinfo {author} {\bibfnamefont {A.}~\bibnamefont {Aspuru-Guzik}},\
  }\bibfield  {title} {\enquote {\bibinfo {title} {The theory of variational
  hybrid quantum-classical algorithms},}\ }\href {\doibase
  10.1088/1367-2630/18/2/023023} {\bibfield  {journal} {\bibinfo  {journal}
  {New Journal of Physics}\ }\textbf {\bibinfo {volume} {18}},\ \bibinfo
  {pages} {023023} (\bibinfo {year} {2016})}\BibitemShut {NoStop}%
\bibitem [{\citenamefont {O’Malley}\ \emph {et~al.}(2016)\citenamefont
  {O’Malley}, \citenamefont {Babbush}, \citenamefont {Kivlichan},
  \citenamefont {Romero}, \citenamefont {McClean}, \citenamefont {Barends},
  \citenamefont {Kelly}, \citenamefont {Roushan}, \citenamefont {Tranter},
  \citenamefont {Ding} \emph {et~al.}}]{o2016scalable}%
  \BibitemOpen
  \bibfield  {author} {\bibinfo {author} {\bibfnamefont {P.~J.}\ \bibnamefont
  {O’Malley}}, \bibinfo {author} {\bibfnamefont {R.}~\bibnamefont {Babbush}},
  \bibinfo {author} {\bibfnamefont {I.~D.}\ \bibnamefont {Kivlichan}}, \bibinfo
  {author} {\bibfnamefont {J.}~\bibnamefont {Romero}}, \bibinfo {author}
  {\bibfnamefont {J.~R.}\ \bibnamefont {McClean}}, \bibinfo {author}
  {\bibfnamefont {R.}~\bibnamefont {Barends}}, \bibinfo {author} {\bibfnamefont
  {J.}~\bibnamefont {Kelly}}, \bibinfo {author} {\bibfnamefont
  {P.}~\bibnamefont {Roushan}}, \bibinfo {author} {\bibfnamefont
  {A.}~\bibnamefont {Tranter}}, \bibinfo {author} {\bibfnamefont
  {N.}~\bibnamefont {Ding}},  \emph {et~al.},\ }\bibfield  {title} {\enquote
  {\bibinfo {title} {Scalable quantum simulation of molecular energies},}\
  }\href {\doibase 10.1103/physrevx.6.031007} {\bibfield  {journal} {\bibinfo
  {journal} {Physical Review X}\ }\textbf {\bibinfo {volume} {6}} (\bibinfo
  {year} {2016}),\ 10.1103/physrevx.6.031007}\BibitemShut {NoStop}%
\bibitem [{\citenamefont {Yuan}\ \emph {et~al.}(2019)\citenamefont {Yuan},
  \citenamefont {Endo}, \citenamefont {Zhao}, \citenamefont {Li},\ and\
  \citenamefont {Benjamin}}]{yuan2019theory}%
  \BibitemOpen
  \bibfield  {author} {\bibinfo {author} {\bibfnamefont {X.}~\bibnamefont
  {Yuan}}, \bibinfo {author} {\bibfnamefont {S.}~\bibnamefont {Endo}}, \bibinfo
  {author} {\bibfnamefont {Q.}~\bibnamefont {Zhao}}, \bibinfo {author}
  {\bibfnamefont {Y.}~\bibnamefont {Li}}, \ and\ \bibinfo {author}
  {\bibfnamefont {S.~C.}\ \bibnamefont {Benjamin}},\ }\bibfield  {title}
  {\enquote {\bibinfo {title} {Theory of variational quantum simulation},}\
  }\href {\doibase 10.22331/q-2019-10-07-191} {\bibfield  {journal} {\bibinfo
  {journal} {Quantum}\ }\textbf {\bibinfo {volume} {3}},\ \bibinfo {pages}
  {191} (\bibinfo {year} {2019})}\BibitemShut {NoStop}%
\bibitem [{\citenamefont {Endo}\ \emph {et~al.}(2020)\citenamefont {Endo},
  \citenamefont {Sun}, \citenamefont {Li}, \citenamefont {Benjamin},\ and\
  \citenamefont {Yuan}}]{endo2020variational}%
  \BibitemOpen
  \bibfield  {author} {\bibinfo {author} {\bibfnamefont {S.}~\bibnamefont
  {Endo}}, \bibinfo {author} {\bibfnamefont {J.}~\bibnamefont {Sun}}, \bibinfo
  {author} {\bibfnamefont {Y.}~\bibnamefont {Li}}, \bibinfo {author}
  {\bibfnamefont {S.~C.}\ \bibnamefont {Benjamin}}, \ and\ \bibinfo {author}
  {\bibfnamefont {X.}~\bibnamefont {Yuan}},\ }\bibfield  {title} {\enquote
  {\bibinfo {title} {Variational quantum simulation of general processes},}\
  }\href {\doibase 10.1103/physrevlett.125.010501} {\bibfield  {journal}
  {\bibinfo  {journal} {Physical Review Letters}\ }\textbf {\bibinfo {volume}
  {125}} (\bibinfo {year} {2020}),\ 10.1103/physrevlett.125.010501}\BibitemShut
  {NoStop}%
\bibitem [{\citenamefont {Cerezo}\ \emph {et~al.}(2021)\citenamefont {Cerezo},
  \citenamefont {Arrasmith}, \citenamefont {Babbush}, \citenamefont {Benjamin},
  \citenamefont {Endo}, \citenamefont {Fujii}, \citenamefont {McClean},
  \citenamefont {Mitarai}, \citenamefont {Yuan}, \citenamefont {Cincio},\ and\
  \citenamefont {Coles}}]{Cerezo_2021}%
  \BibitemOpen
  \bibfield  {author} {\bibinfo {author} {\bibfnamefont {M.}~\bibnamefont
  {Cerezo}}, \bibinfo {author} {\bibfnamefont {A.}~\bibnamefont {Arrasmith}},
  \bibinfo {author} {\bibfnamefont {R.}~\bibnamefont {Babbush}}, \bibinfo
  {author} {\bibfnamefont {S.~C.}\ \bibnamefont {Benjamin}}, \bibinfo {author}
  {\bibfnamefont {S.}~\bibnamefont {Endo}}, \bibinfo {author} {\bibfnamefont
  {K.}~\bibnamefont {Fujii}}, \bibinfo {author} {\bibfnamefont {J.~R.}\
  \bibnamefont {McClean}}, \bibinfo {author} {\bibfnamefont {K.}~\bibnamefont
  {Mitarai}}, \bibinfo {author} {\bibfnamefont {X.}~\bibnamefont {Yuan}},
  \bibinfo {author} {\bibfnamefont {L.}~\bibnamefont {Cincio}}, \ and\ \bibinfo
  {author} {\bibfnamefont {P.~J.}\ \bibnamefont {Coles}},\ }\bibfield  {title}
  {\enquote {\bibinfo {title} {Variational quantum algorithms},}\ }\href
  {\doibase 10.1038/s42254-021-00348-9} {\bibfield  {journal} {\bibinfo
  {journal} {Nature Reviews Physics}\ }\textbf {\bibinfo {volume} {3}},\
  \bibinfo {pages} {625–644} (\bibinfo {year} {2021})}\BibitemShut {NoStop}%
\bibitem [{\citenamefont {Jones}\ \emph {et~al.}(2019)\citenamefont {Jones},
  \citenamefont {Endo}, \citenamefont {McArdle}, \citenamefont {Yuan},\ and\
  \citenamefont {Benjamin}}]{jones2019variational}%
  \BibitemOpen
  \bibfield  {author} {\bibinfo {author} {\bibfnamefont {T.}~\bibnamefont
  {Jones}}, \bibinfo {author} {\bibfnamefont {S.}~\bibnamefont {Endo}},
  \bibinfo {author} {\bibfnamefont {S.}~\bibnamefont {McArdle}}, \bibinfo
  {author} {\bibfnamefont {X.}~\bibnamefont {Yuan}}, \ and\ \bibinfo {author}
  {\bibfnamefont {S.~C.}\ \bibnamefont {Benjamin}},\ }\bibfield  {title}
  {\enquote {\bibinfo {title} {Variational quantum algorithms for discovering
  hamiltonian spectra},}\ }\href {\doibase 10.1103/PhysRevA.99.062304}
  {\bibfield  {journal} {\bibinfo  {journal} {Phys. Rev. A}\ }\textbf {\bibinfo
  {volume} {99}},\ \bibinfo {pages} {062304} (\bibinfo {year}
  {2019})}\BibitemShut {NoStop}%
\bibitem [{\citenamefont {Li}\ \emph {et~al.}(2023)\citenamefont {Li},
  \citenamefont {Mukhopadhyay},\ and\ \citenamefont {Bayat}}]{Li_2023}%
  \BibitemOpen
  \bibfield  {author} {\bibinfo {author} {\bibfnamefont {Q.}~\bibnamefont
  {Li}}, \bibinfo {author} {\bibfnamefont {C.}~\bibnamefont {Mukhopadhyay}}, \
  and\ \bibinfo {author} {\bibfnamefont {A.}~\bibnamefont {Bayat}},\ }\bibfield
   {title} {\enquote {\bibinfo {title} {Fermionic simulators for enhanced
  scalability of variational quantum simulation},}\ }\href {\doibase
  10.1103/PhysRevResearch.5.043175} {\bibfield  {journal} {\bibinfo  {journal}
  {Phys. Rev. Res.}\ }\textbf {\bibinfo {volume} {5}},\ \bibinfo {pages}
  {043175} (\bibinfo {year} {2023})}\BibitemShut {NoStop}%
\bibitem [{\citenamefont {Aliverti-Piuri}\ \emph {et~al.}(2024)\citenamefont
  {Aliverti-Piuri}, \citenamefont {Chatterjee}, \citenamefont {Ding},
  \citenamefont {Liao}, \citenamefont {Liebert},\ and\ \citenamefont
  {Schilling}}]{aliverti2024can}%
  \BibitemOpen
  \bibfield  {author} {\bibinfo {author} {\bibfnamefont {D.}~\bibnamefont
  {Aliverti-Piuri}}, \bibinfo {author} {\bibfnamefont {K.}~\bibnamefont
  {Chatterjee}}, \bibinfo {author} {\bibfnamefont {L.}~\bibnamefont {Ding}},
  \bibinfo {author} {\bibfnamefont {K.}~\bibnamefont {Liao}}, \bibinfo {author}
  {\bibfnamefont {J.}~\bibnamefont {Liebert}}, \ and\ \bibinfo {author}
  {\bibfnamefont {C.}~\bibnamefont {Schilling}},\ }\bibfield  {title} {\enquote
  {\bibinfo {title} {What can quantum information theory offer to quantum
  chemistry?}}\ }\href {\doibase 10.1039/d4fd00059e} {\bibfield  {journal}
  {\bibinfo  {journal} {Faraday Discussions}\ }\textbf {\bibinfo {volume}
  {254}},\ \bibinfo {pages} {76–106} (\bibinfo {year} {2024})}\BibitemShut
  {NoStop}%
\bibitem [{\citenamefont {Brozzi}\ \emph {et~al.}(2026)\citenamefont {Brozzi},
  \citenamefont {Turati}, \citenamefont {Ferrari~Dacrema}, \citenamefont
  {Caruso},\ and\ \citenamefont {Cremonesi}}]{Brozzi_2026}%
  \BibitemOpen
  \bibfield  {author} {\bibinfo {author} {\bibfnamefont {F.}~\bibnamefont
  {Brozzi}}, \bibinfo {author} {\bibfnamefont {G.}~\bibnamefont {Turati}},
  \bibinfo {author} {\bibfnamefont {M.}~\bibnamefont {Ferrari~Dacrema}},
  \bibinfo {author} {\bibfnamefont {F.}~\bibnamefont {Caruso}}, \ and\ \bibinfo
  {author} {\bibfnamefont {P.}~\bibnamefont {Cremonesi}},\ }\bibfield  {title}
  {\enquote {\bibinfo {title} {Hamiltonian expressibility for ansatz selection
  in variational quantum algorithms},}\ }\href {\doibase
  10.1007/s42484-026-00407-3} {\bibfield  {journal} {\bibinfo  {journal}
  {Quantum Machine Intelligence}\ }\textbf {\bibinfo {volume} {8}} (\bibinfo
  {year} {2026}),\ 10.1007/s42484-026-00407-3}\BibitemShut {NoStop}%
\bibitem [{\citenamefont {Nakanishi}\ \emph {et~al.}(2019)\citenamefont
  {Nakanishi}, \citenamefont {Mitarai},\ and\ \citenamefont
  {Fujii}}]{nakanishi2019subspace}%
  \BibitemOpen
  \bibfield  {author} {\bibinfo {author} {\bibfnamefont {K.~M.}\ \bibnamefont
  {Nakanishi}}, \bibinfo {author} {\bibfnamefont {K.}~\bibnamefont {Mitarai}},
  \ and\ \bibinfo {author} {\bibfnamefont {K.}~\bibnamefont {Fujii}},\
  }\bibfield  {title} {\enquote {\bibinfo {title} {Subspace-search variational
  quantum eigensolver for excited states},}\ }\href {\doibase
  10.1103/PhysRevResearch.1.033062} {\bibfield  {journal} {\bibinfo  {journal}
  {Phys. Rev. Res.}\ }\textbf {\bibinfo {volume} {1}},\ \bibinfo {pages}
  {033062} (\bibinfo {year} {2019})}\BibitemShut {NoStop}%
\bibitem [{\citenamefont {Lyu}\ \emph {et~al.}(2023)\citenamefont {Lyu},
  \citenamefont {Xu}, \citenamefont {Yung},\ and\ \citenamefont
  {Bayat}}]{lyu2023symmetry}%
  \BibitemOpen
  \bibfield  {author} {\bibinfo {author} {\bibfnamefont {C.}~\bibnamefont
  {Lyu}}, \bibinfo {author} {\bibfnamefont {X.}~\bibnamefont {Xu}}, \bibinfo
  {author} {\bibfnamefont {M.-H.}\ \bibnamefont {Yung}}, \ and\ \bibinfo
  {author} {\bibfnamefont {A.}~\bibnamefont {Bayat}},\ }\bibfield  {title}
  {\enquote {\bibinfo {title} {Symmetry enhanced variational quantum spin
  eigensolver},}\ }\href {\doibase 10.22331/q-2023-01-19-899} {\bibfield
  {journal} {\bibinfo  {journal} {Quantum}\ }\textbf {\bibinfo {volume} {7}},\
  \bibinfo {pages} {899} (\bibinfo {year} {2023})}\BibitemShut {NoStop}%
\bibitem [{\citenamefont {Han}\ \emph {et~al.}(2024)\citenamefont {Han},
  \citenamefont {Lyu}, \citenamefont {Zhou}, \citenamefont {Yuan},
  \citenamefont {Chu}, \citenamefont {Nuerbolati}, \citenamefont {Jia},
  \citenamefont {Nie}, \citenamefont {Wei}, \citenamefont {Yang}, \citenamefont
  {Zhang}, \citenamefont {Zhang}, \citenamefont {Hu}, \citenamefont {Hu},
  \citenamefont {Li}, \citenamefont {Tan}, \citenamefont {Bayat}, \citenamefont
  {Liu}, \citenamefont {Yan},\ and\ \citenamefont {Yu}}]{han2024multilevel}%
  \BibitemOpen
  \bibfield  {author} {\bibinfo {author} {\bibfnamefont {Z.}~\bibnamefont
  {Han}}, \bibinfo {author} {\bibfnamefont {C.}~\bibnamefont {Lyu}}, \bibinfo
  {author} {\bibfnamefont {Y.}~\bibnamefont {Zhou}}, \bibinfo {author}
  {\bibfnamefont {J.}~\bibnamefont {Yuan}}, \bibinfo {author} {\bibfnamefont
  {J.}~\bibnamefont {Chu}}, \bibinfo {author} {\bibfnamefont {W.}~\bibnamefont
  {Nuerbolati}}, \bibinfo {author} {\bibfnamefont {H.}~\bibnamefont {Jia}},
  \bibinfo {author} {\bibfnamefont {L.}~\bibnamefont {Nie}}, \bibinfo {author}
  {\bibfnamefont {W.}~\bibnamefont {Wei}}, \bibinfo {author} {\bibfnamefont
  {Z.}~\bibnamefont {Yang}}, \bibinfo {author} {\bibfnamefont {L.}~\bibnamefont
  {Zhang}}, \bibinfo {author} {\bibfnamefont {Z.}~\bibnamefont {Zhang}},
  \bibinfo {author} {\bibfnamefont {C.-K.}\ \bibnamefont {Hu}}, \bibinfo
  {author} {\bibfnamefont {L.}~\bibnamefont {Hu}}, \bibinfo {author}
  {\bibfnamefont {J.}~\bibnamefont {Li}}, \bibinfo {author} {\bibfnamefont
  {D.}~\bibnamefont {Tan}}, \bibinfo {author} {\bibfnamefont {A.}~\bibnamefont
  {Bayat}}, \bibinfo {author} {\bibfnamefont {S.}~\bibnamefont {Liu}}, \bibinfo
  {author} {\bibfnamefont {F.}~\bibnamefont {Yan}}, \ and\ \bibinfo {author}
  {\bibfnamefont {D.}~\bibnamefont {Yu}},\ }\bibfield  {title} {\enquote
  {\bibinfo {title} {Multilevel variational spectroscopy using a programmable
  quantum simulator},}\ }\href {\doibase 10.1103/PhysRevResearch.6.013015}
  {\bibfield  {journal} {\bibinfo  {journal} {Phys. Rev. Res.}\ }\textbf
  {\bibinfo {volume} {6}},\ \bibinfo {pages} {013015} (\bibinfo {year}
  {2024})}\BibitemShut {NoStop}%
\bibitem [{\citenamefont {Wu}\ \emph {et~al.}(2021)\citenamefont {Wu},
  \citenamefont {Zhang},\ and\ \citenamefont {Zhai}}]{wu2021scrambling}%
  \BibitemOpen
  \bibfield  {author} {\bibinfo {author} {\bibfnamefont {Y.}~\bibnamefont
  {Wu}}, \bibinfo {author} {\bibfnamefont {P.}~\bibnamefont {Zhang}}, \ and\
  \bibinfo {author} {\bibfnamefont {H.}~\bibnamefont {Zhai}},\ }\bibfield
  {title} {\enquote {\bibinfo {title} {Scrambling ability of quantum neural
  network architectures},}\ }\href {\doibase 10.1103/PhysRevResearch.3.L032057}
  {\bibfield  {journal} {\bibinfo  {journal} {Phys. Rev. Res.}\ }\textbf
  {\bibinfo {volume} {3}},\ \bibinfo {pages} {L032057} (\bibinfo {year}
  {2021})}\BibitemShut {NoStop}%
\bibitem [{\citenamefont {Ding}\ \emph {et~al.}(2024)\citenamefont {Ding},
  \citenamefont {Huang},\ and\ \citenamefont {Yuan}}]{Ding_2024}%
  \BibitemOpen
  \bibfield  {author} {\bibinfo {author} {\bibfnamefont {Q.-M.}\ \bibnamefont
  {Ding}}, \bibinfo {author} {\bibfnamefont {Y.-M.}\ \bibnamefont {Huang}}, \
  and\ \bibinfo {author} {\bibfnamefont {X.}~\bibnamefont {Yuan}},\ }\bibfield
  {title} {\enquote {\bibinfo {title} {Molecular docking via quantum
  approximate optimization algorithm},}\ }\href {\doibase
  10.1103/PhysRevApplied.21.034036} {\bibfield  {journal} {\bibinfo  {journal}
  {Phys. Rev. Appl.}\ }\textbf {\bibinfo {volume} {21}},\ \bibinfo {pages}
  {034036} (\bibinfo {year} {2024})}\BibitemShut {NoStop}%
\bibitem [{\citenamefont {Wu}\ \emph {et~al.}(2023)\citenamefont {Wu},
  \citenamefont {Zhu}, \citenamefont {Bai}, \citenamefont {Wang},\ and\
  \citenamefont {Chiribella}}]{wu2023quantum}%
  \BibitemOpen
  \bibfield  {author} {\bibinfo {author} {\bibfnamefont {Y.-D.}\ \bibnamefont
  {Wu}}, \bibinfo {author} {\bibfnamefont {Y.}~\bibnamefont {Zhu}}, \bibinfo
  {author} {\bibfnamefont {G.}~\bibnamefont {Bai}}, \bibinfo {author}
  {\bibfnamefont {Y.}~\bibnamefont {Wang}}, \ and\ \bibinfo {author}
  {\bibfnamefont {G.}~\bibnamefont {Chiribella}},\ }\bibfield  {title}
  {\enquote {\bibinfo {title} {Quantum similarity testing with convolutional
  neural networks},}\ }\href {\doibase 10.1103/PhysRevLett.130.210601}
  {\bibfield  {journal} {\bibinfo  {journal} {Phys. Rev. Lett.}\ }\textbf
  {\bibinfo {volume} {130}},\ \bibinfo {pages} {210601} (\bibinfo {year}
  {2023})}\BibitemShut {NoStop}%
\bibitem [{\citenamefont {Du}\ \emph {et~al.}(2026)\citenamefont {Du},
  \citenamefont {Zhu}, \citenamefont {Zhang}, \citenamefont {Hsieh},
  \citenamefont {Rebentrost}, \citenamefont {Gao}, \citenamefont {You},
  \citenamefont {Eisert}, \citenamefont {Chiribella}, \citenamefont {Tao},
  \citenamefont {Sanders},\ and\ \citenamefont {Wu}}]{Du_2026}%
  \BibitemOpen
  \bibfield  {author} {\bibinfo {author} {\bibfnamefont {Y.}~\bibnamefont
  {Du}}, \bibinfo {author} {\bibfnamefont {Y.}~\bibnamefont {Zhu}}, \bibinfo
  {author} {\bibfnamefont {Y.-H.}\ \bibnamefont {Zhang}}, \bibinfo {author}
  {\bibfnamefont {M.-H.}\ \bibnamefont {Hsieh}}, \bibinfo {author}
  {\bibfnamefont {P.}~\bibnamefont {Rebentrost}}, \bibinfo {author}
  {\bibfnamefont {W.}~\bibnamefont {Gao}}, \bibinfo {author} {\bibfnamefont
  {Y.-Z.}\ \bibnamefont {You}}, \bibinfo {author} {\bibfnamefont
  {J.}~\bibnamefont {Eisert}}, \bibinfo {author} {\bibfnamefont
  {G.}~\bibnamefont {Chiribella}}, \bibinfo {author} {\bibfnamefont
  {D.}~\bibnamefont {Tao}}, \bibinfo {author} {\bibfnamefont {B.~C.}\
  \bibnamefont {Sanders}}, \ and\ \bibinfo {author} {\bibfnamefont {Y.-D.}\
  \bibnamefont {Wu}},\ }\bibfield  {title} {\enquote {\bibinfo {title}
  {Artificial intelligence for representing and characterizing quantum
  systems},}\ }\href {\doibase 10.1038/s42254-026-00962-5} {\bibfield
  {journal} {\bibinfo  {journal} {Nature Reviews Physics}\ }\textbf {\bibinfo
  {volume} {8}},\ \bibinfo {pages} {579–596} (\bibinfo {year}
  {2026})}\BibitemShut {NoStop}%
\bibitem [{\citenamefont {Wu}\ \emph {et~al.}(2024)\citenamefont {Wu},
  \citenamefont {Yao}, \citenamefont {Zhang},\ and\ \citenamefont
  {Li}}]{wu2024randomness}%
  \BibitemOpen
  \bibfield  {author} {\bibinfo {author} {\bibfnamefont {Y.}~\bibnamefont
  {Wu}}, \bibinfo {author} {\bibfnamefont {J.}~\bibnamefont {Yao}}, \bibinfo
  {author} {\bibfnamefont {P.}~\bibnamefont {Zhang}}, \ and\ \bibinfo {author}
  {\bibfnamefont {X.}~\bibnamefont {Li}},\ }\bibfield  {title} {\enquote
  {\bibinfo {title} {Randomness-enhanced expressivity of quantum neural
  networks},}\ }\href {\doibase 10.1103/PhysRevLett.132.010602} {\bibfield
  {journal} {\bibinfo  {journal} {Phys. Rev. Lett.}\ }\textbf {\bibinfo
  {volume} {132}},\ \bibinfo {pages} {010602} (\bibinfo {year}
  {2024})}\BibitemShut {NoStop}%
\bibitem [{\citenamefont {{Sharifian}}\ and\ \citenamefont
  {{Bayat}}(2026)}]{Sharifian2025hybrid}%
  \BibitemOpen
  \bibfield  {author} {\bibinfo {author} {\bibfnamefont {M.}~\bibnamefont
  {{Sharifian}}}\ and\ \bibinfo {author} {\bibfnamefont {A.}~\bibnamefont
  {{Bayat}}},\ }\bibfield  {title} {\enquote {\bibinfo {title} {{Hybrid boson
  sampling-neural network architecture for enhanced classification}},}\ }\href
  {https://doi.org/10.1038/s41534-026-01321-z} {\bibfield  {journal} {\bibinfo
  {journal} {npj Quantum Information}\ } (\bibinfo {year} {2026})},\ \Eprint
  {http://arxiv.org/abs/2510.13332} {arXiv:2510.13332 [quant-ph]} \BibitemShut
  {NoStop}%
\bibitem [{\citenamefont {Ren}\ \emph {et~al.}(2022)\citenamefont {Ren},
  \citenamefont {Li}, \citenamefont {Xu}, \citenamefont {Wang}, \citenamefont
  {Jiang}, \citenamefont {Jin}, \citenamefont {Zhu}, \citenamefont {Chen},
  \citenamefont {Song}, \citenamefont {Zhang}, \citenamefont {Dong},
  \citenamefont {Zhang}, \citenamefont {Deng}, \citenamefont {Gao},
  \citenamefont {Zhang}, \citenamefont {Wu}, \citenamefont {Zhang},
  \citenamefont {Guo}, \citenamefont {Li}, \citenamefont {Wang}, \citenamefont
  {Biamonte}, \citenamefont {Song}, \citenamefont {Deng},\ and\ \citenamefont
  {Wang}}]{ren2022experimental}%
  \BibitemOpen
  \bibfield  {author} {\bibinfo {author} {\bibfnamefont {W.}~\bibnamefont
  {Ren}}, \bibinfo {author} {\bibfnamefont {W.}~\bibnamefont {Li}}, \bibinfo
  {author} {\bibfnamefont {S.}~\bibnamefont {Xu}}, \bibinfo {author}
  {\bibfnamefont {K.}~\bibnamefont {Wang}}, \bibinfo {author} {\bibfnamefont
  {W.}~\bibnamefont {Jiang}}, \bibinfo {author} {\bibfnamefont
  {F.}~\bibnamefont {Jin}}, \bibinfo {author} {\bibfnamefont {X.}~\bibnamefont
  {Zhu}}, \bibinfo {author} {\bibfnamefont {J.}~\bibnamefont {Chen}}, \bibinfo
  {author} {\bibfnamefont {Z.}~\bibnamefont {Song}}, \bibinfo {author}
  {\bibfnamefont {P.}~\bibnamefont {Zhang}}, \bibinfo {author} {\bibfnamefont
  {H.}~\bibnamefont {Dong}}, \bibinfo {author} {\bibfnamefont {X.}~\bibnamefont
  {Zhang}}, \bibinfo {author} {\bibfnamefont {J.}~\bibnamefont {Deng}},
  \bibinfo {author} {\bibfnamefont {Y.}~\bibnamefont {Gao}}, \bibinfo {author}
  {\bibfnamefont {C.}~\bibnamefont {Zhang}}, \bibinfo {author} {\bibfnamefont
  {Y.}~\bibnamefont {Wu}}, \bibinfo {author} {\bibfnamefont {B.}~\bibnamefont
  {Zhang}}, \bibinfo {author} {\bibfnamefont {Q.}~\bibnamefont {Guo}}, \bibinfo
  {author} {\bibfnamefont {H.}~\bibnamefont {Li}}, \bibinfo {author}
  {\bibfnamefont {Z.}~\bibnamefont {Wang}}, \bibinfo {author} {\bibfnamefont
  {J.}~\bibnamefont {Biamonte}}, \bibinfo {author} {\bibfnamefont
  {C.}~\bibnamefont {Song}}, \bibinfo {author} {\bibfnamefont {D.-L.}\
  \bibnamefont {Deng}}, \ and\ \bibinfo {author} {\bibfnamefont
  {H.}~\bibnamefont {Wang}},\ }\bibfield  {title} {\enquote {\bibinfo {title}
  {Experimental quantum adversarial learning with programmable superconducting
  qubits},}\ }\href {\doibase 10.1038/s43588-022-00351-9} {\bibfield  {journal}
  {\bibinfo  {journal} {Nature Computational Science}\ }\textbf {\bibinfo
  {volume} {2}},\ \bibinfo {pages} {711–717} (\bibinfo {year}
  {2022})}\BibitemShut {NoStop}%
\bibitem [{\citenamefont {Chen}\ \emph {et~al.}(2026)\citenamefont {Chen},
  \citenamefont {Gan}, \citenamefont {Jin}, \citenamefont {Li}, \citenamefont
  {Wang}, \citenamefont {Wu}, \citenamefont {Wang}, \citenamefont {Zhang},
  \citenamefont {Srivastava}, \citenamefont {Chen}, \citenamefont
  {Kulshrestha}, \citenamefont {Liu}, \citenamefont {Mendoza-Arenas},
  \citenamefont {Seshadreesan}, \citenamefont {Upadhyay}, \citenamefont
  {Zhang}, \citenamefont {Zhuang},\ and\ \citenamefont {Liu}}]{chen2026ai}%
  \BibitemOpen
  \bibfield  {author} {\bibinfo {author} {\bibfnamefont {M.}~\bibnamefont
  {Chen}}, \bibinfo {author} {\bibfnamefont {Y.}~\bibnamefont {Gan}}, \bibinfo
  {author} {\bibfnamefont {X.}~\bibnamefont {Jin}}, \bibinfo {author}
  {\bibfnamefont {Y.}~\bibnamefont {Li}}, \bibinfo {author} {\bibfnamefont
  {J.}~\bibnamefont {Wang}}, \bibinfo {author} {\bibfnamefont {Z.}~\bibnamefont
  {Wu}}, \bibinfo {author} {\bibfnamefont {Y.}~\bibnamefont {Wang}}, \bibinfo
  {author} {\bibfnamefont {B.}~\bibnamefont {Zhang}}, \bibinfo {author}
  {\bibfnamefont {P.}~\bibnamefont {Srivastava}}, \bibinfo {author}
  {\bibfnamefont {T.}~\bibnamefont {Chen}}, \bibinfo {author} {\bibfnamefont
  {A.}~\bibnamefont {Kulshrestha}}, \bibinfo {author} {\bibfnamefont
  {Y.}~\bibnamefont {Liu}}, \bibinfo {author} {\bibfnamefont {J.~J.}\
  \bibnamefont {Mendoza-Arenas}}, \bibinfo {author} {\bibfnamefont {K.~P.}\
  \bibnamefont {Seshadreesan}}, \bibinfo {author} {\bibfnamefont
  {S.}~\bibnamefont {Upadhyay}}, \bibinfo {author} {\bibfnamefont
  {X.}~\bibnamefont {Zhang}}, \bibinfo {author} {\bibfnamefont
  {Q.}~\bibnamefont {Zhuang}}, \ and\ \bibinfo {author} {\bibfnamefont
  {J.}~\bibnamefont {Liu}},\ }\href {https://arxiv.org/abs/2607.00365}
  {\enquote {\bibinfo {title} {When ai meets quantum information: A
  comprehensive review},}\ } (\bibinfo {year} {2026}),\ \Eprint
  {http://arxiv.org/abs/2607.00365} {arXiv:2607.00365 [quant-ph]} \BibitemShut
  {NoStop}%
\bibitem [{\citenamefont {Hu}\ \emph {et~al.}(2019)\citenamefont {Hu},
  \citenamefont {Wu}, \citenamefont {Cai}, \citenamefont {Ma}, \citenamefont
  {Mu}, \citenamefont {Xu}, \citenamefont {Wang}, \citenamefont {Song},
  \citenamefont {Deng}, \citenamefont {Zou},\ and\ \citenamefont
  {Sun}}]{hu2019quantum}%
  \BibitemOpen
  \bibfield  {author} {\bibinfo {author} {\bibfnamefont {L.}~\bibnamefont
  {Hu}}, \bibinfo {author} {\bibfnamefont {S.-H.}\ \bibnamefont {Wu}}, \bibinfo
  {author} {\bibfnamefont {W.}~\bibnamefont {Cai}}, \bibinfo {author}
  {\bibfnamefont {Y.}~\bibnamefont {Ma}}, \bibinfo {author} {\bibfnamefont
  {X.}~\bibnamefont {Mu}}, \bibinfo {author} {\bibfnamefont {Y.}~\bibnamefont
  {Xu}}, \bibinfo {author} {\bibfnamefont {H.}~\bibnamefont {Wang}}, \bibinfo
  {author} {\bibfnamefont {Y.}~\bibnamefont {Song}}, \bibinfo {author}
  {\bibfnamefont {D.-L.}\ \bibnamefont {Deng}}, \bibinfo {author}
  {\bibfnamefont {C.-L.}\ \bibnamefont {Zou}}, \ and\ \bibinfo {author}
  {\bibfnamefont {L.}~\bibnamefont {Sun}},\ }\bibfield  {title} {\enquote
  {\bibinfo {title} {Quantum generative adversarial learning in a
  superconducting quantum circuit},}\ }\href {\doibase 10.1126/sciadv.aav2761}
  {\bibfield  {journal} {\bibinfo  {journal} {Science Advances}\ }\textbf
  {\bibinfo {volume} {5}} (\bibinfo {year} {2019}),\
  10.1126/sciadv.aav2761}\BibitemShut {NoStop}%
\bibitem [{\citenamefont {Li}\ \emph {et~al.}(2024)\citenamefont {Li},
  \citenamefont {Huang}, \citenamefont {Hou}, \citenamefont {Li}, \citenamefont
  {Wang},\ and\ \citenamefont {Bayat}}]{li2024ensemblelearning}%
  \BibitemOpen
  \bibfield  {author} {\bibinfo {author} {\bibfnamefont {Q.}~\bibnamefont
  {Li}}, \bibinfo {author} {\bibfnamefont {Y.}~\bibnamefont {Huang}}, \bibinfo
  {author} {\bibfnamefont {X.}~\bibnamefont {Hou}}, \bibinfo {author}
  {\bibfnamefont {Y.}~\bibnamefont {Li}}, \bibinfo {author} {\bibfnamefont
  {X.}~\bibnamefont {Wang}}, \ and\ \bibinfo {author} {\bibfnamefont
  {A.}~\bibnamefont {Bayat}},\ }\bibfield  {title} {\enquote {\bibinfo {title}
  {Ensemble-learning error mitigation for variational quantum shallow-circuit
  classifiers},}\ }\href {\doibase 10.1103/PhysRevResearch.6.013027} {\bibfield
   {journal} {\bibinfo  {journal} {Phys. Rev. Res.}\ }\textbf {\bibinfo
  {volume} {6}},\ \bibinfo {pages} {013027} (\bibinfo {year}
  {2024})}\BibitemShut {NoStop}%
\bibitem [{\citenamefont {Farhi}\ \emph {et~al.}(2014)\citenamefont {Farhi},
  \citenamefont {Goldstone},\ and\ \citenamefont
  {Gutmann}}]{farhi2014quantumapproximateoptimizationalgorithm}%
  \BibitemOpen
  \bibfield  {author} {\bibinfo {author} {\bibfnamefont {E.}~\bibnamefont
  {Farhi}}, \bibinfo {author} {\bibfnamefont {J.}~\bibnamefont {Goldstone}}, \
  and\ \bibinfo {author} {\bibfnamefont {S.}~\bibnamefont {Gutmann}},\ }\href
  {https://arxiv.org/abs/1411.4028} {\enquote {\bibinfo {title} {A quantum
  approximate optimization algorithm},}\ } (\bibinfo {year} {2014}),\ \Eprint
  {http://arxiv.org/abs/1411.4028} {arXiv:1411.4028 [quant-ph]} \BibitemShut
  {NoStop}%
\bibitem [{\citenamefont {Zhou}\ \emph {et~al.}(2020)\citenamefont {Zhou},
  \citenamefont {Wang}, \citenamefont {Choi}, \citenamefont {Pichler},\ and\
  \citenamefont {Lukin}}]{Zhou_2020}%
  \BibitemOpen
  \bibfield  {author} {\bibinfo {author} {\bibfnamefont {L.}~\bibnamefont
  {Zhou}}, \bibinfo {author} {\bibfnamefont {S.-T.}\ \bibnamefont {Wang}},
  \bibinfo {author} {\bibfnamefont {S.}~\bibnamefont {Choi}}, \bibinfo {author}
  {\bibfnamefont {H.}~\bibnamefont {Pichler}}, \ and\ \bibinfo {author}
  {\bibfnamefont {M.~D.}\ \bibnamefont {Lukin}},\ }\bibfield  {title} {\enquote
  {\bibinfo {title} {Quantum approximate optimization algorithm: Performance,
  mechanism, and implementation on near-term devices},}\ }\href {\doibase
  10.1103/physrevx.10.021067} {\bibfield  {journal} {\bibinfo  {journal}
  {Physical Review X}\ }\textbf {\bibinfo {volume} {10}} (\bibinfo {year}
  {2020}),\ 10.1103/physrevx.10.021067}\BibitemShut {NoStop}%
\bibitem [{\citenamefont {McClean}\ \emph {et~al.}(2018)\citenamefont
  {McClean}, \citenamefont {Boixo}, \citenamefont {Smelyanskiy}, \citenamefont
  {Babbush},\ and\ \citenamefont {Neven}}]{McClean_2018}%
  \BibitemOpen
  \bibfield  {author} {\bibinfo {author} {\bibfnamefont {J.~R.}\ \bibnamefont
  {McClean}}, \bibinfo {author} {\bibfnamefont {S.}~\bibnamefont {Boixo}},
  \bibinfo {author} {\bibfnamefont {V.~N.}\ \bibnamefont {Smelyanskiy}},
  \bibinfo {author} {\bibfnamefont {R.}~\bibnamefont {Babbush}}, \ and\
  \bibinfo {author} {\bibfnamefont {H.}~\bibnamefont {Neven}},\ }\bibfield
  {title} {\enquote {\bibinfo {title} {Barren plateaus in quantum neural
  network training landscapes},}\ }\href {\doibase 10.1038/s41467-018-07090-4}
  {\bibfield  {journal} {\bibinfo  {journal} {Nature Communications}\ }\textbf
  {\bibinfo {volume} {9}} (\bibinfo {year} {2018}),\
  10.1038/s41467-018-07090-4}\BibitemShut {NoStop}%
\bibitem [{\citenamefont {Fujii}\ and\ \citenamefont
  {Nakajima}(2017)}]{fujii2017harnessing}%
  \BibitemOpen
  \bibfield  {author} {\bibinfo {author} {\bibfnamefont {K.}~\bibnamefont
  {Fujii}}\ and\ \bibinfo {author} {\bibfnamefont {K.}~\bibnamefont
  {Nakajima}},\ }\bibfield  {title} {\enquote {\bibinfo {title} {Harnessing
  disordered-ensemble quantum dynamics for machine learning},}\ }\href
  {\doibase 10.1103/PhysRevApplied.8.024030} {\bibfield  {journal} {\bibinfo
  {journal} {Phys. Rev. Appl.}\ }\textbf {\bibinfo {volume} {8}},\ \bibinfo
  {pages} {024030} (\bibinfo {year} {2017})}\BibitemShut {NoStop}%
\bibitem [{\citenamefont {Ghosh}\ \emph {et~al.}(2021)\citenamefont {Ghosh},
  \citenamefont {Nakajima}, \citenamefont {Krisnanda}, \citenamefont {Fujii},\
  and\ \citenamefont {Liew}}]{ghosh2021quantum}%
  \BibitemOpen
  \bibfield  {author} {\bibinfo {author} {\bibfnamefont {S.}~\bibnamefont
  {Ghosh}}, \bibinfo {author} {\bibfnamefont {K.}~\bibnamefont {Nakajima}},
  \bibinfo {author} {\bibfnamefont {T.}~\bibnamefont {Krisnanda}}, \bibinfo
  {author} {\bibfnamefont {K.}~\bibnamefont {Fujii}}, \ and\ \bibinfo {author}
  {\bibfnamefont {T.~C.}\ \bibnamefont {Liew}},\ }\bibfield  {title} {\enquote
  {\bibinfo {title} {Quantum neuromorphic computing with reservoir computing
  networks},}\ }\href {https://doi.org/10.1002/qute.202100053} {\bibfield
  {journal} {\bibinfo  {journal} {Advanced Quantum Technologies}\ }\textbf
  {\bibinfo {volume} {4}},\ \bibinfo {pages} {2100053} (\bibinfo {year}
  {2021})}\BibitemShut {NoStop}%
\bibitem [{\citenamefont {Kobayashi}\ \emph {et~al.}(2024)\citenamefont
  {Kobayashi}, \citenamefont {Fujii},\ and\ \citenamefont
  {Yamamoto}}]{kobayashi2024feedback}%
  \BibitemOpen
  \bibfield  {author} {\bibinfo {author} {\bibfnamefont {K.}~\bibnamefont
  {Kobayashi}}, \bibinfo {author} {\bibfnamefont {K.}~\bibnamefont {Fujii}}, \
  and\ \bibinfo {author} {\bibfnamefont {N.}~\bibnamefont {Yamamoto}},\
  }\bibfield  {title} {\enquote {\bibinfo {title} {Feedback-driven quantum
  reservoir computing for time-series analysis},}\ }\href {\doibase
  10.1103/PRXQuantum.5.040325} {\bibfield  {journal} {\bibinfo  {journal} {PRX
  Quantum}\ }\textbf {\bibinfo {volume} {5}},\ \bibinfo {pages} {040325}
  (\bibinfo {year} {2024})}\BibitemShut {NoStop}%
\bibitem [{\citenamefont {Das}\ \emph {et~al.}(2026)\citenamefont {Das},
  \citenamefont {Giorgi},\ and\ \citenamefont {Zambrini}}]{das2026quatum}%
  \BibitemOpen
  \bibfield  {author} {\bibinfo {author} {\bibfnamefont {S.}~\bibnamefont
  {Das}}, \bibinfo {author} {\bibfnamefont {G.~L.}\ \bibnamefont {Giorgi}}, \
  and\ \bibinfo {author} {\bibfnamefont {R.}~\bibnamefont {Zambrini}},\
  }\bibfield  {title} {\enquote {\bibinfo {title} {Quantum reservoir computing
  in jaynes-cummings models: Nonlinear memory and time-series prediction},}\
  }\href {\doibase 10.1103/ffd3-ytbt} {\bibfield  {journal} {\bibinfo
  {journal} {Phys. Rev. Res.}\ }\textbf {\bibinfo {volume} {8}},\ \bibinfo
  {pages} {023148} (\bibinfo {year} {2026})}\BibitemShut {NoStop}%
\bibitem [{\citenamefont {Zia}\ \emph {et~al.}(2025)\citenamefont {Zia},
  \citenamefont {Innocenti}, \citenamefont {Minati}, \citenamefont {Lorenzo},
  \citenamefont {Suprano}, \citenamefont {Di~Bartolo}, \citenamefont
  {Spagnolo}, \citenamefont {Giordani}, \citenamefont {Cimini}, \citenamefont
  {Palma}, \citenamefont {Ferraro}, \citenamefont {Sciarrino},\ and\
  \citenamefont {Paternostro}}]{zia2025quantum}%
  \BibitemOpen
  \bibfield  {author} {\bibinfo {author} {\bibfnamefont {D.}~\bibnamefont
  {Zia}}, \bibinfo {author} {\bibfnamefont {L.}~\bibnamefont {Innocenti}},
  \bibinfo {author} {\bibfnamefont {G.}~\bibnamefont {Minati}}, \bibinfo
  {author} {\bibfnamefont {S.}~\bibnamefont {Lorenzo}}, \bibinfo {author}
  {\bibfnamefont {A.}~\bibnamefont {Suprano}}, \bibinfo {author} {\bibfnamefont
  {R.}~\bibnamefont {Di~Bartolo}}, \bibinfo {author} {\bibfnamefont
  {N.}~\bibnamefont {Spagnolo}}, \bibinfo {author} {\bibfnamefont
  {T.}~\bibnamefont {Giordani}}, \bibinfo {author} {\bibfnamefont
  {V.}~\bibnamefont {Cimini}}, \bibinfo {author} {\bibfnamefont {G.~M.}\
  \bibnamefont {Palma}}, \bibinfo {author} {\bibfnamefont {A.}~\bibnamefont
  {Ferraro}}, \bibinfo {author} {\bibfnamefont {F.}~\bibnamefont {Sciarrino}},
  \ and\ \bibinfo {author} {\bibfnamefont {M.}~\bibnamefont {Paternostro}},\
  }\bibfield  {title} {\enquote {\bibinfo {title} {Quantum reservoir computing
  for photonic entanglement witnessing},}\ }\href {\doibase
  10.1126/sciadv.ady7987} {\bibfield  {journal} {\bibinfo  {journal} {Science
  Advances}\ }\textbf {\bibinfo {volume} {11}} (\bibinfo {year} {2025}),\
  10.1126/sciadv.ady7987}\BibitemShut {NoStop}%
\bibitem [{\citenamefont {Sannia}\ \emph
  {et~al.}(2026{\natexlab{a}})\citenamefont {Sannia}, \citenamefont
  {Rodríguez}, \citenamefont {Giorgi},\ and\ \citenamefont
  {Zambrini}}]{Sannia_2026}%
  \BibitemOpen
  \bibfield  {author} {\bibinfo {author} {\bibfnamefont {A.}~\bibnamefont
  {Sannia}}, \bibinfo {author} {\bibfnamefont {R.~R.}\ \bibnamefont
  {Rodríguez}}, \bibinfo {author} {\bibfnamefont {G.~L.}\ \bibnamefont
  {Giorgi}}, \ and\ \bibinfo {author} {\bibfnamefont {R.}~\bibnamefont
  {Zambrini}},\ }\bibfield  {title} {\enquote {\bibinfo {title}
  {Non-markovianity and memory enhancement in quantum reservoir computing},}\
  }\href {\doibase 10.1038/s41534-026-01257-4} {\bibfield  {journal} {\bibinfo
  {journal} {npj Quantum Information}\ }\textbf {\bibinfo {volume} {12}}
  (\bibinfo {year} {2026}{\natexlab{a}}),\
  10.1038/s41534-026-01257-4}\BibitemShut {NoStop}%
\bibitem [{\citenamefont {Palacios}\ \emph {et~al.}(2024)\citenamefont
  {Palacios}, \citenamefont {Martínez-Peña}, \citenamefont {Soriano},
  \citenamefont {Giorgi},\ and\ \citenamefont {Zambrini}}]{Palacios_2024}%
  \BibitemOpen
  \bibfield  {author} {\bibinfo {author} {\bibfnamefont {A.}~\bibnamefont
  {Palacios}}, \bibinfo {author} {\bibfnamefont {R.}~\bibnamefont
  {Martínez-Peña}}, \bibinfo {author} {\bibfnamefont {M.~C.}\ \bibnamefont
  {Soriano}}, \bibinfo {author} {\bibfnamefont {G.~L.}\ \bibnamefont {Giorgi}},
  \ and\ \bibinfo {author} {\bibfnamefont {R.}~\bibnamefont {Zambrini}},\
  }\bibfield  {title} {\enquote {\bibinfo {title} {Role of coherence in
  many-body quantum reservoir computing},}\ }\href {\doibase
  10.1038/s42005-024-01859-4} {\bibfield  {journal} {\bibinfo  {journal}
  {Communications Physics}\ }\textbf {\bibinfo {volume} {7}} (\bibinfo {year}
  {2024}),\ 10.1038/s42005-024-01859-4}\BibitemShut {NoStop}%
\bibitem [{\citenamefont {Mujal}\ \emph {et~al.}(2023)\citenamefont {Mujal},
  \citenamefont {Martínez-Peña}, \citenamefont {Giorgi}, \citenamefont
  {Soriano},\ and\ \citenamefont {Zambrini}}]{Mujal_2023}%
  \BibitemOpen
  \bibfield  {author} {\bibinfo {author} {\bibfnamefont {P.}~\bibnamefont
  {Mujal}}, \bibinfo {author} {\bibfnamefont {R.}~\bibnamefont
  {Martínez-Peña}}, \bibinfo {author} {\bibfnamefont {G.~L.}\ \bibnamefont
  {Giorgi}}, \bibinfo {author} {\bibfnamefont {M.~C.}\ \bibnamefont {Soriano}},
  \ and\ \bibinfo {author} {\bibfnamefont {R.}~\bibnamefont {Zambrini}},\
  }\bibfield  {title} {\enquote {\bibinfo {title} {Time-series quantum
  reservoir computing with weak and projective measurements},}\ }\href
  {\doibase 10.1038/s41534-023-00682-z} {\bibfield  {journal} {\bibinfo
  {journal} {npj Quantum Information}\ }\textbf {\bibinfo {volume} {9}}
  (\bibinfo {year} {2023}),\ 10.1038/s41534-023-00682-z}\BibitemShut {NoStop}%
\bibitem [{\citenamefont {Xia}\ \emph {et~al.}(2023{\natexlab{a}})\citenamefont
  {Xia}, \citenamefont {Zou}, \citenamefont {Qiu}, \citenamefont {Chen},
  \citenamefont {Zhu}, \citenamefont {Li}, \citenamefont {Deng},\ and\
  \citenamefont {Li}}]{Xia_2023}%
  \BibitemOpen
  \bibfield  {author} {\bibinfo {author} {\bibfnamefont {W.}~\bibnamefont
  {Xia}}, \bibinfo {author} {\bibfnamefont {J.}~\bibnamefont {Zou}}, \bibinfo
  {author} {\bibfnamefont {X.}~\bibnamefont {Qiu}}, \bibinfo {author}
  {\bibfnamefont {F.}~\bibnamefont {Chen}}, \bibinfo {author} {\bibfnamefont
  {B.}~\bibnamefont {Zhu}}, \bibinfo {author} {\bibfnamefont {C.}~\bibnamefont
  {Li}}, \bibinfo {author} {\bibfnamefont {D.-L.}\ \bibnamefont {Deng}}, \ and\
  \bibinfo {author} {\bibfnamefont {X.}~\bibnamefont {Li}},\ }\bibfield
  {title} {\enquote {\bibinfo {title} {Configured quantum reservoir computing
  for multi-task machine learning},}\ }\href {\doibase
  https://doi.org/10.1016/j.scib.2023.08.040} {\bibfield  {journal} {\bibinfo
  {journal} {Science Bulletin}\ }\textbf {\bibinfo {volume} {68}},\ \bibinfo
  {pages} {2321--2329} (\bibinfo {year} {2023}{\natexlab{a}})}\BibitemShut
  {NoStop}%
\bibitem [{\citenamefont {Suzuki}\ \emph {et~al.}(2022)\citenamefont {Suzuki},
  \citenamefont {Gao}, \citenamefont {Pradel}, \citenamefont {Yasuoka},\ and\
  \citenamefont {Yamamoto}}]{suzuki2022natural}%
  \BibitemOpen
  \bibfield  {author} {\bibinfo {author} {\bibfnamefont {Y.}~\bibnamefont
  {Suzuki}}, \bibinfo {author} {\bibfnamefont {Q.}~\bibnamefont {Gao}},
  \bibinfo {author} {\bibfnamefont {K.~C.}\ \bibnamefont {Pradel}}, \bibinfo
  {author} {\bibfnamefont {K.}~\bibnamefont {Yasuoka}}, \ and\ \bibinfo
  {author} {\bibfnamefont {N.}~\bibnamefont {Yamamoto}},\ }\bibfield  {title}
  {\enquote {\bibinfo {title} {Natural quantum reservoir computing for temporal
  information processing},}\ }\href
  {https://doi.org/10.1038/s41598-022-05061-w} {\bibfield  {journal} {\bibinfo
  {journal} {Scientific reports}\ }\textbf {\bibinfo {volume} {12}},\ \bibinfo
  {pages} {1353} (\bibinfo {year} {2022})}\BibitemShut {NoStop}%
\bibitem [{\citenamefont {Mart{\'{\i}}nez-Pe{\~n}a}\ and\ \citenamefont
  {Ortega}(2023)}]{pena2023quantum}%
  \BibitemOpen
  \bibfield  {author} {\bibinfo {author} {\bibfnamefont {R.}~\bibnamefont
  {Mart{\'{\i}}nez-Pe{\~n}a}}\ and\ \bibinfo {author} {\bibfnamefont {J.-P.}\
  \bibnamefont {Ortega}},\ }\bibfield  {title} {\enquote {\bibinfo {title}
  {Quantum reservoir computing in finite dimensions},}\ }\href {\doibase
  10.1103/PhysRevE.107.035306} {\bibfield  {journal} {\bibinfo  {journal}
  {Phys. Rev. E}\ }\textbf {\bibinfo {volume} {107}},\ \bibinfo {pages}
  {035306} (\bibinfo {year} {2023})}\BibitemShut {NoStop}%
\bibitem [{\citenamefont {Askari}\ \emph {et~al.}(2025)\citenamefont {Askari},
  \citenamefont {Kora},\ and\ \citenamefont
  {Simon}}]{askari2025spinnetworkquantumreservoircomputing}%
  \BibitemOpen
  \bibfield  {author} {\bibinfo {author} {\bibfnamefont {S.}~\bibnamefont
  {Askari}}, \bibinfo {author} {\bibfnamefont {Y.}~\bibnamefont {Kora}}, \ and\
  \bibinfo {author} {\bibfnamefont {C.}~\bibnamefont {Simon}},\ }\href
  {https://arxiv.org/abs/2511.04900} {\enquote {\bibinfo {title} {Spin-network
  quantum reservoir computing with distributed inputs: The role of
  entanglement},}\ } (\bibinfo {year} {2025}),\ \Eprint
  {http://arxiv.org/abs/2511.04900} {arXiv:2511.04900 [quant-ph]} \BibitemShut
  {NoStop}%
\bibitem [{\citenamefont {Tariq}\ \emph {et~al.}(2026)\citenamefont {Tariq},
  \citenamefont {Talha}, \citenamefont {Ali}, \citenamefont {Diyan},\ and\
  \citenamefont {Chatzinotas}}]{tariq2026quantumreservoircomputingrecent}%
  \BibitemOpen
  \bibfield  {author} {\bibinfo {author} {\bibfnamefont {S.}~\bibnamefont
  {Tariq}}, \bibinfo {author} {\bibfnamefont {M.}~\bibnamefont {Talha}},
  \bibinfo {author} {\bibfnamefont {A.}~\bibnamefont {Ali}}, \bibinfo {author}
  {\bibfnamefont {M.}~\bibnamefont {Diyan}}, \ and\ \bibinfo {author}
  {\bibfnamefont {S.}~\bibnamefont {Chatzinotas}},\ }\href
  {https://arxiv.org/abs/2607.18552} {\enquote {\bibinfo {title} {Quantum
  reservoir computing: Recent advances and future directions},}\ } (\bibinfo
  {year} {2026}),\ \Eprint {http://arxiv.org/abs/2607.18552} {arXiv:2607.18552
  [quant-ph]} \BibitemShut {NoStop}%
\bibitem [{\citenamefont {Spagnolo}\ \emph {et~al.}(2022)\citenamefont
  {Spagnolo}, \citenamefont {Morris}, \citenamefont {Piacentini}, \citenamefont
  {Antesberger}, \citenamefont {Massa}, \citenamefont {Crespi}, \citenamefont
  {Ceccarelli}, \citenamefont {Osellame},\ and\ \citenamefont
  {Walther}}]{Spagnolo_2022}%
  \BibitemOpen
  \bibfield  {author} {\bibinfo {author} {\bibfnamefont {M.}~\bibnamefont
  {Spagnolo}}, \bibinfo {author} {\bibfnamefont {J.}~\bibnamefont {Morris}},
  \bibinfo {author} {\bibfnamefont {S.}~\bibnamefont {Piacentini}}, \bibinfo
  {author} {\bibfnamefont {M.}~\bibnamefont {Antesberger}}, \bibinfo {author}
  {\bibfnamefont {F.}~\bibnamefont {Massa}}, \bibinfo {author} {\bibfnamefont
  {A.}~\bibnamefont {Crespi}}, \bibinfo {author} {\bibfnamefont
  {F.}~\bibnamefont {Ceccarelli}}, \bibinfo {author} {\bibfnamefont
  {R.}~\bibnamefont {Osellame}}, \ and\ \bibinfo {author} {\bibfnamefont
  {P.}~\bibnamefont {Walther}},\ }\bibfield  {title} {\enquote {\bibinfo
  {title} {Experimental photonic quantum memristor},}\ }\href {\doibase
  10.1038/s41566-022-00973-5} {\bibfield  {journal} {\bibinfo  {journal}
  {Nature Photonics}\ }\textbf {\bibinfo {volume} {16}},\ \bibinfo {pages}
  {318–323} (\bibinfo {year} {2022})}\BibitemShut {NoStop}%
\bibitem [{\citenamefont {Paparelle}\ \emph {et~al.}(2026)\citenamefont
  {Paparelle}, \citenamefont {Henaff}, \citenamefont {Garcia-Beni},
  \citenamefont {Gillet}, \citenamefont {Montesinos}, \citenamefont {Giorgi},
  \citenamefont {Soriano}, \citenamefont {Zambrini},\ and\ \citenamefont
  {Parigi}}]{paparelle2026experimental}%
  \BibitemOpen
  \bibfield  {author} {\bibinfo {author} {\bibfnamefont {I.}~\bibnamefont
  {Paparelle}}, \bibinfo {author} {\bibfnamefont {J.}~\bibnamefont {Henaff}},
  \bibinfo {author} {\bibfnamefont {J.}~\bibnamefont {Garcia-Beni}}, \bibinfo
  {author} {\bibfnamefont {{\'E}.}~\bibnamefont {Gillet}}, \bibinfo {author}
  {\bibfnamefont {D.}~\bibnamefont {Montesinos}}, \bibinfo {author}
  {\bibfnamefont {G.~L.}\ \bibnamefont {Giorgi}}, \bibinfo {author}
  {\bibfnamefont {M.~C.}\ \bibnamefont {Soriano}}, \bibinfo {author}
  {\bibfnamefont {R.}~\bibnamefont {Zambrini}}, \ and\ \bibinfo {author}
  {\bibfnamefont {V.}~\bibnamefont {Parigi}},\ }\bibfield  {title} {\enquote
  {\bibinfo {title} {Experimental memory control in continuous-variable optical
  quantum reservoir computing},}\ }\href
  {https://doi.org/10.1038/s41566-026-01880-9} {\bibfield  {journal} {\bibinfo
  {journal} {Nature Photonics}\ }\textbf {\bibinfo {volume} {20}},\ \bibinfo
  {pages} {413--420} (\bibinfo {year} {2026})}\BibitemShut {NoStop}%
\bibitem [{\citenamefont {Xia}\ \emph {et~al.}(2022)\citenamefont {Xia},
  \citenamefont {Zou}, \citenamefont {Qiu},\ and\ \citenamefont
  {Li}}]{xia2022reservoir}%
  \BibitemOpen
  \bibfield  {author} {\bibinfo {author} {\bibfnamefont {W.}~\bibnamefont
  {Xia}}, \bibinfo {author} {\bibfnamefont {J.}~\bibnamefont {Zou}}, \bibinfo
  {author} {\bibfnamefont {X.}~\bibnamefont {Qiu}}, \ and\ \bibinfo {author}
  {\bibfnamefont {X.}~\bibnamefont {Li}},\ }\bibfield  {title} {\enquote
  {\bibinfo {title} {The reservoir learning power across quantum many-body
  localization transition},}\ }\href {\doibase 10.1007/s11467-022-1158-1}
  {\bibfield  {journal} {\bibinfo  {journal} {Frontiers of Physics}\ }\textbf
  {\bibinfo {volume} {17}} (\bibinfo {year} {2022}),\
  10.1007/s11467-022-1158-1}\BibitemShut {NoStop}%
\bibitem [{\citenamefont {Hou}\ \emph {et~al.}(2026{\natexlab{a}})\citenamefont
  {Hou}, \citenamefont {Hua}, \citenamefont {Wu}, \citenamefont {Xia},
  \citenamefont {Chen}, \citenamefont {Li}, \citenamefont {Li}, \citenamefont
  {Peng},\ and\ \citenamefont {Du}}]{hou2026highaccuracy}%
  \BibitemOpen
  \bibfield  {author} {\bibinfo {author} {\bibfnamefont {Y.}~\bibnamefont
  {Hou}}, \bibinfo {author} {\bibfnamefont {J.}~\bibnamefont {Hua}}, \bibinfo
  {author} {\bibfnamefont {Z.}~\bibnamefont {Wu}}, \bibinfo {author}
  {\bibfnamefont {W.}~\bibnamefont {Xia}}, \bibinfo {author} {\bibfnamefont
  {Y.}~\bibnamefont {Chen}}, \bibinfo {author} {\bibfnamefont {X.}~\bibnamefont
  {Li}}, \bibinfo {author} {\bibfnamefont {Z.}~\bibnamefont {Li}}, \bibinfo
  {author} {\bibfnamefont {X.}~\bibnamefont {Peng}}, \ and\ \bibinfo {author}
  {\bibfnamefont {J.}~\bibnamefont {Du}},\ }\bibfield  {title} {\enquote
  {\bibinfo {title} {High-accuracy temporal prediction via experimental quantum
  reservoir computing in correlated spins},}\ }\href {\doibase
  10.1103/r8ww-qw7j} {\bibfield  {journal} {\bibinfo  {journal} {Phys. Rev.
  Lett.}\ }\textbf {\bibinfo {volume} {136}},\ \bibinfo {pages} {120602}
  (\bibinfo {year} {2026}{\natexlab{a}})}\BibitemShut {NoStop}%
\bibitem [{\citenamefont {Xia}\ \emph {et~al.}(2023{\natexlab{b}})\citenamefont
  {Xia}, \citenamefont {Zou}, \citenamefont {Qiu}, \citenamefont {Chen},
  \citenamefont {Zhu}, \citenamefont {Li}, \citenamefont {Deng},\ and\
  \citenamefont {Li}}]{xia2023configured}%
  \BibitemOpen
  \bibfield  {author} {\bibinfo {author} {\bibfnamefont {W.}~\bibnamefont
  {Xia}}, \bibinfo {author} {\bibfnamefont {J.}~\bibnamefont {Zou}}, \bibinfo
  {author} {\bibfnamefont {X.}~\bibnamefont {Qiu}}, \bibinfo {author}
  {\bibfnamefont {F.}~\bibnamefont {Chen}}, \bibinfo {author} {\bibfnamefont
  {B.}~\bibnamefont {Zhu}}, \bibinfo {author} {\bibfnamefont {C.}~\bibnamefont
  {Li}}, \bibinfo {author} {\bibfnamefont {D.-L.}\ \bibnamefont {Deng}}, \ and\
  \bibinfo {author} {\bibfnamefont {X.}~\bibnamefont {Li}},\ }\bibfield
  {title} {\enquote {\bibinfo {title} {Configured quantum reservoir computing
  for multi-task machine learning},}\ }\href {\doibase
  10.1016/j.scib.2023.08.040} {\bibfield  {journal} {\bibinfo  {journal}
  {Science Bulletin}\ }\textbf {\bibinfo {volume} {68}},\ \bibinfo {pages}
  {2321--2329} (\bibinfo {year} {2023}{\natexlab{b}})}\BibitemShut {NoStop}%
\bibitem [{\citenamefont {Innocenti}\ \emph {et~al.}(2023)\citenamefont
  {Innocenti}, \citenamefont {Lorenzo}, \citenamefont {Palmisano},
  \citenamefont {Ferraro}, \citenamefont {Paternostro},\ and\ \citenamefont
  {Palma}}]{Innocenti_2023}%
  \BibitemOpen
  \bibfield  {author} {\bibinfo {author} {\bibfnamefont {L.}~\bibnamefont
  {Innocenti}}, \bibinfo {author} {\bibfnamefont {S.}~\bibnamefont {Lorenzo}},
  \bibinfo {author} {\bibfnamefont {I.}~\bibnamefont {Palmisano}}, \bibinfo
  {author} {\bibfnamefont {A.}~\bibnamefont {Ferraro}}, \bibinfo {author}
  {\bibfnamefont {M.}~\bibnamefont {Paternostro}}, \ and\ \bibinfo {author}
  {\bibfnamefont {G.~M.}\ \bibnamefont {Palma}},\ }\bibfield  {title} {\enquote
  {\bibinfo {title} {Potential and limitations of quantum extreme learning
  machines},}\ }\href {\doibase 10.1038/s42005-023-01233-w} {\bibfield
  {journal} {\bibinfo  {journal} {Communications Physics}\ }\textbf {\bibinfo
  {volume} {6}} (\bibinfo {year} {2023}),\
  10.1038/s42005-023-01233-w}\BibitemShut {NoStop}%
\bibitem [{\citenamefont {Xiong}\ \emph {et~al.}(2025)\citenamefont {Xiong},
  \citenamefont {Facelli}, \citenamefont {Sahebi}, \citenamefont {Agnel},
  \citenamefont {Chotibut}, \citenamefont {Thanasilp},\ and\ \citenamefont
  {Holmes}}]{xiong2025fundamental}%
  \BibitemOpen
  \bibfield  {author} {\bibinfo {author} {\bibfnamefont {W.}~\bibnamefont
  {Xiong}}, \bibinfo {author} {\bibfnamefont {G.}~\bibnamefont {Facelli}},
  \bibinfo {author} {\bibfnamefont {M.}~\bibnamefont {Sahebi}}, \bibinfo
  {author} {\bibfnamefont {O.}~\bibnamefont {Agnel}}, \bibinfo {author}
  {\bibfnamefont {T.}~\bibnamefont {Chotibut}}, \bibinfo {author}
  {\bibfnamefont {S.}~\bibnamefont {Thanasilp}}, \ and\ \bibinfo {author}
  {\bibfnamefont {Z.}~\bibnamefont {Holmes}},\ }\bibfield  {title} {\enquote
  {\bibinfo {title} {On fundamental aspects of quantum extreme learning
  machines},}\ }\href {\doibase 10.1007/s42484-025-00239-7} {\bibfield
  {journal} {\bibinfo  {journal} {Quantum Machine Intelligence}\ }\textbf
  {\bibinfo {volume} {7}} (\bibinfo {year} {2025}),\
  10.1007/s42484-025-00239-7}\BibitemShut {NoStop}%
\bibitem [{\citenamefont {De~Lorenzis}\ \emph {et~al.}(2025)\citenamefont
  {De~Lorenzis}, \citenamefont {Casado}, \citenamefont {Estarellas},
  \citenamefont {Lo~Gullo}, \citenamefont {Lux}, \citenamefont {Plastina},
  \citenamefont {Riera},\ and\ \citenamefont {Settino}}]{de2025harnessing}%
  \BibitemOpen
  \bibfield  {author} {\bibinfo {author} {\bibfnamefont {A.}~\bibnamefont
  {De~Lorenzis}}, \bibinfo {author} {\bibfnamefont {M.}~\bibnamefont {Casado}},
  \bibinfo {author} {\bibfnamefont {M.}~\bibnamefont {Estarellas}}, \bibinfo
  {author} {\bibfnamefont {N.}~\bibnamefont {Lo~Gullo}}, \bibinfo {author}
  {\bibfnamefont {T.}~\bibnamefont {Lux}}, \bibinfo {author} {\bibfnamefont
  {F.}~\bibnamefont {Plastina}}, \bibinfo {author} {\bibfnamefont
  {A.}~\bibnamefont {Riera}}, \ and\ \bibinfo {author} {\bibfnamefont
  {J.}~\bibnamefont {Settino}},\ }\bibfield  {title} {\enquote {\bibinfo
  {title} {Harnessing quantum extreme learning machines for image
  classification},}\ }\href {DOI:
  https://doi.org/10.1103/PhysRevApplied.23.044024} {\bibfield  {journal}
  {\bibinfo  {journal} {Physical Review Applied}\ }\textbf {\bibinfo {volume}
  {23}},\ \bibinfo {pages} {044024} (\bibinfo {year} {2025})}\BibitemShut
  {NoStop}%
\bibitem [{\citenamefont {Suprano}\ \emph {et~al.}(2024)\citenamefont
  {Suprano}, \citenamefont {Zia}, \citenamefont {Innocenti}, \citenamefont
  {Lorenzo}, \citenamefont {Cimini}, \citenamefont {Giordani}, \citenamefont
  {Palmisano}, \citenamefont {Polino}, \citenamefont {Spagnolo}, \citenamefont
  {Sciarrino}, \citenamefont {Palma}, \citenamefont {Ferraro},\ and\
  \citenamefont {Paternostro}}]{Suprano_2024}%
  \BibitemOpen
  \bibfield  {author} {\bibinfo {author} {\bibfnamefont {A.}~\bibnamefont
  {Suprano}}, \bibinfo {author} {\bibfnamefont {D.}~\bibnamefont {Zia}},
  \bibinfo {author} {\bibfnamefont {L.}~\bibnamefont {Innocenti}}, \bibinfo
  {author} {\bibfnamefont {S.}~\bibnamefont {Lorenzo}}, \bibinfo {author}
  {\bibfnamefont {V.}~\bibnamefont {Cimini}}, \bibinfo {author} {\bibfnamefont
  {T.}~\bibnamefont {Giordani}}, \bibinfo {author} {\bibfnamefont
  {I.}~\bibnamefont {Palmisano}}, \bibinfo {author} {\bibfnamefont
  {E.}~\bibnamefont {Polino}}, \bibinfo {author} {\bibfnamefont
  {N.}~\bibnamefont {Spagnolo}}, \bibinfo {author} {\bibfnamefont
  {F.}~\bibnamefont {Sciarrino}}, \bibinfo {author} {\bibfnamefont {G.~M.}\
  \bibnamefont {Palma}}, \bibinfo {author} {\bibfnamefont {A.}~\bibnamefont
  {Ferraro}}, \ and\ \bibinfo {author} {\bibfnamefont {M.}~\bibnamefont
  {Paternostro}},\ }\bibfield  {title} {\enquote {\bibinfo {title}
  {Experimental property reconstruction in a photonic quantum extreme learning
  machine},}\ }\href {\doibase 10.1103/PhysRevLett.132.160802} {\bibfield
  {journal} {\bibinfo  {journal} {Phys. Rev. Lett.}\ }\textbf {\bibinfo
  {volume} {132}},\ \bibinfo {pages} {160802} (\bibinfo {year}
  {2024})}\BibitemShut {NoStop}%
\bibitem [{\citenamefont {Solanki}\ and\ \citenamefont
  {Pham}(2026)}]{Solanki_2026}%
  \BibitemOpen
  \bibfield  {author} {\bibinfo {author} {\bibfnamefont {P.~D.}\ \bibnamefont
  {Solanki}}\ and\ \bibinfo {author} {\bibfnamefont {A.}~\bibnamefont {Pham}},\
  }\bibfield  {title} {\enquote {\bibinfo {title} {Harnessing quantum dynamics
  for robust and scalable quantum extreme learning machines},}\ }\href
  {\doibase 10.1103/2pm1-wrl6} {\bibfield  {journal} {\bibinfo  {journal}
  {Physical Review A}\ }\textbf {\bibinfo {volume} {113}} (\bibinfo {year}
  {2026}),\ 10.1103/2pm1-wrl6}\BibitemShut {NoStop}%
\bibitem [{\citenamefont {Joly}\ \emph {et~al.}(2025)\citenamefont {Joly},
  \citenamefont {Makowski}, \citenamefont {Courme}, \citenamefont
  {Porstendorfer}, \citenamefont {Wilksen}, \citenamefont {Charbon},
  \citenamefont {Gies}, \citenamefont {Defienne},\ and\ \citenamefont
  {Gigan}}]{joly2025harnessingphotonindistinguishabilityquantum}%
  \BibitemOpen
  \bibfield  {author} {\bibinfo {author} {\bibfnamefont {M.}~\bibnamefont
  {Joly}}, \bibinfo {author} {\bibfnamefont {A.}~\bibnamefont {Makowski}},
  \bibinfo {author} {\bibfnamefont {B.}~\bibnamefont {Courme}}, \bibinfo
  {author} {\bibfnamefont {L.}~\bibnamefont {Porstendorfer}}, \bibinfo {author}
  {\bibfnamefont {S.}~\bibnamefont {Wilksen}}, \bibinfo {author} {\bibfnamefont
  {E.}~\bibnamefont {Charbon}}, \bibinfo {author} {\bibfnamefont
  {C.}~\bibnamefont {Gies}}, \bibinfo {author} {\bibfnamefont {H.}~\bibnamefont
  {Defienne}}, \ and\ \bibinfo {author} {\bibfnamefont {S.}~\bibnamefont
  {Gigan}},\ }\href {https://arxiv.org/abs/2505.11238} {\enquote {\bibinfo
  {title} {Harnessing photon indistinguishability in quantum extreme learning
  machines},}\ } (\bibinfo {year} {2025}),\ \Eprint
  {http://arxiv.org/abs/2505.11238} {arXiv:2505.11238 [quant-ph]} \BibitemShut
  {NoStop}%
\bibitem [{\citenamefont {De~Lorenzis}\ \emph {et~al.}(2026)\citenamefont
  {De~Lorenzis}, \citenamefont {Casado}, \citenamefont {Lo~Gullo},
  \citenamefont {Lux}, \citenamefont {Plastina},\ and\ \citenamefont
  {Riera}}]{de2026entanglement}%
  \BibitemOpen
  \bibfield  {author} {\bibinfo {author} {\bibfnamefont {A.}~\bibnamefont
  {De~Lorenzis}}, \bibinfo {author} {\bibfnamefont {M.}~\bibnamefont {Casado}},
  \bibinfo {author} {\bibfnamefont {N.}~\bibnamefont {Lo~Gullo}}, \bibinfo
  {author} {\bibfnamefont {T.}~\bibnamefont {Lux}}, \bibinfo {author}
  {\bibfnamefont {F.}~\bibnamefont {Plastina}}, \ and\ \bibinfo {author}
  {\bibfnamefont {A.}~\bibnamefont {Riera}},\ }\bibfield  {title} {\enquote
  {\bibinfo {title} {Entanglement and classical simulability in quantum
  extreme-learning machines},}\ }\href {\doibase 10.1103/891x-ltcy} {\bibfield
  {journal} {\bibinfo  {journal} {Phys. Rev. Appl.}\ }\textbf {\bibinfo
  {volume} {25}},\ \bibinfo {pages} {054052} (\bibinfo {year}
  {2026})}\BibitemShut {NoStop}%
\bibitem [{\citenamefont {Assil}\ \emph {et~al.}(2026)\citenamefont {Assil},
  \citenamefont {El~Allati},\ and\ \citenamefont {Giorgi}}]{Assil_2026}%
  \BibitemOpen
  \bibfield  {author} {\bibinfo {author} {\bibfnamefont {H.}~\bibnamefont
  {Assil}}, \bibinfo {author} {\bibfnamefont {A.}~\bibnamefont {El~Allati}}, \
  and\ \bibinfo {author} {\bibfnamefont {G.~L.}\ \bibnamefont {Giorgi}},\
  }\bibfield  {title} {\enquote {\bibinfo {title} {Memory-enhanced quantum
  extreme learning machines for characterizing non-markovian dynamics},}\
  }\href {\doibase 10.1103/ycvz-lzgl} {\bibfield  {journal} {\bibinfo
  {journal} {Physical Review Research}\ }\textbf {\bibinfo {volume} {8}}
  (\bibinfo {year} {2026}),\ 10.1103/ycvz-lzgl}\BibitemShut {NoStop}%
\bibitem [{\citenamefont {Cao}\ \emph {et~al.}(2026)\citenamefont {Cao},
  \citenamefont {Xia},\ and\ \citenamefont {Li}}]{cao2026theory}%
  \BibitemOpen
  \bibfield  {author} {\bibinfo {author} {\bibfnamefont {S.}~\bibnamefont
  {Cao}}, \bibinfo {author} {\bibfnamefont {W.}~\bibnamefont {Xia}}, \ and\
  \bibinfo {author} {\bibfnamefont {X.}~\bibnamefont {Li}},\ }\bibfield
  {title} {\enquote {\bibinfo {title} {Theory of the edge-of-chaos advantage in
  quantum reservoir computing},}\ }\href {https://arxiv.org/abs/2610.05213} {\
  (\bibinfo {year} {2026})},\ \Eprint {http://arxiv.org/abs/2610.05213}
  {arXiv:2610.05213 [quant-ph]} \BibitemShut {NoStop}%
\bibitem [{\citenamefont {Wang}\ \emph {et~al.}(2024)\citenamefont {Wang},
  \citenamefont {Ali}, \citenamefont {Arrieta}, \citenamefont {Arcaini},\ and\
  \citenamefont {Arratibel}}]{wang2024application}%
  \BibitemOpen
  \bibfield  {author} {\bibinfo {author} {\bibfnamefont {X.}~\bibnamefont
  {Wang}}, \bibinfo {author} {\bibfnamefont {S.}~\bibnamefont {Ali}}, \bibinfo
  {author} {\bibfnamefont {A.}~\bibnamefont {Arrieta}}, \bibinfo {author}
  {\bibfnamefont {P.}~\bibnamefont {Arcaini}}, \ and\ \bibinfo {author}
  {\bibfnamefont {M.}~\bibnamefont {Arratibel}},\ }\href
  {https://arxiv.org/abs/2402.12777} {\enquote {\bibinfo {title} {Application
  of quantum extreme learning machines for qos prediction of elevators'
  software in an industrial context},}\ } (\bibinfo {year} {2024}),\ \Eprint
  {http://arxiv.org/abs/2402.12777} {arXiv:2402.12777 [cs.SE]} \BibitemShut
  {NoStop}%
\bibitem [{\citenamefont {Maier}\ \emph {et~al.}(2026)\citenamefont {Maier},
  \citenamefont {Spannowsky},\ and\ \citenamefont {Williams}}]{Maier_2026}%
  \BibitemOpen
  \bibfield  {author} {\bibinfo {author} {\bibfnamefont {B.}~\bibnamefont
  {Maier}}, \bibinfo {author} {\bibfnamefont {M.}~\bibnamefont {Spannowsky}}, \
  and\ \bibinfo {author} {\bibfnamefont {S.}~\bibnamefont {Williams}},\
  }\bibfield  {title} {\enquote {\bibinfo {title} {Continuous-variable photonic
  quantum extreme learning machines for fast collider-data selection},}\ }\href
  {\doibase 10.1140/epjqt/s40507-026-00507-w} {\bibfield  {journal} {\bibinfo
  {journal} {EPJ Quantum Technology}\ }\textbf {\bibinfo {volume} {13}}
  (\bibinfo {year} {2026}),\ 10.1140/epjqt/s40507-026-00507-w}\BibitemShut
  {NoStop}%
\bibitem [{\citenamefont {Vetrano}\ \emph {et~al.}(2025)\citenamefont
  {Vetrano}, \citenamefont {Lo~Monaco}, \citenamefont {Innocenti},
  \citenamefont {Lorenzo},\ and\ \citenamefont {Palma}}]{Vetrano_2025}%
  \BibitemOpen
  \bibfield  {author} {\bibinfo {author} {\bibfnamefont {M.}~\bibnamefont
  {Vetrano}}, \bibinfo {author} {\bibfnamefont {G.}~\bibnamefont {Lo~Monaco}},
  \bibinfo {author} {\bibfnamefont {L.}~\bibnamefont {Innocenti}}, \bibinfo
  {author} {\bibfnamefont {S.}~\bibnamefont {Lorenzo}}, \ and\ \bibinfo
  {author} {\bibfnamefont {G.~M.}\ \bibnamefont {Palma}},\ }\bibfield  {title}
  {\enquote {\bibinfo {title} {State estimation with quantum extreme learning
  machines beyond the scrambling time},}\ }\href {\doibase
  10.1038/s41534-024-00927-5} {\bibfield  {journal} {\bibinfo  {journal} {npj
  Quantum Information}\ }\textbf {\bibinfo {volume} {11}} (\bibinfo {year}
  {2025}),\ 10.1038/s41534-024-00927-5}\BibitemShut {NoStop}%
\bibitem [{\citenamefont {Li}\ \emph {et~al.}(2026{\natexlab{a}})\citenamefont
  {Li}, \citenamefont {Mukhopadhyay}, \citenamefont {Minati},\ and\
  \citenamefont {Bayat}}]{li2026quantumchaotic}%
  \BibitemOpen
  \bibfield  {author} {\bibinfo {author} {\bibfnamefont {Q.}~\bibnamefont
  {Li}}, \bibinfo {author} {\bibfnamefont {C.}~\bibnamefont {Mukhopadhyay}},
  \bibinfo {author} {\bibfnamefont {L.}~\bibnamefont {Minati}}, \ and\ \bibinfo
  {author} {\bibfnamefont {A.}~\bibnamefont {Bayat}},\ }\bibfield  {title}
  {\enquote {\bibinfo {title} {Quantum reservoir computing for predicting and
  characterizing chaotic maps},}\ }\href {\doibase 10.1103/qwvv-xqfq}
  {\bibfield  {journal} {\bibinfo  {journal} {Phys. Rev. Res.}\ }\textbf
  {\bibinfo {volume} {8}},\ \bibinfo {pages} {013304} (\bibinfo {year}
  {2026}{\natexlab{a}})}\BibitemShut {NoStop}%
\bibitem [{\citenamefont {Li}\ \emph {et~al.}(2026{\natexlab{b}})\citenamefont
  {Li}, \citenamefont {Mukhopadhyay}, \citenamefont {Bayat},\ and\
  \citenamefont {Habibnia}}]{li2026quantumrv}%
  \BibitemOpen
  \bibfield  {author} {\bibinfo {author} {\bibfnamefont {Q.}~\bibnamefont
  {Li}}, \bibinfo {author} {\bibfnamefont {C.}~\bibnamefont {Mukhopadhyay}},
  \bibinfo {author} {\bibfnamefont {A.}~\bibnamefont {Bayat}}, \ and\ \bibinfo
  {author} {\bibfnamefont {A.}~\bibnamefont {Habibnia}},\ }\bibfield  {title}
  {\enquote {\bibinfo {title} {Quantum reservoir computing for realized
  volatility forecasting},}\ }\href {\doibase 10.1103/rbj7-4wnq} {\bibfield
  {journal} {\bibinfo  {journal} {Phys. Rev. Res.}\ }\textbf {\bibinfo {volume}
  {8}},\ \bibinfo {pages} {023028} (\bibinfo {year}
  {2026}{\natexlab{b}})}\BibitemShut {NoStop}%
\bibitem [{\citenamefont {Sannia}\ \emph
  {et~al.}(2026{\natexlab{b}})\citenamefont {Sannia}, \citenamefont {Menta},
  \citenamefont {Sathe}, \citenamefont {Santis}, \citenamefont {Giovannetti},
  \citenamefont {García-Pintos}, \citenamefont {Venturelli}, \citenamefont
  {Giorgi}, \citenamefont {Zambrini},\ and\ \citenamefont
  {Caravelli}}]{sannia2026temporalinformationprocessing4500qubit}%
  \BibitemOpen
  \bibfield  {author} {\bibinfo {author} {\bibfnamefont {A.}~\bibnamefont
  {Sannia}}, \bibinfo {author} {\bibfnamefont {R.}~\bibnamefont {Menta}},
  \bibinfo {author} {\bibfnamefont {P.}~\bibnamefont {Sathe}}, \bibinfo
  {author} {\bibfnamefont {D.~D.}\ \bibnamefont {Santis}}, \bibinfo {author}
  {\bibfnamefont {V.}~\bibnamefont {Giovannetti}}, \bibinfo {author}
  {\bibfnamefont {L.~P.}\ \bibnamefont {García-Pintos}}, \bibinfo {author}
  {\bibfnamefont {D.}~\bibnamefont {Venturelli}}, \bibinfo {author}
  {\bibfnamefont {G.~L.}\ \bibnamefont {Giorgi}}, \bibinfo {author}
  {\bibfnamefont {R.}~\bibnamefont {Zambrini}}, \ and\ \bibinfo {author}
  {\bibfnamefont {F.}~\bibnamefont {Caravelli}},\ }\href
  {https://arxiv.org/abs/2609.19308} {\enquote {\bibinfo {title} {Temporal
  information processing on a 4,500-qubit quantum annealer},}\ } (\bibinfo
  {year} {2026}{\natexlab{b}}),\ \Eprint {http://arxiv.org/abs/2609.19308}
  {arXiv:2609.19308 [quant-ph]} \BibitemShut {NoStop}%
\bibitem [{\citenamefont {Hou}\ \emph {et~al.}(2026{\natexlab{b}})\citenamefont
  {Hou}, \citenamefont {Hua}, \citenamefont {Wu}, \citenamefont {Xia},
  \citenamefont {Chen}, \citenamefont {Li}, \citenamefont {Li}, \citenamefont
  {Peng},\ and\ \citenamefont {Du}}]{hou2026high}%
  \BibitemOpen
  \bibfield  {author} {\bibinfo {author} {\bibfnamefont {Y.}~\bibnamefont
  {Hou}}, \bibinfo {author} {\bibfnamefont {J.}~\bibnamefont {Hua}}, \bibinfo
  {author} {\bibfnamefont {Z.}~\bibnamefont {Wu}}, \bibinfo {author}
  {\bibfnamefont {W.}~\bibnamefont {Xia}}, \bibinfo {author} {\bibfnamefont
  {Y.}~\bibnamefont {Chen}}, \bibinfo {author} {\bibfnamefont {X.}~\bibnamefont
  {Li}}, \bibinfo {author} {\bibfnamefont {Z.}~\bibnamefont {Li}}, \bibinfo
  {author} {\bibfnamefont {X.}~\bibnamefont {Peng}}, \ and\ \bibinfo {author}
  {\bibfnamefont {J.}~\bibnamefont {Du}},\ }\bibfield  {title} {\enquote
  {\bibinfo {title} {High-accuracy temporal prediction via experimental quantum
  reservoir computing in correlated spins},}\ }\href {\doibase
  10.1103/r8ww-qw7j} {\bibfield  {journal} {\bibinfo  {journal} {Physical
  Review Letters}\ }\textbf {\bibinfo {volume} {136}} (\bibinfo {year}
  {2026}{\natexlab{b}}),\ 10.1103/r8ww-qw7j}\BibitemShut {NoStop}%
\bibitem [{\citenamefont {Steinegger}\ and\ \citenamefont
  {R{\"a}th}(2025)}]{steinegger2025predicting}%
  \BibitemOpen
  \bibfield  {author} {\bibinfo {author} {\bibfnamefont {J.}~\bibnamefont
  {Steinegger}}\ and\ \bibinfo {author} {\bibfnamefont {C.}~\bibnamefont
  {R{\"a}th}},\ }\bibfield  {title} {\enquote {\bibinfo {title} {Predicting
  three-dimensional chaotic systems with four qubit quantum systems},}\ }\href
  {https://doi.org/10.1038/s41598-025-87768-0} {\bibfield  {journal} {\bibinfo
  {journal} {Scientific Reports}\ }\textbf {\bibinfo {volume} {15}},\ \bibinfo
  {pages} {6201} (\bibinfo {year} {2025})}\BibitemShut {NoStop}%
\bibitem [{\citenamefont {Fry}\ \emph {et~al.}(2023)\citenamefont {Fry},
  \citenamefont {Deshmukh}, \citenamefont {Chen}, \citenamefont {Rastunkov},\
  and\ \citenamefont {Markov}}]{fry2023optimizing}%
  \BibitemOpen
  \bibfield  {author} {\bibinfo {author} {\bibfnamefont {D.}~\bibnamefont
  {Fry}}, \bibinfo {author} {\bibfnamefont {A.}~\bibnamefont {Deshmukh}},
  \bibinfo {author} {\bibfnamefont {S.~Y.-C.}\ \bibnamefont {Chen}}, \bibinfo
  {author} {\bibfnamefont {V.}~\bibnamefont {Rastunkov}}, \ and\ \bibinfo
  {author} {\bibfnamefont {V.}~\bibnamefont {Markov}},\ }\bibfield  {title}
  {\enquote {\bibinfo {title} {Optimizing quantum noise-induced reservoir
  computing for nonlinear and chaotic time series prediction},}\ }\href
  {https://doi.org/10.1038/s41598-023-45015-4} {\bibfield  {journal} {\bibinfo
  {journal} {Scientific Reports}\ }\textbf {\bibinfo {volume} {13}},\ \bibinfo
  {pages} {19326} (\bibinfo {year} {2023})}\BibitemShut {NoStop}%
\bibitem [{\citenamefont {Ahmed}\ \emph {et~al.}(2025)\citenamefont {Ahmed},
  \citenamefont {Tennie},\ and\ \citenamefont {Magri}}]{ahmed2025robust}%
  \BibitemOpen
  \bibfield  {author} {\bibinfo {author} {\bibfnamefont {O.}~\bibnamefont
  {Ahmed}}, \bibinfo {author} {\bibfnamefont {F.}~\bibnamefont {Tennie}}, \
  and\ \bibinfo {author} {\bibfnamefont {L.}~\bibnamefont {Magri}},\ }\bibfield
   {title} {\enquote {\bibinfo {title} {Robust quantum reservoir computers for
  forecasting chaotic dynamics: generalized synchronization and stability},}\
  }\href {\doibase 10.1098/rspa.2025.0550} {\bibfield  {journal} {\bibinfo
  {journal} {Proceedings of the Royal Society A: Mathematical, Physical and
  Engineering Sciences}\ }\textbf {\bibinfo {volume} {481}},\ \bibinfo {pages}
  {20250550} (\bibinfo {year} {2025})}\BibitemShut {NoStop}%
\bibitem [{\citenamefont {Sornsaeng}\ \emph {et~al.}(2024)\citenamefont
  {Sornsaeng}, \citenamefont {Dangniam},\ and\ \citenamefont
  {Chotibut}}]{Sornsaeng_2024}%
  \BibitemOpen
  \bibfield  {author} {\bibinfo {author} {\bibfnamefont {A.}~\bibnamefont
  {Sornsaeng}}, \bibinfo {author} {\bibfnamefont {N.}~\bibnamefont {Dangniam}},
  \ and\ \bibinfo {author} {\bibfnamefont {T.}~\bibnamefont {Chotibut}},\
  }\bibfield  {title} {\enquote {\bibinfo {title} {Quantum next generation
  reservoir computing: an efficient quantum algorithm for forecasting quantum
  dynamics},}\ }\href {\doibase 10.1007/s42484-024-00188-7} {\bibfield
  {journal} {\bibinfo  {journal} {Quantum Machine Intelligence}\ }\textbf
  {\bibinfo {volume} {6}} (\bibinfo {year} {2024}),\
  10.1007/s42484-024-00188-7}\BibitemShut {NoStop}%
\bibitem [{\citenamefont {Fellner}\ \emph {et~al.}(2026)\citenamefont
  {Fellner}, \citenamefont {Merklinger},\ and\ \citenamefont
  {Holm}}]{fellner2026multivariatequantumreservoircomputing}%
  \BibitemOpen
  \bibfield  {author} {\bibinfo {author} {\bibfnamefont {T.}~\bibnamefont
  {Fellner}}, \bibinfo {author} {\bibfnamefont {J.}~\bibnamefont {Merklinger}},
  \ and\ \bibinfo {author} {\bibfnamefont {C.}~\bibnamefont {Holm}},\ }\href
  {https://arxiv.org/abs/2604.08427} {\enquote {\bibinfo {title} {Multivariate
  quantum reservoir computing with discrete and continuous variable systems},}\
  } (\bibinfo {year} {2026}),\ \Eprint {http://arxiv.org/abs/2604.08427}
  {arXiv:2604.08427 [quant-ph]} \BibitemShut {NoStop}%
\bibitem [{\citenamefont {Li}\ \emph {et~al.}(2026{\natexlab{c}})\citenamefont
  {Li}, \citenamefont {Huo}, \citenamefont {Zhao},\ and\ \citenamefont
  {Duan}}]{Li2026higher}%
  \BibitemOpen
  \bibfield  {author} {\bibinfo {author} {\bibfnamefont {X.}~\bibnamefont
  {Li}}, \bibinfo {author} {\bibfnamefont {C.}~\bibnamefont {Huo}}, \bibinfo
  {author} {\bibfnamefont {C.}~\bibnamefont {Zhao}}, \ and\ \bibinfo {author}
  {\bibfnamefont {X.}~\bibnamefont {Duan}},\ }\bibfield  {title} {\enquote
  {\bibinfo {title} {Higher-order quantum reservoir computing for forecasting
  complex dynamics},}\ }\href {\doibase 10.3934/dcdsi.A260101} {\bibfield
  {journal} {\bibinfo  {journal} {Discrete and Continuous Dynamical Systems -
  I: Intelligence}\ }\textbf {\bibinfo {volume} {1}} (\bibinfo {year}
  {2026}{\natexlab{c}}),\ 10.3934/dcdsi.A260101}\BibitemShut {NoStop}%
\bibitem [{\citenamefont {Sprott}\ and\ \citenamefont
  {Thio}(2022)}]{sprottbook}%
  \BibitemOpen
  \bibfield  {author} {\bibinfo {author} {\bibfnamefont {J.~C.}\ \bibnamefont
  {Sprott}}\ and\ \bibinfo {author} {\bibfnamefont {W.~J.-C.}\ \bibnamefont
  {Thio}},\ }\href {\doibase 10.1142/12362} {\emph {\bibinfo {title} {Elegant
  Circuits}}}\ (\bibinfo  {publisher} {World Scientific},\ \bibinfo {year}
  {2022})\ \Eprint
  {http://arxiv.org/abs/https://www.worldscientific.com/doi/pdf/10.1142/12362}
  {https://www.worldscientific.com/doi/pdf/10.1142/12362} \BibitemShut
  {NoStop}%
\bibitem [{\citenamefont {Minati}\ \emph {et~al.}(2022)\citenamefont {Minati},
  \citenamefont {Innocenti}, \citenamefont {Mijatovic}, \citenamefont {Ito},\
  and\ \citenamefont {Frasca}}]{minati2022mechanisms}%
  \BibitemOpen
  \bibfield  {author} {\bibinfo {author} {\bibfnamefont {L.}~\bibnamefont
  {Minati}}, \bibinfo {author} {\bibfnamefont {G.}~\bibnamefont {Innocenti}},
  \bibinfo {author} {\bibfnamefont {G.}~\bibnamefont {Mijatovic}}, \bibinfo
  {author} {\bibfnamefont {H.}~\bibnamefont {Ito}}, \ and\ \bibinfo {author}
  {\bibfnamefont {M.}~\bibnamefont {Frasca}},\ }\bibfield  {title} {\enquote
  {\bibinfo {title} {Mechanisms of chaos generation in an atypical
  single-transistor oscillator},}\ }\href {\doibase
  10.1016/j.chaos.2022.111878} {\bibfield  {journal} {\bibinfo  {journal}
  {Chaos, Solitons \& Fractals}\ }\textbf {\bibinfo {volume} {157}},\ \bibinfo
  {pages} {111878} (\bibinfo {year} {2022})}\BibitemShut {NoStop}%
\bibitem [{\citenamefont {Takens}(2006)}]{takens2006detecting}%
  \BibitemOpen
  \bibfield  {author} {\bibinfo {author} {\bibfnamefont {F.}~\bibnamefont
  {Takens}},\ }\bibfield  {title} {\enquote {\bibinfo {title} {Detecting
  strange attractors in turbulence},}\ }\href {\doibase
  https://doi.org/10.1007/BFb0091924} {\bibfield  {journal} {\bibinfo
  {journal} {Dynamical Systems and Turbulence, Warwick}\ ,\ \bibinfo {pages}
  {366--381}} (\bibinfo {year} {2006})}\BibitemShut {NoStop}%
\bibitem [{\citenamefont {Hochreiter}\ and\ \citenamefont
  {Schmidhuber}(1997)}]{hochreiter1997long}%
  \BibitemOpen
  \bibfield  {author} {\bibinfo {author} {\bibfnamefont {S.}~\bibnamefont
  {Hochreiter}}\ and\ \bibinfo {author} {\bibfnamefont {J.}~\bibnamefont
  {Schmidhuber}},\ }\bibfield  {title} {\enquote {\bibinfo {title} {Long
  short-term memory},}\ }\href {\doibase 10.1162/neco.1997.9.8.1735} {\bibfield
   {journal} {\bibinfo  {journal} {Neural Computation}\ }\textbf {\bibinfo
  {volume} {9}},\ \bibinfo {pages} {1735--1780} (\bibinfo {year}
  {1997})}\BibitemShut {NoStop}%
\bibitem [{zen(2026)}]{zenodo}%
  \BibitemOpen
  \bibfield  {title} {\enquote {\bibinfo {title} {Experimental time series
  recorded from the analog electronic circuits},}\ }\href {\doibase
  10.5281/zenodo.22857591} {\bibfield  {journal} {\bibinfo  {journal} {Zenodo,
  Online resource}\ } (\bibinfo {year} {2026}),\ 10.5281/zenodo.22857591},\
  \bibinfo {note} {available online}\BibitemShut {NoStop}%
\end{thebibliography}%

\clearpage
\onecolumngrid

\appendix
\renewcommand{\thefigure}{S\arabic{figure}}
\setcounter{figure}{0}

\section*{Supplementary Information}

\subsection*{Uniform discretization with one-to-one prediction}

In the main text, uniform discretization is primarily considered in the four-to-one setting, where the past values of all four signals are applied to the setup and prediction of the future value of a selected target signal is done. Here, for completeness, we consider uniform discretization in a one-to-one setting, in which each individual signal is predicted exclusively from its own past values. This provides a complementary assessment of the prediction capability of the model using the temporal history of each individual signal, without exploiting correlations with the other dynamical variables.

The data preparation and sampling procedure are identical to those used for the four-to-one uniform discretization results in the main text. In particular, for each signal, the temporal separation \(m\) between the encoded samples is determined from the \(1/e\) crossing of its autocorrelation function. The same training and testing intervals (\(5\text{--}210~\mu\mathrm{s}\) and
\(210\text{--}300~\mu\mathrm{s}\), respectively) and normalization procedure are also employed. For the one-to-one setting, the input window for a target signal \(u[n]\) therefore takes the form of
\(		\bm{\mathcal{X}}[n]
{=}
\left[
u[n{-}m],
u[n{-}2m],
\ldots,
u[n{-}N_{\mathrm{steps}}m]
\right]^{\mathsf{T}}\),
and the QEL model predicts the future value \(\tilde{u}[n]\). Since only one signal is encoded, the total number of qubits is \(N_q{=}N_{\mathrm{steps}}{+}N_{\mathrm{hid}}\),
rather than \(N_q{=}4N_{\mathrm{steps}}{+}N_{\mathrm{hid}}\) as in the four-to-one setting.

Figure~\ref{fig:timeseries1to1} shows the best predictions obtained with the one-to-one uniform discretization scheme. For each signal, the prediction shown corresponds to the Hamiltonian realization with the lowest RMSE within the optimal reservoir configuration. The optimal configuration is determined from the parameter sweep by varying \(N_{\mathrm{steps}}\) and \(N_{\mathrm{hid}}\), with \(N_{\mathrm{hid}}{=}1,\ldots,10\), while restricting the total reservoir size to \(N_q{\leq}11\). The left and right columns correspond to the theoretical and experimental signals, respectively. Panels (a,b), (c,d), (e,f), and (g,h) show the predictions for \(u_{C_1}\), \(u_{C_2}\), \(i_{L_1}\), and \(i_{L_2}\), respectively. The reservoir-configuration sweep used to identify these optimal configurations is summarized in Fig.~\ref{fig:heatmap1to1}. For each combination of \(N_{\mathrm{steps}}\) and \(N_{\mathrm{hid}}\), 100 independent realizations of the reservoir Hamiltonian are generated, and the RMSE shown in the heatmap is averaged over these realizations. For each signal, the configuration yielding the lowest average RMSE is identified as the optimal configuration used to obtain the corresponding prediction in Fig.~\ref{fig:timeseries1to1}. Although there is a consistent improvement in the prediction by increasing the number of qubits, the four-to-one scenario has a better performance due to exploiting the interconnected features of the signals together.

\begin{figure}[h]
	\centering
	\scalebox{0.92}{%
		\begin{minipage}{\textwidth}
			\centering
			%		\vspace{-20pt}
			% ==================== Shared legend ====================
			\includegraphics[
			width=0.25\textwidth,
			trim=0 0 0 0,
			clip
			]{figures/nonmaplike_legend.png}
			
			\vspace{-18pt}
			
			% ==================== Left column: Theory ====================
			\begin{minipage}[t]{0.48\textwidth}
				\vspace{0pt}
				\begin{tcolorbox}[
					enhanced,
					colback=white,
					colframe=black,
					boxrule=0.6pt,
					arc=0pt,
					left=3pt,
					right=3pt,
					top=7pt,
					bottom=3pt,
					title=Theory,
					fonttitle=\small\bfseries,
					coltitle=black,
					colbacktitle=white,
					attach boxed title to top left={
						xshift=8pt,
						yshift=-2mm
					},
					boxed title style={
						colback=white,
						colframe=white,
						boxrule=0pt,
						left=2pt,
						right=2pt,
						top=0pt,
						bottom=0pt
					}
					]
					\centering
					
					% ---------- u_C1 ----------
					\begin{overpic}[
						width=\textwidth
						]{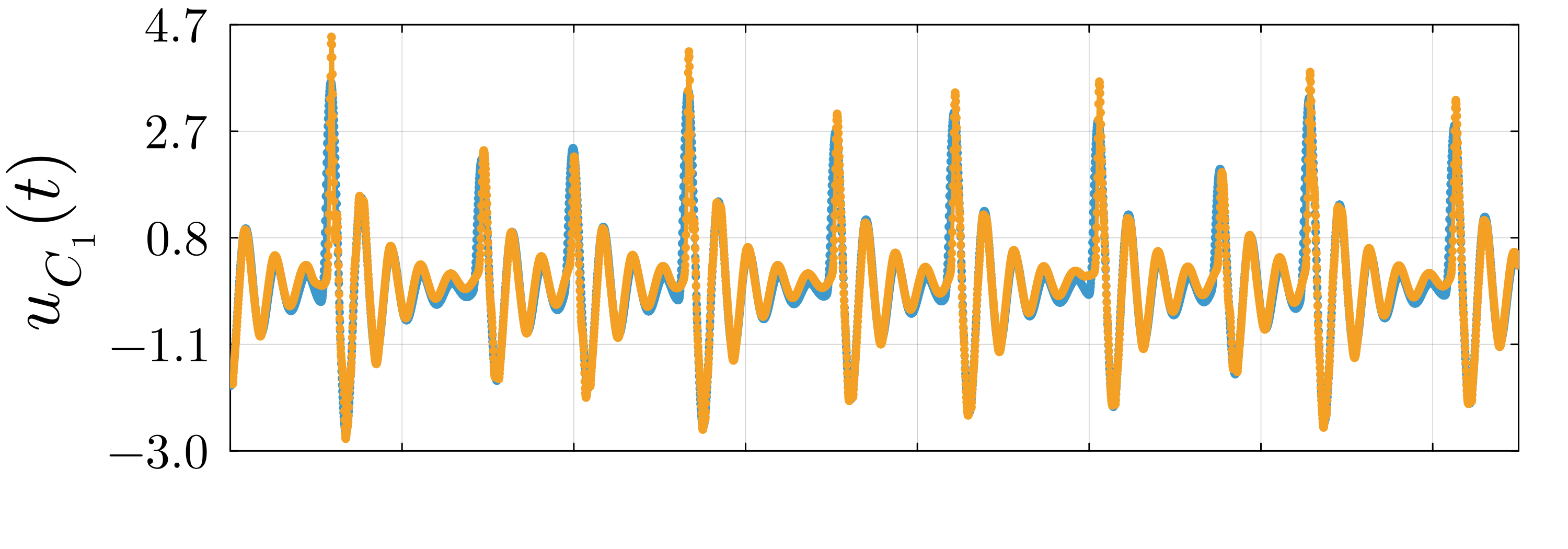}
						\put(-2,32){{(a)}}
					\end{overpic}
					
					\vspace{-5pt}
					
					% ---------- u_C2 ----------
					\begin{overpic}[
						width=\textwidth
						]{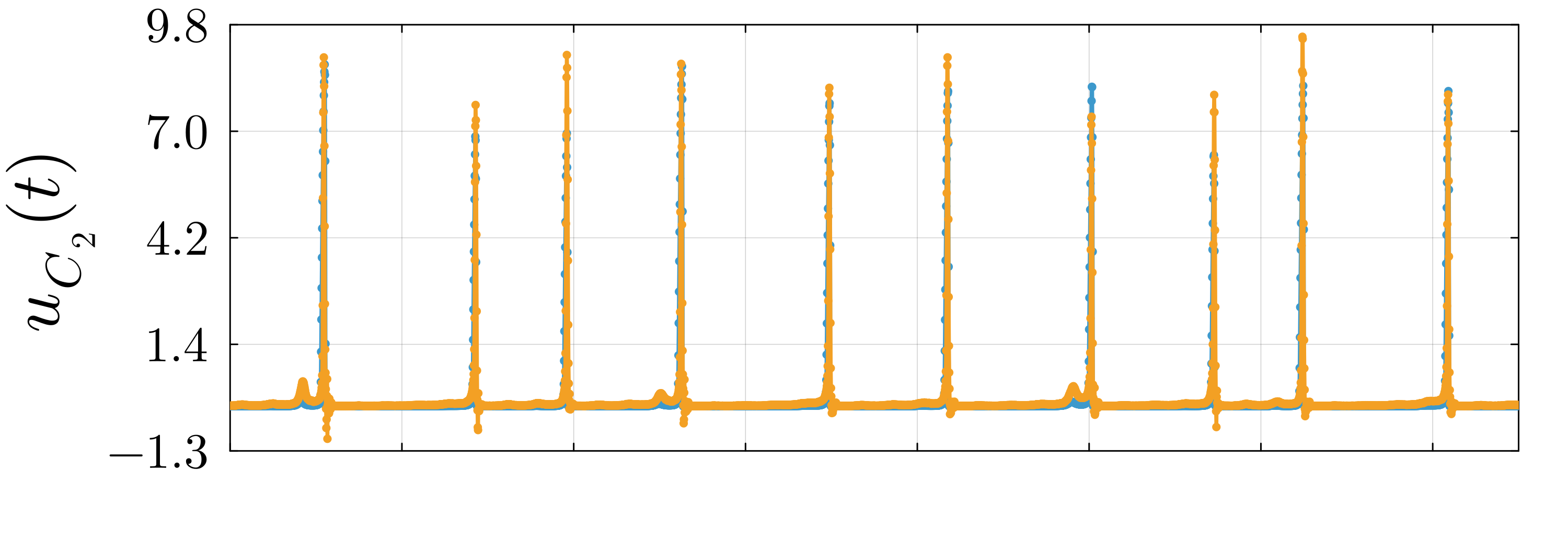}
						\put(-2,32){{(c)}}
					\end{overpic}
					
					\vspace{-5pt}
					
					% ---------- i_L1 ----------
					\begin{overpic}[
						width=\textwidth
						]{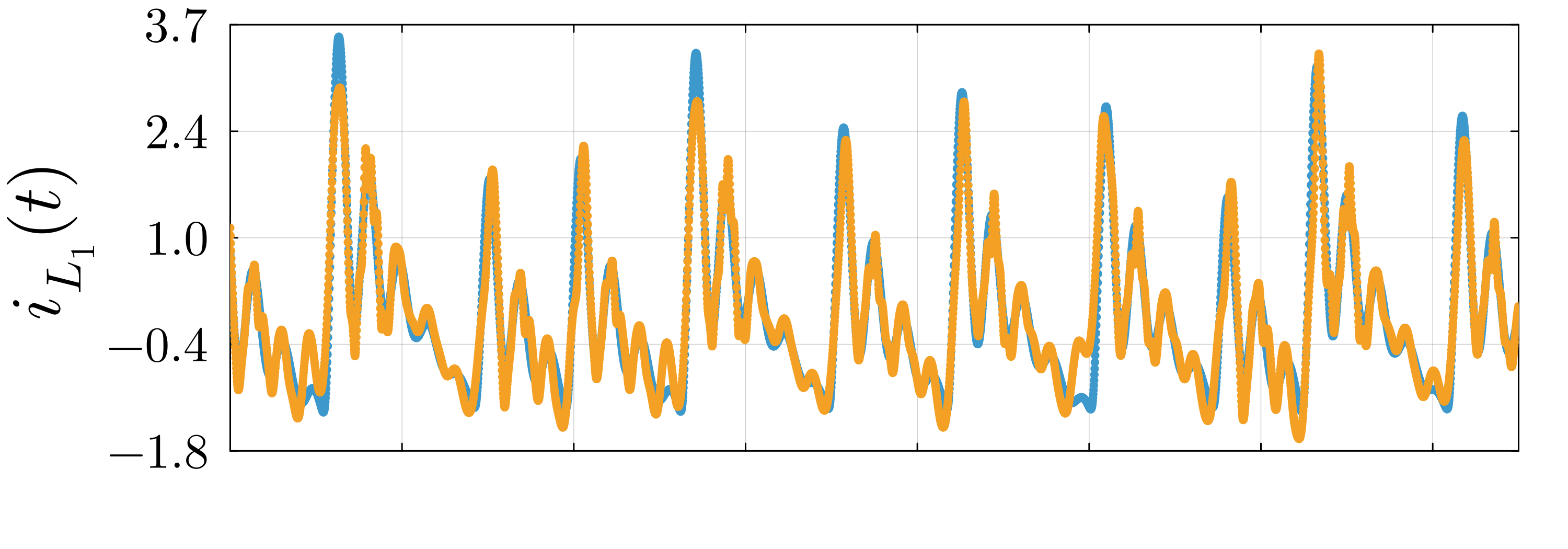}
						\put(-2,32){{(e)}}
					\end{overpic}
					
					\vspace{-5pt}
					
					% ---------- i_L2 ----------
					\begin{overpic}[
						width=\textwidth
						]{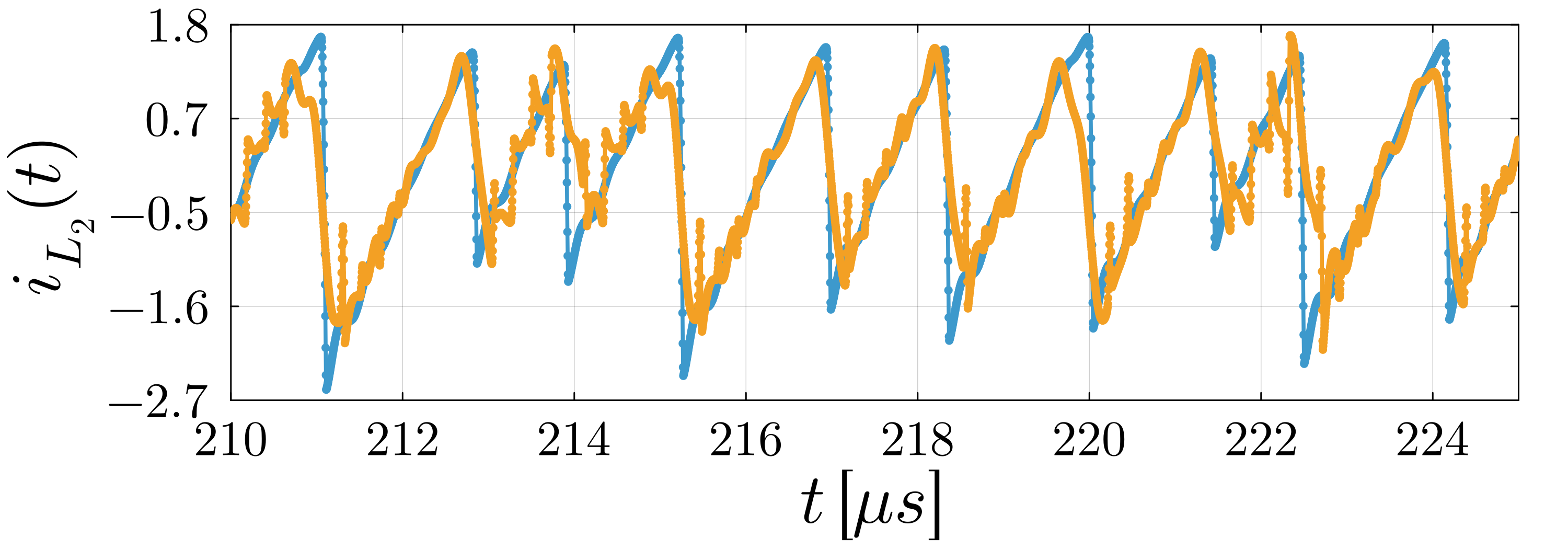}
						\put(-2,32){{(g)}}
					\end{overpic}
					
				\end{tcolorbox}
			\end{minipage}
			\hfill
			% ==================== Right column: Experiment ====================
			\begin{minipage}[t]{0.48\textwidth}
				\vspace{0pt}
				\begin{tcolorbox}[
					enhanced,
					colback=white,
					colframe=black,
					boxrule=0.6pt,
					arc=0pt,
					left=3pt,
					right=3pt,
					top=7pt,
					bottom=3pt,
					title=Experiment,
					fonttitle=\small\bfseries,
					coltitle=black,
					colbacktitle=white,
					attach boxed title to top left={
						xshift=8pt,
						yshift=-2mm
					},
					boxed title style={
						colback=white,
						colframe=white,
						boxrule=0pt,
						left=2pt,
						right=2pt,
						top=0pt,
						bottom=0pt
					}
					]
					\centering
					
					% ---------- u_C1 ----------
					\begin{overpic}[
						width=\textwidth
						]{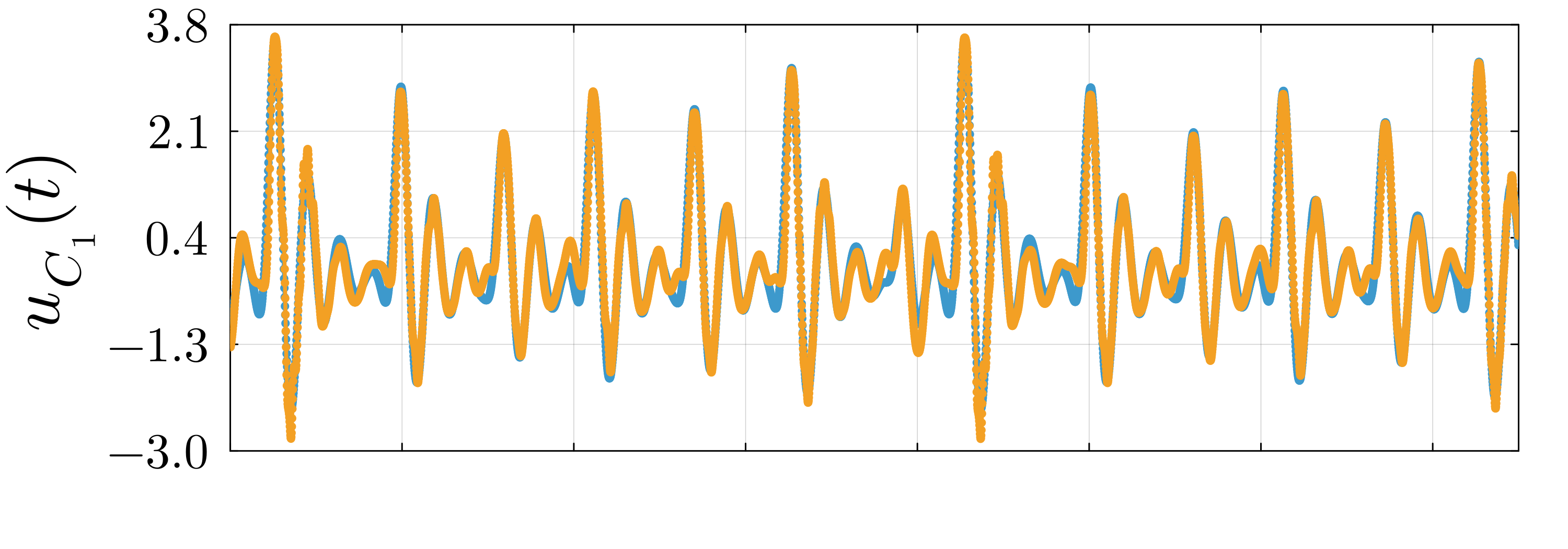}
						\put(-2,32){{(b)}}
					\end{overpic}
					
					\vspace{-5pt}
					
					% ---------- u_C2 ----------
					\begin{overpic}[
						width=\textwidth
						]{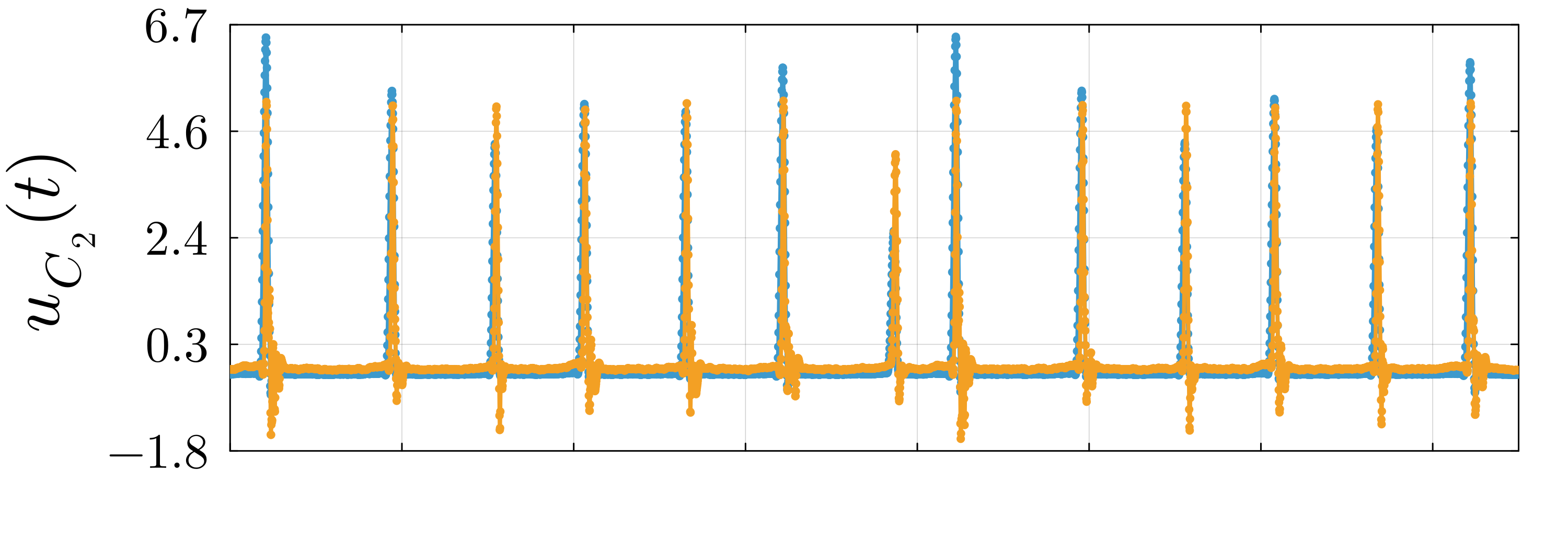}
						\put(-2,32){{(d)}}
					\end{overpic}
					
					\vspace{-5pt}
					
					% ---------- i_L1 ----------
					\begin{overpic}[
						width=\textwidth
						]{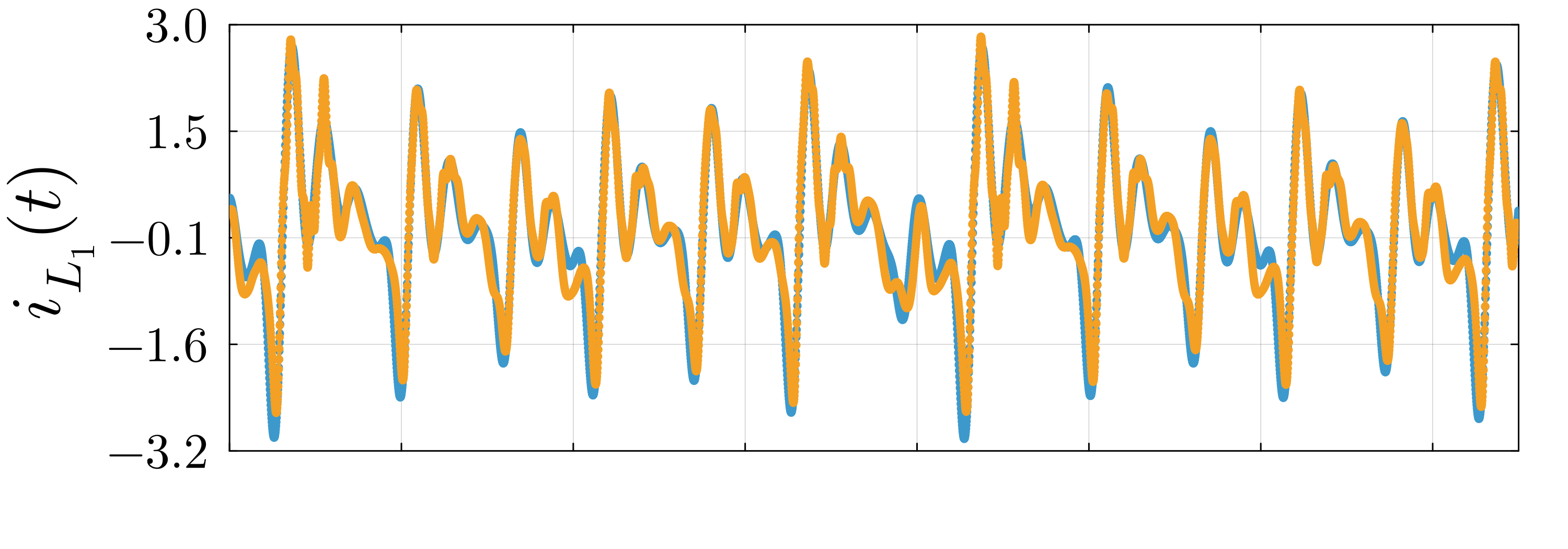}
						\put(-2,32){{(f)}}
					\end{overpic}
					
					\vspace{-5pt}
					
					% ---------- i_L2 ----------
					\begin{overpic}[
						width=\textwidth
						]{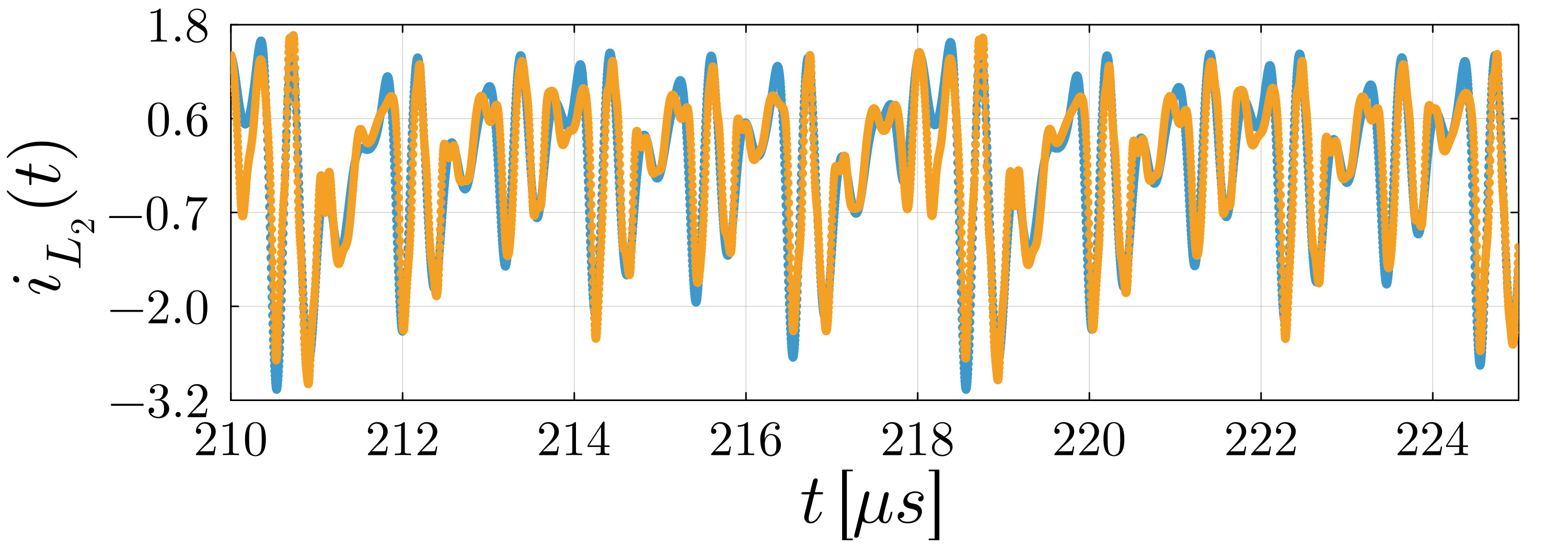}
						\put(-2,32){{(h)}}
					\end{overpic}
					
				\end{tcolorbox}
			\end{minipage}
		\end{minipage}%
	}
	\caption{
		Prediction results of uniform discretization with one-to-one prediction setting.
		The left and right columns show the theoretical and
		experimental results, respectively:
		(a,b) $u_{C_1}$,
		(c,d) $u_{C_2}$,
		(e,f) $i_{L_1}$, and
		(g,h) $i_{L_2}$.
	}
	\label{fig:timeseries1to1}
\end{figure}

\begin{figure}[t]
	\centering
	\begin{minipage}{1.0\textwidth}

		\centering
		% =========================================================
		% Theory group
		% =========================================================
		\begin{tcolorbox}[
			enhanced,
			colback=white,
			colframe=black,
			boxrule=0.6pt,
			arc=0pt,
			left=3pt,
			right=3pt,
			top=1pt,
			bottom=-8pt,
			title=Theory,
			fonttitle=\small\bfseries,
			coltitle=black,
			colbacktitle=white,
			attach boxed title to top left={
				xshift=8pt,
				yshift=-2mm
			},
			boxed title style={
				colback=white,
				colframe=white,
				boxrule=0pt,
				left=2pt,
				right=2pt,
				top=0pt,
				bottom=0pt
			}
			]
			\centering
			
			\begin{subfigure}{.235\linewidth}
				\centering
				\begin{overpic}[width=\textwidth]{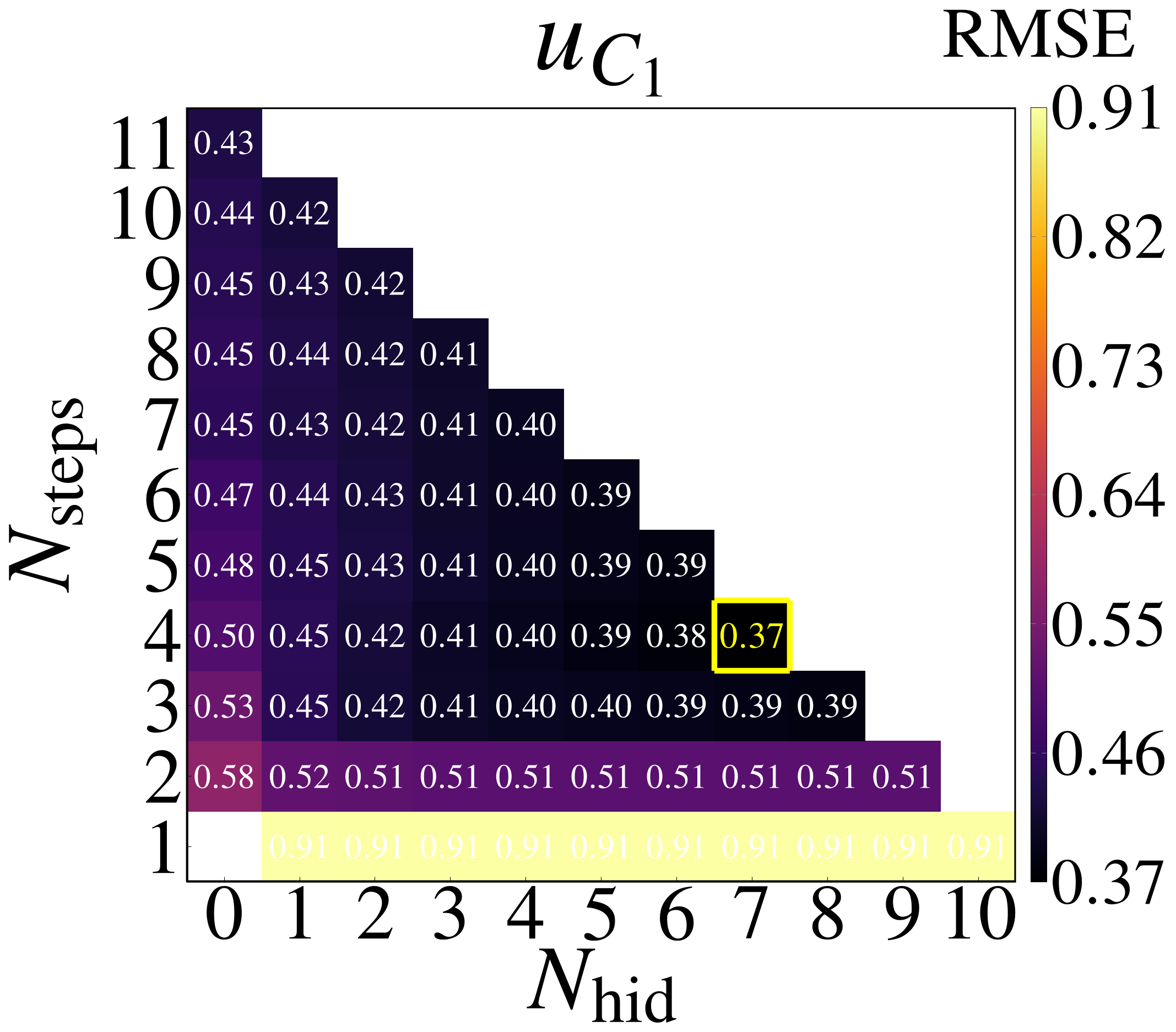}
					\put(-4,82.4){{(a)}}
				\end{overpic}
				\label{fig:u1heatmap4to1theory_maplike}
			\end{subfigure}
			\hfill
			\begin{subfigure}{.235\linewidth}
				\centering
				\begin{overpic}[width=\textwidth]{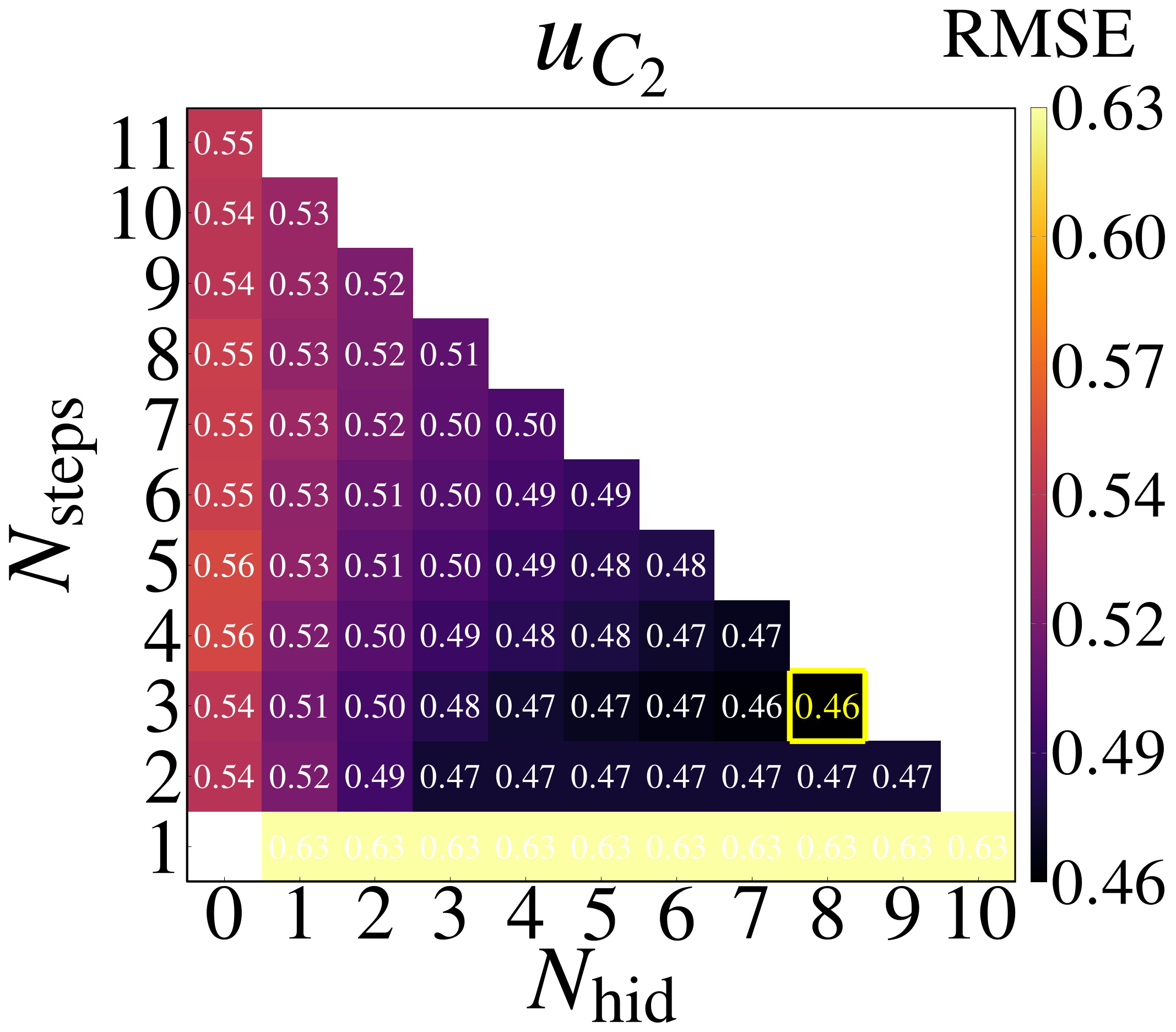}
					\put(-4,82.4){{(c)}}
				\end{overpic}
				\label{fig:u2heatmap4to1theory_maplike}
			\end{subfigure}
			\hfill
			\begin{subfigure}{.235\linewidth}
				\centering
				\begin{overpic}[width=\textwidth]{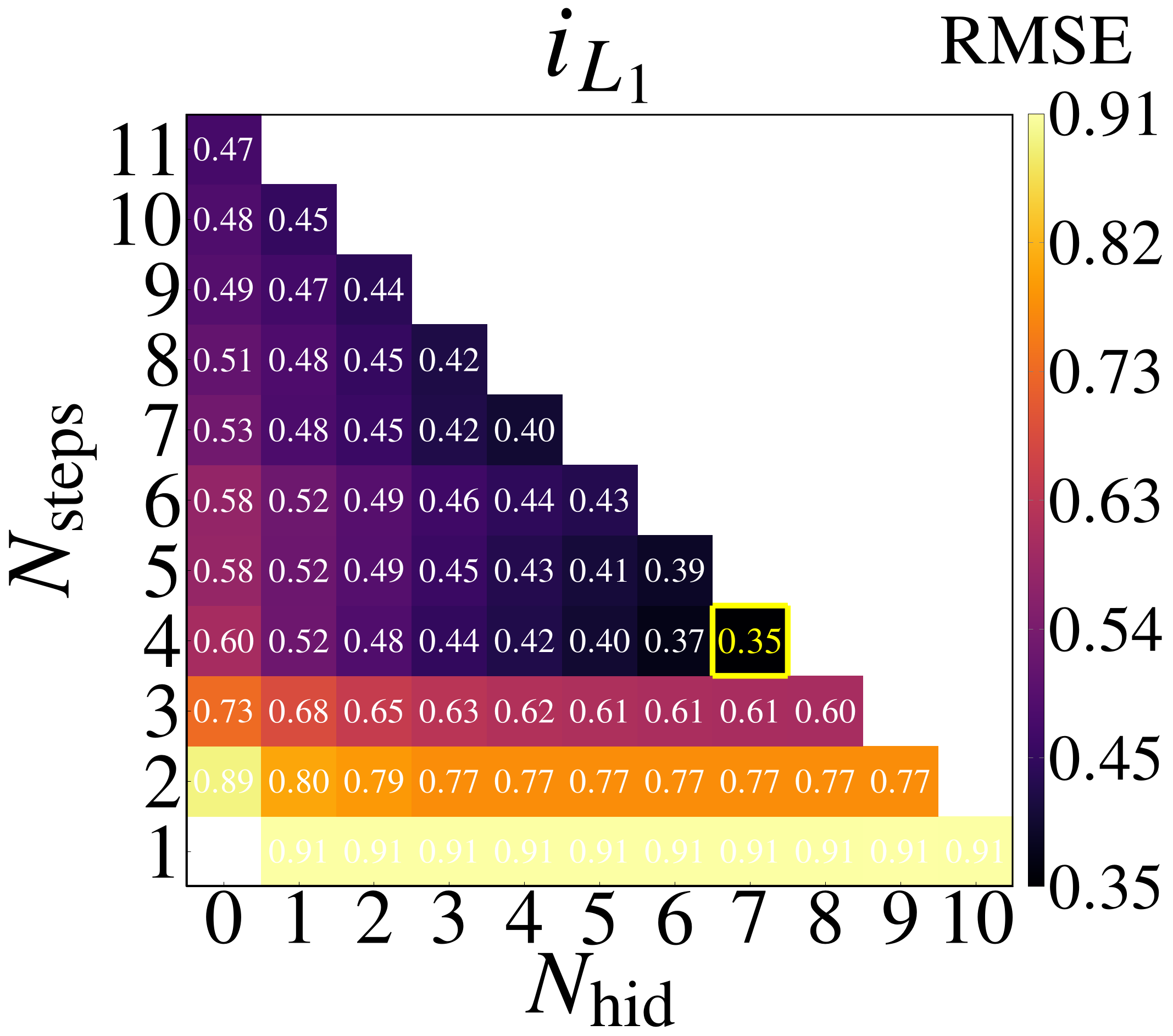}
					\put(-4,83){{(e)}}
				\end{overpic}
				\label{fig:i1heatmap4to1theory_maplike}
			\end{subfigure}
			\hfill
			\begin{subfigure}{.235\linewidth}
				\centering
				\begin{overpic}[width=\textwidth]{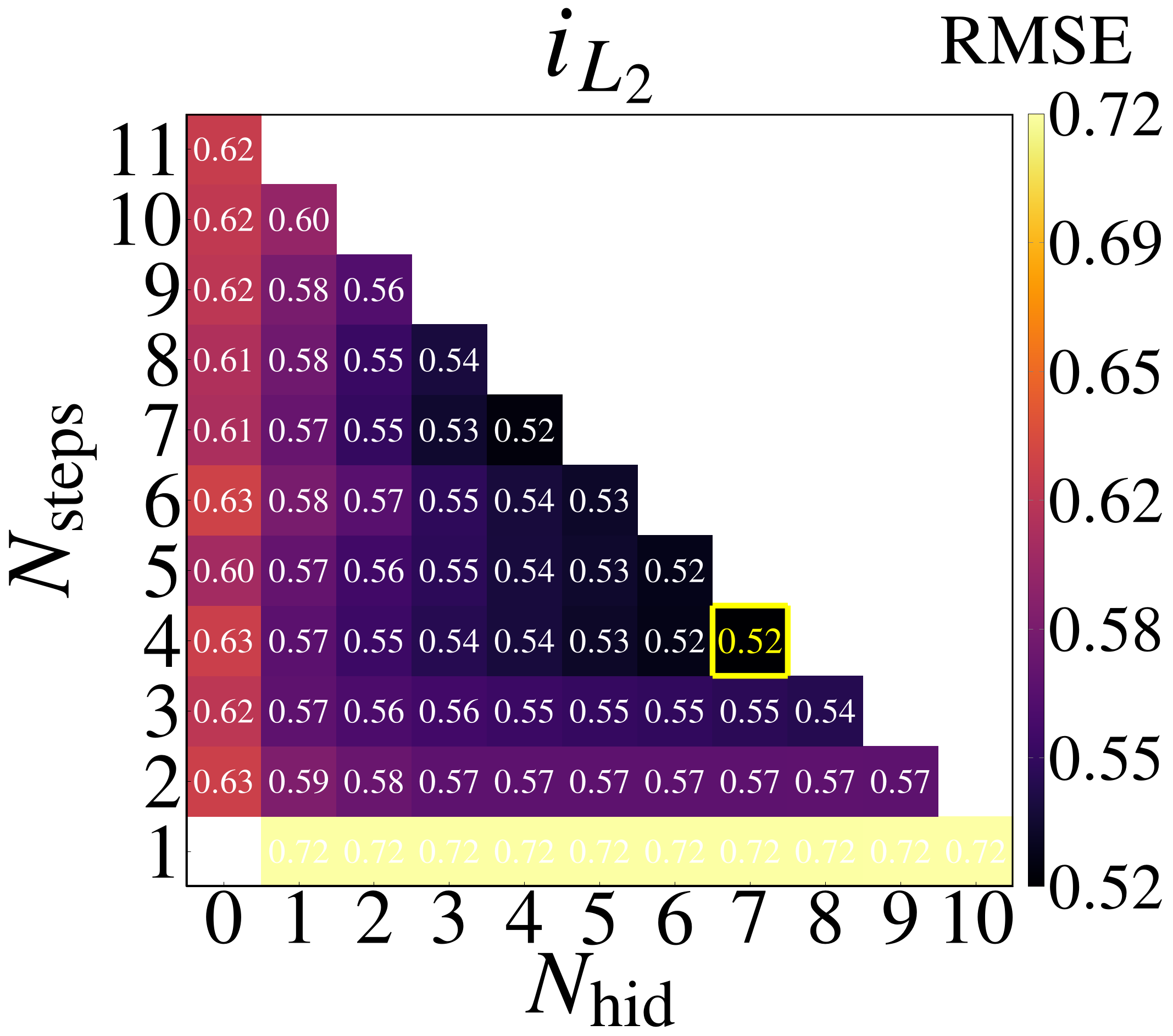}
					\put(-4,83){{(g)}}
				\end{overpic}
				\label{fig:i2heatmap4to1theory_maplike}
			\end{subfigure}
		\end{tcolorbox}
		
		\vspace{-0.25cm}
		
		% =========================================================
		% Experiment group
		% =========================================================
		\begin{tcolorbox}[
			enhanced,
			colback=white,
			colframe=black,
			boxrule=0.6pt,
			arc=0pt,
			left=3pt,
			right=3pt,
			top=1pt,
			bottom=-8pt,
			title=Experiment,
			fonttitle=\small\bfseries,
			coltitle=black,
			colbacktitle=white,
			attach boxed title to top left={
				xshift=8pt,
				yshift=-2mm
			},
			boxed title style={
				colback=white,
				colframe=white,
				boxrule=0pt,
				left=2pt,
				right=2pt,
				top=0pt,
				bottom=0pt
			}
			]
			\centering
			
			\begin{subfigure}{.235\linewidth}
				\centering
				\begin{overpic}[width=\textwidth]{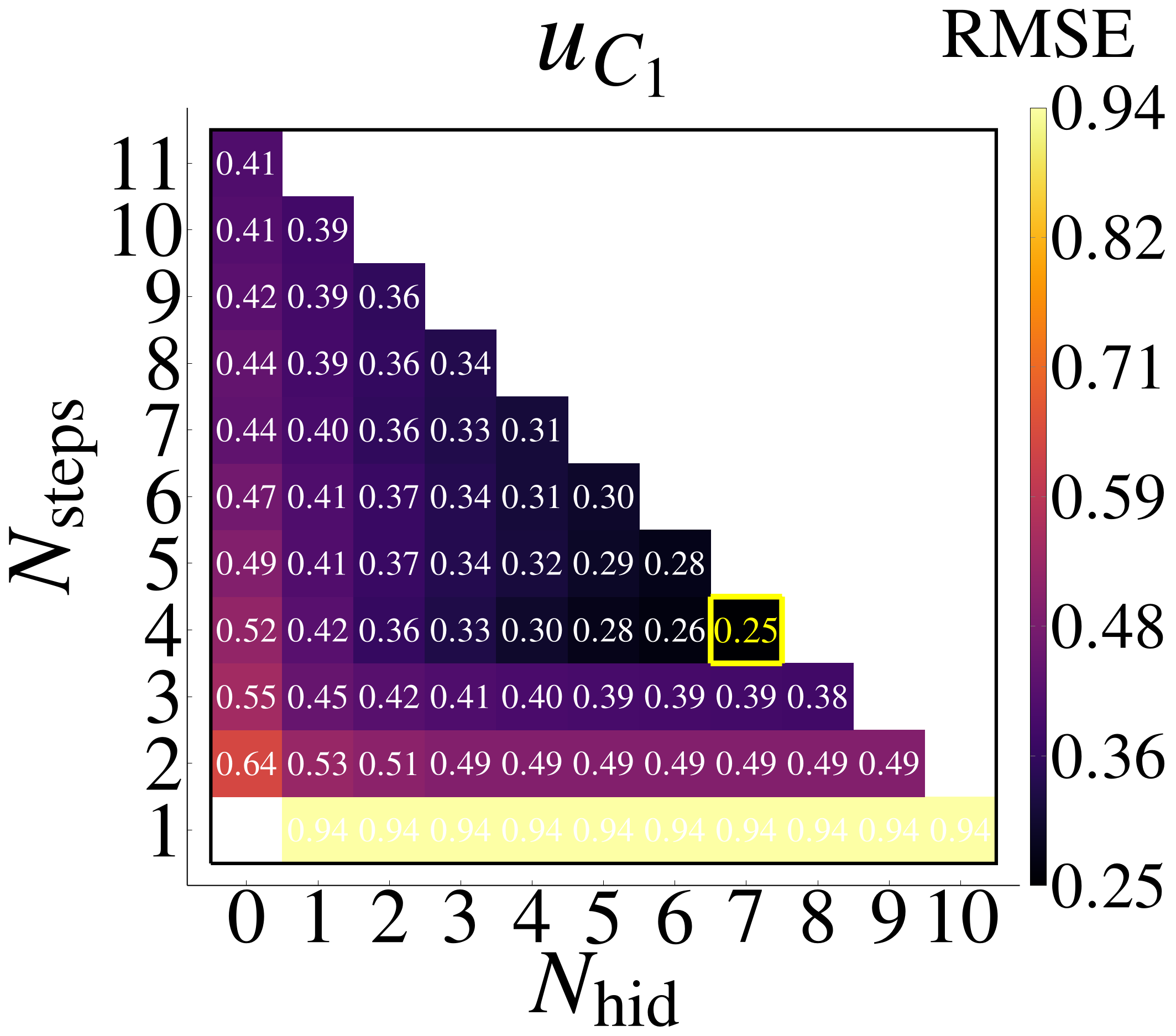}
					\put(-4,82.4){{(b)}}
				\end{overpic}
				\label{fig:u1heatmap4to1experiment_maplike}
			\end{subfigure}
			\hfill
			\begin{subfigure}{.235\linewidth}
				\centering
				\begin{overpic}[width=\textwidth]{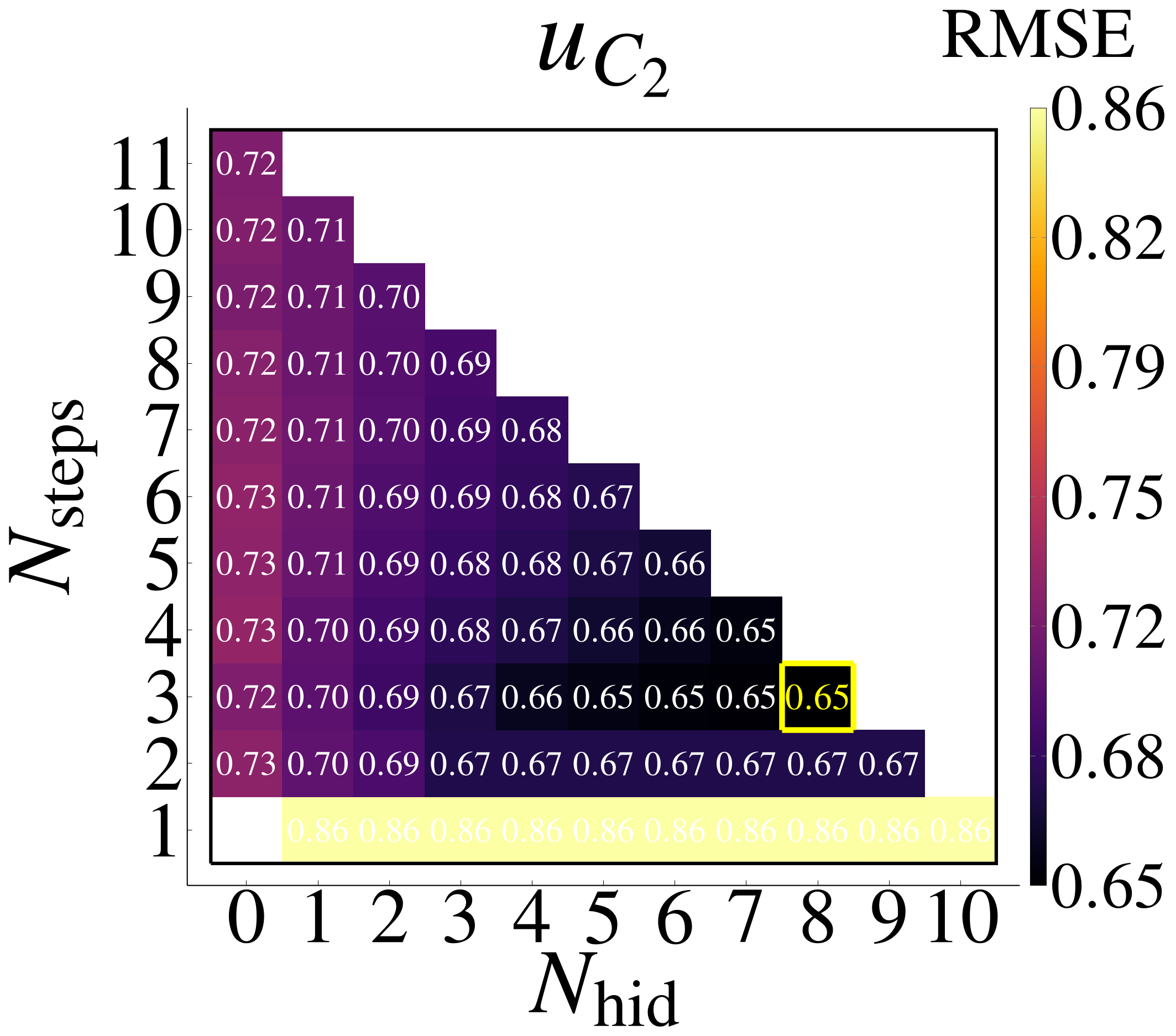}
					\put(-4,82.4){{(d)}}
				\end{overpic}
				\label{fig:u2heatmap4to1experiment_maplike}
			\end{subfigure}
			\hfill
			\begin{subfigure}{.235\linewidth}
				\centering
				\begin{overpic}[width=\textwidth]{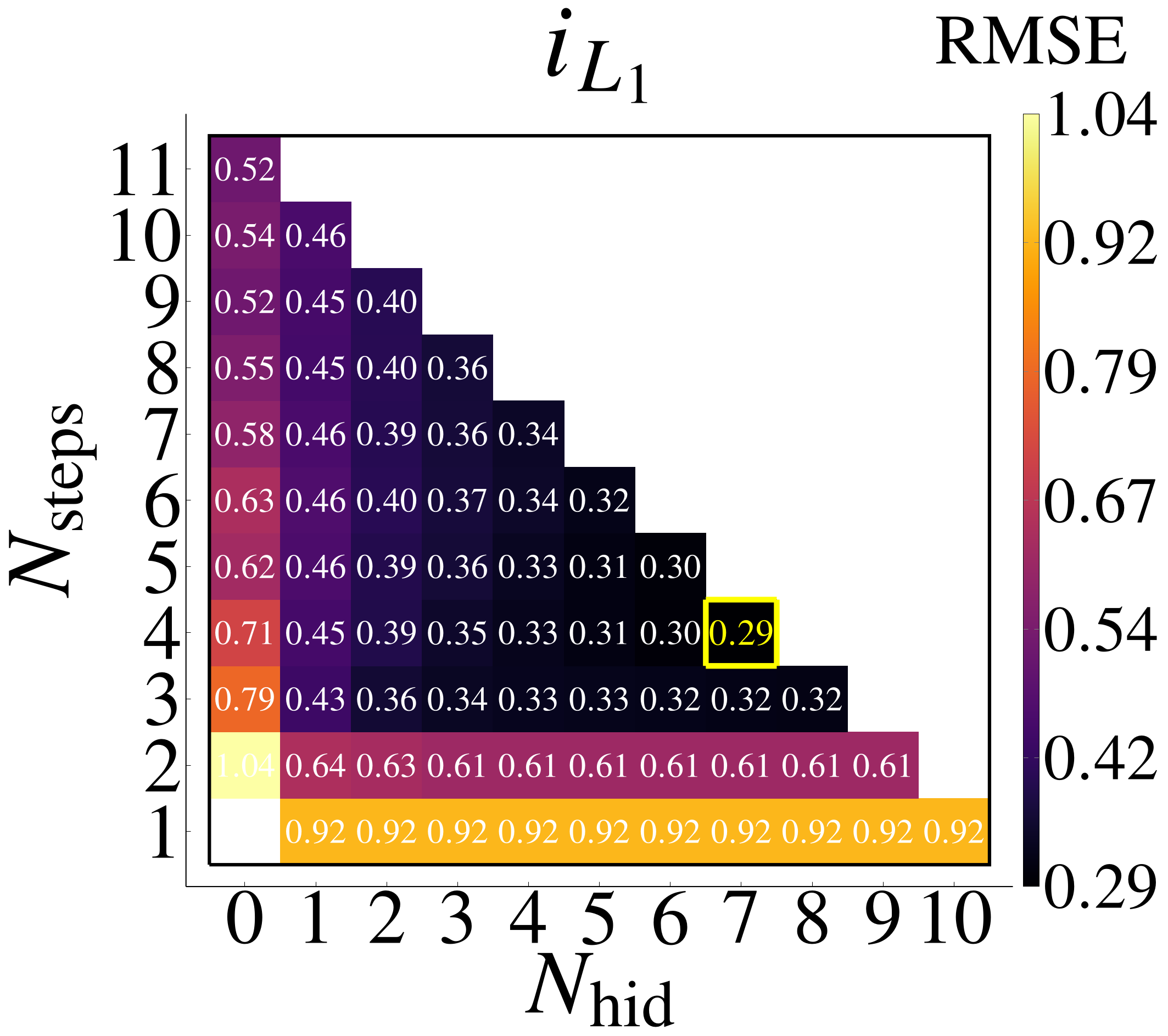}
					\put(-4,82.4){{(f)}}
				\end{overpic}
				\label{fig:i1heatmap4to1experiment_maplike}
			\end{subfigure}
			\hfill
			\begin{subfigure}{.235\linewidth}
				\centering
				\begin{overpic}[width=\textwidth]{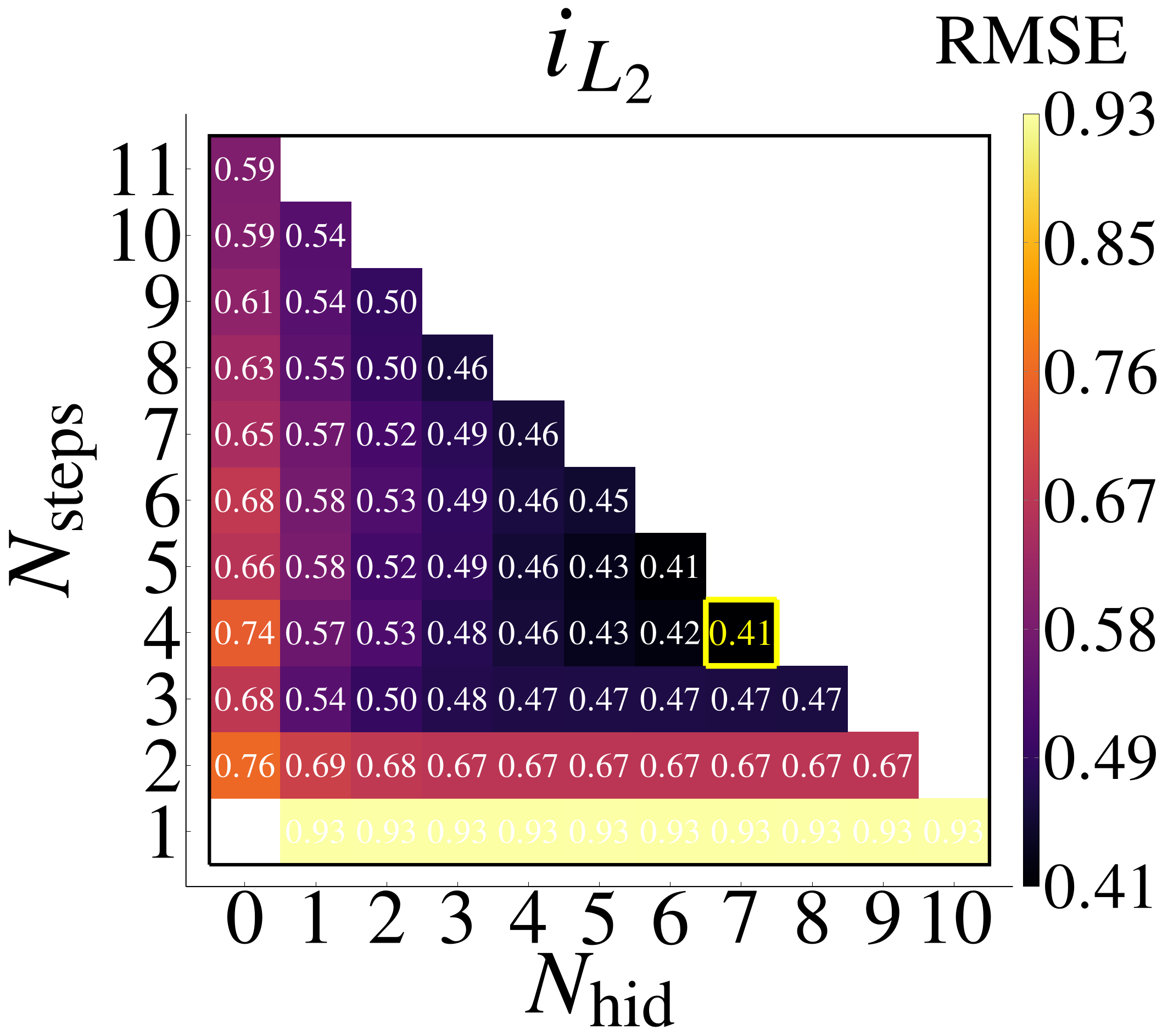}
					\put(-4,82.4){{(h)}}
				\end{overpic}
				\label{fig:i2heatmap4to1experiment_maplike}
			\end{subfigure}
		\end{tcolorbox}
	\end{minipage}
	\caption{Performance heatmap for uniform discretization with  one-to-one prediction setting.
		The top row shows the theoretical results and the bottom row shows the experimental results.
		Panels (a,b), (c,d), (e,f), and (g,h) correspond to $u_{C_1}$, $u_{C_2}$, $i_{L_1}$, and $i_{L_2}$, respectively.}
	\label{fig:heatmap1to1}
\end{figure}

\end{document}